\documentclass[twocolumn,superscriptaddress,amsmath,amssymb,aps,floatfix]{revtex4-1}

\usepackage{graphicx}
\usepackage{slashed}
\usepackage{epstopdf}
\usepackage{bm}
\usepackage{hyperref}
\usepackage{graphicx,color}
\usepackage{multirow}
\usepackage{amsmath,amstext,array}
\usepackage[normalem]{ulem}

\newcolumntype{C}{>{$}c<{$}}

\begin{document}
\title{ 
Challenges with Internal Photons in Constructive QED}
\author{Neil Christensen}
\email{nchris3@ilstu.edu}
\author{Harold Diaz-Quiroz}
\affiliation{Department of Physics, Illinois State University, Normal, IL 61790}
\author{Bryan Field}
\email{New email address: bryan.field@science.doe.gov}
\affiliation{Department of Physics, Farmingdale State College, Farmingdale, NY, 11735}

\author{Justin Hayward}
\author{John Miles}
\affiliation{Department of Physics, Illinois State University, Normal, IL 61790}

\date{\today}

\begin{abstract}
We find the correct spinor amplitude for a simple photon-mediated process and show that, in contrast, the result for the same process using the standard constructive techniques do not agree with Feynman diagrams when the fermions are massive.  Along the way, we analyze the $x$ factor used in photon vertices, we work out the spinor shifts for massive particles when the momenta are analytically continued and we consider the large $z$ limit of the amplitudes in this paper and show that the photon-mediated process does not vanish in this limit for any choice of two of its momenta.  For comparison with the photon-mediated process, we also describe two processes with external photons that are mediated by massive particles.  In both cases, we show that the current techniques are sufficient and that the final results agree with Feynman diagrams.  We also demonstrate that by using a massive photon in our calculations and taking the massless limit at the end, we can achieve agreement with Feynman diagrams in all the processes discussed here, including the photon-mediated amplitudes.
\end{abstract}

\maketitle

%\tableofcontents

For nearly a century, Feynman diagrams resulting from field theory have, in principle, given a complete solution to the calculation of perturbative scattering amplitudes.  However, with experiments reaching ever higher collision energies and ever greater precision in their measurements of the resulting final state particles, the calculation of the relevant and required high-multiplicity and higher-loop Feynman diagrams has become challenging and sometimes impossible, even for computers.  Partly as a result of this, some have begun looking into alternative ways of doing these amplitude calculations and have discovered some remarkable, and at times astounding, simplifications to both the final formulas as well as the intermediate recipes that give them.  Among them, one of the most profound was the discovery that the maximally helicity-violating amplitude for gluons could be written as a single term on the back of an envelope no matter how many thousands or millions of Feynman diagrams would be required to achieve the same result \cite{Parke:1986gb}.  Another is the discovery of a complete set of recursion relations for building up any gluon amplitude with any helicity combination using a simple set of 3-point vertices and an on-shell combination technique that completely bypasses both field theory and Feynman diagrams, removes the gauge symmetry and the need for gauge invariance which is trivially satisfied, produces a final result which is many orders of magnitude simpler than the Feynman-diagram result, yet equals it exactly for all energies \cite{Britto:2005fq}.  With these results and others, ``constructive'' techniques, that bypass field theory, have become increasingly important for calculations of massless scattering amplitudes \cite{Gastmans:1990xh, Dixon:1996wi, Cachazo:2004kj, Britto:2004ap, Bern:2007dw, Berger:2008sj, Bern:2008qj, Elvang:2013cua}. 

In order to extend this methodology to massive theories such as the standard model (SM), the authors of \cite{Arkani-Hamed:2017jhn} generalized the concept of a helicity spinor to a spin spinor, an object that transformed under a product of the spin little group, rather than the helicity little group, as well as the Lorentz group.  This allowed them to write generalized 3-point vertices for particles of any mass and any spin and describe the constructive process of combining these vertices with propagators to obtain 4-point amplitudes and beyond.  In principle, with this breakthrough, it appeared that it was now possible to apply the full constructive apparatus to massive theories such as the SM and a flurry of calculations were performed with their generalization.  A selection follows.  In \cite{Ochirov:2018uyq}, helicity amplitudes for QCD with massive quarks were performed.  The on-shell constructability of Born amplitudes was investigated using these methods in  \cite{Franken:2019wqr}.  In \cite{Aoude:2019tzn,Durieux:2019eor}, on-shell amplitudes in standard model effective theory are considered.  Some comments on massive spinors were made in \cite{Heuson:2019yej}.  The Higgs mechanism was studied in this formalism in \cite{Bachu:2019ehv,Balkin:2021dko}.  Higher-dimensional operators were considered in \cite{Durieux:2019siw}.  A discussion in the context of gravity can be found in \cite{Falkowski:2020mjq}.  Renormalization of higher-dimensional operators is discussed in \cite{Baratella:2020lzz}.  Four-point contact terms are considered in \cite{Durieux:2020gip}.  In \cite{Alves:2021rjc}, the neutrino sector was studied.  Aspects of gauge invariance were studied in \cite{Liu:2022alx}.  Dark matter was considered in \cite{Salla:2022dxc}.  Simple decay amplitudes of the SM are considered in \cite{Christensen:2019mch}.     

In addition to these papers, in \cite{Christensen:2018zcq} we catalogued the complete set of 3-point vertices in the SM with the intention to begin calculating its 4- (and higher-) point amplitudes.  Unfortunately, as we began calculations in the electroweak sector, we kept running into difficulties with diagrams that contained internal photon lines.  In particular, we were unable to achieve agreement with Feynman diagrams when internal massless photons were involved, using these constructive techniques.  As we attempted to resolve the discrepancy, we studied the photon vertices and the $x$ factor involved in these vertices, we generalized the analytic continuation of the momenta and its accompanying shift in the spin spinors, we looked at the asymptotic limit of the amplitudes when the complex parameter $z\to\infty$ and performed many calculations and comparisons.  In order to further clarify the challenge, we simplified to the simplest theory with this challenge, quantum electrodynamics (QED), and the simplest $4$-point amplitudes that contained the photon, and although we learned much and further developed the spinor shift, none of these extensions were able to bring the constructive amplitude for our simple internal-photon-mediated processes into agreement with Feynman diagrams.  On the other hand, we did find agreement with Feynman diagrams when the photon was an external line with the standard techniques and we separately found that if we first gave the photon a mass and calculated the amplitudes in a purely massive theory and then took the massless-photon limit at the end, we could find agreement with Feynman diagrams for all our processes.  This note describes the progress we made in understanding a simple photon-mediated process as well as the tools involved.  To summarize, we find that the present constructive tools are sufficient for purely massless theories and appear to be sufficient for purely massive theories, but are not sufficient to calculate all amplitudes that contain both massive and massless particles.  In particular, we find that the tools are sufficient when the massless particles are on the external lines but not necessarily when they are on internal lines.  In order to clearly and explicitly show where the challenges are, we consider the simplest theory with both massive and massless particles, namely QED, and we discuss the simplest $4$-point amplitudes in this theory.  We report the correct spinor amplitudes, describe ways of finding them, show where the current published results and methods are in agreement and also where the current published results and methods are not in agreement.  

In particular, we analyze three simple processes. The first is a process with an internal photon, namely $e\bar{e}\mu\bar{\mu}$, where all particles are taken to be incoming throughout this article.  We choose this over $e\bar{e}e\bar{e}$ because it only involves one diagram and therefore cannot involve a cancellation between diagrams.  Nevertheless, we do note how to obtain the result for the process $e\bar{e}e\bar{e}$ from the result for $e\bar{e}\mu\bar{\mu}$.  We will find that the present tools are unable to correctly obtain this amplitude and we will discuss how we obtained it using an intermediately massive photon.  Our next process will be one with only one external photon.  Since QED does not actually have such a process at four points, we include an external Higgs as a minor extension of QED that allows us to calculate the process $e\gamma\bar{e}h$.  We find that the current methods are able to correctly obtain this amplitude, but discuss other ways as well.  Finally, we calculate the process $e\bar{e}\gamma\gamma$, and once again show that the current tools are sufficient and discuss other ways again.

In order to do this, we first develop our tools.  In Sec.~\ref{sec:main:MA}, we describe the calculation of QED amplitudes with a massive photon and take the massless limit at the end of the calculation.  We show that this method works for all the amplitudes of this note, including the process $e\bar{e}\mu\bar{\mu}$ with an internal photon, where the purely massless methods do not work.  In Sec.~\ref{sec:main:x}, we briefly review the $x$ factor and its identities that are used in the massless-photon calculations.  In Sec.~\ref{sec:main:constructive no Shift}, we calculate our amplitudes using $x$  and the standard techniques and show that we do not obtain the correct result for $e\bar{e}\mu\bar{\mu}$, but we do obtain the correct result for $e\gamma\bar{e}h$ and $e\bar{e}\gamma\gamma$.  In Sec.~\ref{sec:main:spinor shifts}, we analytically continue two of the momenta.  In the purely massless theory, this is an essential ingredient to this constructive method.  However, the spinor shifts that accompany the complex momentum were only known for the helicity spinors.  In this section, we generalize this spinor shift to the spin spinors of massive theories.  In Sec.~\ref{sec:main:large z}, we consider the large $z$ limit of the amplitudes, where $z$ is the complex number in the analytic continuation of the momenta.  We find that the process $e\bar{e}\mu\bar{\mu}$ does not vanish for any choice of momentum complexification and that perhaps this is the reason the standard tools do not succeed in finding the correct result.  On the other hand, the processes $e\gamma\bar{e}h$ and $e\bar{e}\gamma\gamma$ do vanish for a variety of complex momenta, which is likely why the standard method works for them.  In Sec.~\ref{sec:main:constructive}, we recalculate the amplitudes using $x$ and the standard tools of constructive theory, but with the addition of the spinor shifts just described.  We find that these spinor shifts remove the ambiguity in the process $e\bar{e}\mu\bar{\mu}$, but that it still is not in agreement with the correct result.  On the other hand, we find that the processes $e\gamma\bar{e}h$ and $e\bar{e}\gamma\gamma$ still give correct amplitudes when using the spinor shifts.  In Sec.~\ref{sec:main:amplitudes}, we further discuss the amplitudes.  We summarize and conclude in Sec.~\ref{sec:conclusions}.

In the main section of this paper, we have tried to keep the details to a minimum to aid readability and to see the big picture.  However, we have also written a series of appendices that give a great deal more of the details for the interested reader.  These appendices follow the structure of Sec.~\ref{sec:main:main}.  In App.~\ref{sec:Massive Photon}, we calculate the amplitudes using a massive photon and take the massless limit.  We do this for $e\bar{e}\mu\bar{\mu}$ in App.~\ref{sec:eemm spinor amp calculation}, for $e\gamma\bar{e}h$ in App.~\ref{sec:eAeh spinor amp calculation} and for $e\bar{e}\gamma\gamma$ in App.~\ref{sec:Massive eeAA}.  

In App.~\ref{app:x}, we review $x$ and derive the identities used in the calculations in this paper.  In App.~\ref{app:Calc AHH Spinor Amps No Shifts}, we use $x$ and the standard methods and describe the derivations of the amplitudes in detail, including $e\bar{e}\mu\bar{\mu}$, where these methods do not give the correct amplitude.
In App.~\ref{app:analytic continuation}, we review the analytic continuation of the external momenta and the spinor shifts in the massless case.  We then extend this to the massive case.  We first consider the shift $\lbrack i,j\rangle$, where both external particles are massless but the internal particle is massive in App.~\ref{app:complexification 2}.  We then turn to the shift $\lbrack\mathbf{i},j\rangle$, where particle $i$ is massive, particle $j$ is massless and the internal line is massive in App.~\ref{app:[i,j> massive massless}, followed by the shift $\lbrack i,\mathbf{j}\rangle$ where particle $i$ is massless, particle $j$ is massive and the internal line is massive.  Finally, in App.~\ref{app:[i,j> both massive}, we derive the shift $\lbrack\mathbf{i,j}\rangle$, where both external particles are massive but the internal line is massless.  We discuss why we don't describe the shift when both external particles are massive as well as the internal line App.~\ref{app:[i,j> both massive massive internal line}.  The formulas are too complicated to be useful and the already described shifts are sufficient to cover all the cases.  We follow this with a derivation of the large $z$ behavior for the amplitudes with all the possible momentum shifts in App.~\ref{app:large z behavior}.  In particular, we show that there are no shifts that cause the amplitude for $e\bar{e}\mu\bar{\mu}$ to vanish at large $z$, while we do find a variety of shifts with large-$z$ vanishing amplitudes for $e\gamma^{\pm}\bar{e}h$, $e\bar{e}\gamma^{\pm}\gamma^{\pm}$ and $e\bar{e}\gamma^{\pm}\gamma^{\mp}$.  In the final App.~\ref{app:Calc AHH Spinor Amps}, we use the spinor shifts and calculate the constructive amplitudes in detail.  For $e\bar{e}\mu\bar{\mu}$ in App.~\ref{sec:Constructive Massless eemm}, we show how it fails, while in Apps.~\ref{sec:Constructive Massless heAe}, \ref{app:eeA+A+} and \ref{app:eeA+A-}, we show that the amplitudes for $e\gamma^+\bar{e}h$, $e\bar{e}\gamma^+\gamma^+$ and $e\bar{e}\gamma^+\gamma^-$ continue to work when including the spinor shifts.

\section{\label{sec:main:main}Tools, Calculations and 4-Point Amplitudes of QED}

\subsection{\label{sec:main:MA}The Massless Limit of a Massive Theory}
Although we were able to find the amplitudes for $e\gamma\bar{e}h$ and $e\bar{e}\gamma\gamma$ using the $x$-factor vertices of \cite{Arkani-Hamed:2017jhn}, as we will see, we were unable to find the amplitude for $e\bar{e}\mu\bar{\mu}$ using this method (see App.~\ref{sec:Constructive Massless eemm}.)  However, we were able to find all the amplitudes if we replaced the massless photon with a massive photon and took the massless limit at the end.  In fact, we have never been unsuccessful calculating an amplitude if all the particles, both internal and external, are massive.  Moreover, when all the particles are massive, we have not needed to analytically continue the momenta or shift the spinors to get the correct results in any of our calculations so far (including \cite{Christensen:2019mch} as well as the amplitudes in this section).  On the other hand, when using a massive photon, we must calculate and add all the possible diagrams and there aren't any shortcuts that bypass diagrams as there are using a massless photon from the beginning.  Since this is the only method that always gives correct results, we detail it first and discuss the $x$-factor method using a massless photon in the next section.

In this section, we will describe these massive calculations and demonstrate how to calculate the amplitudes of this paper using the massless limit.  
Most of the apparatus to do this is already present as we will see, but first we must modify the photon vertices to their massive form.  The electron- (muon-) photon vertex is just like the $Z$-boson vertex, but non chiral, and given by
\begin{equation}
    \mathcal{M}_{\textrm{eeA}} = 
    \frac{e}{M_A}\left(\langle\mathbf{31}\rangle\lbrack\mathbf{23}\rbrack+\lbrack\mathbf{31}\rbrack\langle\mathbf{23}\rangle\right)\ .
    \label{eq:M_eeA massive vertex}
\end{equation}
There are two unphysical aspects introduced by this form of the vertex.  The first is that the photon is now taken as a spin-1 object and therefore has one extra (unphysical) degree of freedom, somewhat akin to the 2 extra (unphysical) degrees of freedom in field theory.  Furthermore, this unphysical degree of freedom must vanish by the end of the calculation for physical amplitudes, just like it does for field theory (and is proven by the Ward identities in field theory).  Though we don't have a proof that this always occurs in this spinor formulation, we have found that they do cancel in the calculations we have thus far performed.  That is, the final physical amplitudes agree with the results from Feynman diagrams.

Unfortunately, by adding an unphysical degree of freedom, it seems that we have ruined some of the motivation for this constructive formulation of particle physics.  Indeed, although this is true, we believe there might still be some reasons to pursue this approach anyway.  First of all, we see that we have only introduced one unphysical degree of freedom rather than two, so this formulation is still an improvement in principle, although it is still not fully satisfactory.  But, moreover, we might still find that the final resulting amplitudes are superior in their economy and insight compared to the results of Feynman diagrams.  This last point will only be known after many more amplitudes, including loop amplitudes, are worked out and compared to their field-theory equivalents.  At this point, we can only say that it looks promising, but is not conclusive.  

The second point is that this vertex introduces division by the mass of the photon, which will be taken to zero, potentially introducing a singularity.  Once again, as in the case of the extra unphysical degree of freedom, we believe that this singularity is removable and that it will always exactly cancel by the end of the calculation.  As with the unphysical degree of freedom, we do not have a proof of this for all calculations, but we have found it to be true for the physical calculations of this paper, as we will see.

In order to calculate the amplitude, we begin by calculating the massive amplitude using the usual rules.  We then Taylor expand the photon spinors using the formulas given in \cite{Arkani-Hamed:2017jhn} and \cite{Christensen:2019mch}.  For convenience, we copy the required expansions here to quadratic order in the photon mass over its energy,
\begin{equation}
 \langle\mathbf{ij}\rangle
 \langle\mathbf{ik}\rangle = 
    \left[\begin{array}{c}
    \Bigl( 1 - m_i^2/(4E_i^2) \Bigr)
    \langle i\mathbf{j}\rangle\langle i\mathbf{k}\rangle  \\
    m_i
    \Bigl(\langle i\mathbf{j}\rangle\langle \zeta_i^-\mathbf{k}\rangle
         +\langle \zeta_i^-\mathbf{j}\rangle\langle i\mathbf{k}\rangle
    \Bigr) /(2\sqrt{2E_i})  \\
     m_i^2
    \Bigl( \langle \zeta_i^-\mathbf{j}\rangle\langle \zeta_i^-\mathbf{k}\rangle \Bigr)/(2E_i)
    \end{array}\right] 
    %+ \mathcal{O}\left(m_i^3\right)
    \ ,
    \label{eq:<ij><ik> expansion}
\end{equation}
\begin{equation}
    \langle \mathbf{ij} \rangle
    \lbrack \mathbf{ik} \rbrack = 
    \left[\begin{array}{c}
    -m_i
       \langle i\mathbf{j}\rangle
       \lbrack \tilde{\zeta}_i^+\mathbf{k}\rbrack / ( \sqrt{2E_i} ) \\ \\
     \left( 1 - m_i^2/(4E_i^2) \right)
        \langle i\mathbf{j} \rangle
        \lbrack i\mathbf{k} \rbrack \\ 
    -m_i^2
        \langle \zeta_i^- \mathbf{j} \rangle
        \lbrack \tilde{\zeta}_i^+ \mathbf{k} \rbrack / ( 2 E_i ) \\ \\
    -m_i
        \langle \zeta_i^-\mathbf{j}\rangle
        \lbrack i\mathbf{k}\rbrack / ( \sqrt{2E_i} )
    \end{array}\right] 
    %+ \mathcal{O}\left(m_i^3\right)
    \ ,
    \label{eq:<ij>[ik] expansion}
\end{equation}
and
\begin{equation}
    \lbrack \mathbf{ij} \rbrack
    \lbrack \mathbf{ik} \rbrack = 
    \left[\begin{array}{c}
    m_i^2 \Bigl(
    \lbrack\tilde{\zeta}_i^+\mathbf{j}\rbrack
    \lbrack\tilde{\zeta}_i^+\mathbf{k}\rbrack \Bigr)/(2 E_i) \\
    m_i
      \left( \lbrack i\mathbf{j}\rbrack
             \lbrack\tilde{\zeta}_i^+\mathbf{k}\rbrack + 
             \lbrack\tilde{\zeta}_i^+\mathbf{j}\rbrack
             \lbrack i\mathbf{k}\rbrack \right)/(2\sqrt{2E_i}) \\
    \left( 1 - m_i^2/(4E_i^2) \right)
       \lbrack i\mathbf{j}\rbrack
       \lbrack i\mathbf{k}\rbrack
    \end{array}\right]
    %+ \mathcal{O}\left(m_i^3\right)
    \ ,
    \label{eq:[ij][ik] expansion}
\end{equation}
where particle $i$ is the photon, the top rows are for the negative helicity case, the middle are for 0 helicity and the bottom are for positive helicity.  As we can see from this expression, if we can simply unbold a spinor and have the right helicity, there is no need for a $\zeta$ or $\tilde{\zeta}$.  If unbolding produces the wrong helicity, on the other hand, we need to replace one or both brackets with a $\zeta$ or $\tilde{\zeta}$, since they act like the opposite helicity.  Furthermore, each $\zeta$ or $\tilde{\zeta}$ also comes with a power of $M/\sqrt{2E}$ and increases the order of the leading term in the expansion.

Once we have Taylor expanded the amplitude, we algebraically simplify the expressions at each order in the mass of the photon, $M_A$.  At this point, all division by $M_A$ drops out of the expression.  We then take the $M_A\to0$ limit giving our final amplitude.  We will now describe the outline of these steps for the major amplitudes of this paper and give greater detail in App.~\ref{sec:Massive Photon}.

The process $e\bar{e}\mu\bar{\mu}$ is begun by multiplying the two photon vertices, symmetrizing the photon spin index, contracting the photon spin index and dividing by the propagator denominator.  We have
\begin{align}
    \mathcal{M}%^{ee\mu\mu} 
    &=
    \frac{e^2}{2 M_A^2\left(s\!-\!M_A^2\right)}
    \left(  \langle\mathbf{1P_{34}}^{I}\rangle
            \lbrack\mathbf{2P_{34}}^{J}\rbrack \!+\! 
            \lbrack\mathbf{1P_{34}}^{I}\rbrack
            \langle\mathbf{2P_{34}}^{J}\rangle
    \right) \nonumber \\
    &\times
     \Big( 
       \langle\mathbf{3P_{12}}_{I}\rangle
       \lbrack\mathbf{4P_{12}}_{J}\rbrack + 
       \lbrack\mathbf{3P_{12}}_{I}\rbrack
       \langle\mathbf{4P_{12}}_{J}\rangle \nonumber \\ 
    & \qquad
      +\langle\mathbf{3P_{12}}_{J}\rangle
       \lbrack\mathbf{4P_{12}}_{I}\rbrack + 
       \lbrack\mathbf{3P_{12}}_{J}\rbrack
       \langle\mathbf{4P_{12}}_{I}\rangle
     \Big) ,
\end{align}
where $P_{ij}=p_i+p_j$ and we have symmetrized the photon index since it is spin 1.
We next expand this expression, reverse the momentum $P_{34}=-P_{12}$ and use the contraction rules from \cite{Christensen:2019mch} to replace the spinors containing $P_{12}$ with the momentum and masses.  We follow this with a series of Schouten identities, mass identities and momentum conservation.  At the end, the amplitude reduces to
\begin{equation}
  \mathcal{M}%^{ee\mu\mu} 
    = 
    e^2\frac{
    \langle\mathbf{13}\rangle\lbrack\mathbf{24}\rbrack
    +\lbrack\mathbf{13}\rbrack\langle\mathbf{24}\rangle
    +\lbrack\mathbf{14}\rbrack\langle\mathbf{23}\rangle
    +\langle\mathbf{14}\rangle\lbrack\mathbf{23}\rbrack
    }{\left(s-M_A^2\right)}
    \ .
\end{equation}
Finally, we take the massless limit $M_A\to0$ to obtain the final amplitude for this process
\begin{equation}
    \mathcal{M}%_{ee\mu\mu} 
    = 
    \frac{e^2}{s}\left(
    \langle\mathbf{13}\rangle\lbrack\mathbf{24}\rbrack
    +\lbrack\mathbf{13}\rbrack\langle\mathbf{24}\rangle
    +\lbrack\mathbf{14}\rbrack\langle\mathbf{23}\rangle
    +\langle\mathbf{14}\rangle\lbrack\mathbf{23}\rbrack
    \right)\ .
    \label{eq:main:eemm massive}
\end{equation}
Further details can be found in App.~\ref{sec:eemm spinor amp calculation}.  We will discuss this amplitude further in Sec.~\ref{sec:eemm}.

We would next like to discuss a process with one external photon.  In order to do this, we must introduce another neutral particle.  Although this isn't strictly a QED amplitude, we will add the Higgs boson and calculate $e\gamma\bar{e}h$.  In order to do this, we will introduce the Higgs vertex from \cite{Christensen:2018zcq} which is $\mathcal{M}_{eeh} = -\frac{m_e}{v}\left(\langle\mathbf{12}\rangle+\lbrack\mathbf{12}\rbrack\right)$,
where $v=2M_Ws_W/e$, $M_W$ is the mass of the W boson and $s_W$ is the sin of the Weinberg angle.  This amplitude has two diagrams, with the $s$- and $t$-channel propagators.  Combining these with the propagator denominators, contracting the spin indices and using the spin-contraction identities gives us
\begin{align}
  \mathcal{M}^{(s)} &= \frac{e \, m_e}{v(s-m_e^2)} 
    \Biggl(
    \frac{m_e}{M_A}
          \left(  \lbrack\mathbf{12}\rbrack\langle\mathbf{23}\rangle +
                  \langle\mathbf{12}\rangle\lbrack\mathbf{23}\rbrack
          \right) \nonumber \\
    &+\frac{
         \left(   \lbrack\mathbf{12}\rbrack
                  \langle\mathbf{2}\lvert p_1\rvert\mathbf{3}\rbrack +
                  \langle\mathbf{12}\rangle
                  \lbrack\mathbf{2}\lvert p_1\rvert\mathbf{3}\rangle
         \right)
         }{M_A} \nonumber \\
    &+ 
         \left(   \lbrack\mathbf{12}\rbrack
                  \lbrack\mathbf{23}\rbrack +
                  \langle\mathbf{12}\rangle
                  \langle\mathbf{23}\rangle \right) 
    \Biggr)% \nonumber \\
\end{align}
\begin{align}
    \mathcal{M}^{(u)} &= \frac{e \, m_e}{v M_A (u-m_e^2)} \Biggl(
    2 m_e\left(
      \langle\mathbf{23}\rangle\lbrack\mathbf{12}\rbrack
     +\langle\mathbf{12}\rangle \lbrack\mathbf{23}\rbrack \right) \nonumber\\
    &+ \lbrack\mathbf{1}\lvert p_4\rvert\mathbf{2}\rangle 
       \lbrack\mathbf{23}\rbrack +
       \langle\mathbf{1}\lvert p_4\rvert\mathbf{2}\rbrack 
       \langle\mathbf{23}\rangle
    \Biggr) \ .
\end{align}
At this point, we must choose a helicity for the photon.  We will describe the positive helicity, but the negative helicity is analogous.  We Taylor expand the spinors using Eqs.~(\ref{eq:<ij><ik> expansion}) through (\ref{eq:[ij][ik] expansion}) and combine the two diagrams.  We follow this by a series of identities and simplification to obtain
\begin{align}
    \mathcal{M}^+_{eAeh} 
    &=
    \frac{e m_e}
         {v\left(s-m_e^2\right)\left(u-m_e^2\right)}
    \Big( 
      m_h ^{2} \lbrack\mathbf{1}2\rbrack
               \lbrack2\mathbf{3}\rbrack \nonumber \\
    &- m_e \lbrack \mathbf{1}2 \rbrack
           \lbrack 2\lvert p_{4} \rvert \mathbf{3} \rangle 
     + m_e \lbrack 2\mathbf{3}\rbrack 
           \lbrack 2\lvert p_4 \rvert \mathbf{1} \rangle \nonumber \\
    &- \langle \mathbf{1} \mathbf{3} \rangle 
       \lbrack 2\lvert p_{3} p_{4} \rvert 2 \rbrack \Big) \ .
    \label{eq:main:M_eAeh}
\end{align}
Further details for this amplitude can be found in App.~\ref{sec:eAeh spinor amp calculation}.  We will discuss this amplitude further in Sec.~\ref{sec:M_eAeh}.

The process $e\bar{e}\gamma\gamma$ has two diagrams, a $t$- and a $u$-channel diagram.  Connecting the photon vertex to an electron line in the two ways, and contracting the spin indices gives the contributions
\begin{align}
  \mathcal{M}^{(t)} 
   &= -\frac{e^2 M_A 
       \left( \langle\mathbf{1}\mathbf{3}\rangle 
              \lbrack\mathbf{2}\mathbf{4}\rbrack 
              \langle\mathbf{3}\mathbf{4}\rangle 
            + \lbrack\mathbf{1}\mathbf{3}\rbrack
              \langle\mathbf{2}\mathbf{4}\rangle 
              \lbrack\mathbf{3}\mathbf{4}\rbrack 
       \right)}{M_A^2 \left(t-m_e^2 \right)} \nonumber \\
    & -\frac{e^2 m_e 
       \left( \langle\mathbf{1}\mathbf{3}\rangle 
              \langle\mathbf{2}\mathbf{4}\rangle 
              \lbrack\mathbf{3}\mathbf{4}\rbrack 
            + \lbrack\mathbf{1}\mathbf{3}\rbrack 
              \lbrack\mathbf{2}\mathbf{4}\rbrack 
              \langle\mathbf{3}\mathbf{4}\rangle
       \right)}{M_A^2 \left(t-m_e^2 \right)} \nonumber \\
    & -\frac{e^2 
       \left( \langle\mathbf{1}\mathbf{3}\rangle 
              \lbrack\mathbf{2}\mathbf{4}\rbrack 
              \lbrack\mathbf{3}\lvert p_{1} \rvert \mathbf{4}\rangle 
            + \lbrack\mathbf{1}\mathbf{3}\rbrack
              \langle\mathbf{2}\mathbf{4}\rangle 
              \lbrack\mathbf{4}\lvert p_{1} \rvert \mathbf{3}\rangle 
       \right)}{M_A^2 \left(t-m_e^2 \right)} \nonumber \\
  \mathcal{M}^{(u)} 
    &= \frac{e^2 M_A 
       \left( \langle\mathbf{1}\mathbf{4}\rangle
              \lbrack\mathbf{2}\mathbf{3}\rbrack
              \langle\mathbf{3}\mathbf{4}\rangle  
            + \lbrack\mathbf{1}\mathbf{4}\rbrack
              \langle\mathbf{2}\mathbf{3}\rangle  
              \lbrack\mathbf{3}\mathbf{4}\rbrack 
       \right)}{M_A^2 \left(u-m_e^2 \right)} \nonumber \\
    & +\frac{e^2 m_e 
       \left( \langle\mathbf{1}\mathbf{4}\rangle 
              \langle\mathbf{2}\mathbf{3}\rangle 
              \lbrack\mathbf{3}\mathbf{4}\rbrack 
            + \lbrack\mathbf{1}\mathbf{4}\rbrack 
              \lbrack\mathbf{2}\mathbf{3}\rbrack 
              \langle\mathbf{3}\mathbf{4}\rangle
       \right)}{M_A^2 \left(u-m_e^2 \right)} \nonumber \\
    & -\frac{e^2 
       \left( \langle\mathbf{1}\mathbf{4}\rangle 
              \lbrack\mathbf{2}\mathbf{3}\rbrack 
              \lbrack\mathbf{4}\lvert p_{1} \rvert \mathbf{3}\rangle 
            + \lbrack\mathbf{1}\mathbf{4}\rbrack
              \langle\mathbf{2}\mathbf{3}\rangle 
              \lbrack\mathbf{3}\lvert p_{1} \rvert \mathbf{4}\rangle 
       \right)}{M_A^2 \left(u-m_e^2 \right)}.
\end{align}
Taylor expanding, combining, applying identities, simplifying and taking the massless limit brings the amplitude to the final form.  For the $++$ helicity amplitude, the amplitude reduces to only one term.  It is
\begin{equation}
    \mathcal{M}^{++}_{eeAA} = 
    \frac{e^2  m_e \lbrack34\rbrack ^{2} \langle\mathbf{1}\mathbf{2}\rangle }
    {\left(t-m_e^2\right)\left(u-m_e^2\right)} \ .
    \label{eq:main:M_eeA+A+}
\end{equation}
In the case of the $+-$ helicity, the final amplitude is nearly as simple.  It is given by
\begin{equation}
    \mathcal{M}_{eeAA}^{+-} = 
    \frac{e^2 \left(\langle\mathbf{2}4\rangle\lbrack\mathbf{1}3\rbrack +\langle\mathbf{1}4\rangle\lbrack\mathbf{2}3\rbrack \right)\lbrack3\lvert p_{1} \rvert 4\rangle}{\left(t-m_e^2\right)\left(u-m_e^2\right)}\ . 
    \label{eq:main:M_eeA+A- massive}
\end{equation}
We give greater details of all these calculations in App.~\ref{sec:Massive eeAA}.  We further discuss these amplitudes in Secs.~\ref{sec:M_eeA+A+} and \ref{sec:M_eeA+A-}.

Although we were able to successfully calculate all the physical amplitudes using the massless limit of a massive-photon theory, there are several reasons we would like achieve a fully successful massless theory that bypasses the massive intermediate.  As we already mentioned at the opening of this section, using a massive photon requires adding an unphysical degree of freedom, which must cancel or vanish in the massless limit.  It would be much preferable to only involve physical features throughout the calculation and that is the hope of the constructive amplitude approach.  Additionally, we see that when we calculate the amplitude with a massive photon, we are required to include every diagram that we would include in a Feynman-diagram calculation and there is no improvement in the economy of the calculation.  When using a massless photon from the beginning, on the other hand, as we will see in the coming sections, it only requires one of the diagrams.  In fact, each diagram gives the same final result and the multiple diagrams are redundant.  This is similar to the same feature in purely massless theories, such as gluodynamics, where only one diagram is required to obtain the final amplitude.  In fact, examples like these suggest that, in some cases, the requirement of multiple diagrams with the same internal state is a shortcoming of Feynman diagrams, even for partly massive theories, and of constructive theories with intermediate massive particles, such as the massive photon discussed in this section.  It appears that adding unphysical degrees of freedom has a consequence of adding unnecessary complications and unphysical aspects that must be cancelled.  Therefore, an approach that never introduces these unphysical aspects would be preferable.

\subsection{\label{sec:main:x}The $x$ Factor}

The $x$ factor was a required introduction for 3-point vertices that contained one massless particle and two massive particles of the same mass as described in \cite{Arkani-Hamed:2017jhn}.  We review its definition and properties in App.~\ref{app:x}.  In order to compare with experiments or Feynman diagrams, we must replace $x$ with expressions involving spinors, masses and other momentum invariants.  

In the context of QED, the photon electron vertex is given by
\begin{align}
\mathcal{M}_{e\bar{e}\gamma^+} = x_{12}\langle\mathbf{12}\rangle  
&&
\mathcal{M}_{e\bar{e}\gamma^-} = \tilde{x}_{12}\lbrack\mathbf{12}\rbrack 
\ .
\end{align}
The $x$ and the $\tilde{x}$ transform as helicity-$+1$ and -$-1$ objects respectively.  They are defined in terms of the symmetrized identities
\begin{align}
    x_{12}|3\rangle &= \frac{1}{2m}\left(p_{2}-p_{1}\right)|3\rbrack
    \label{eq:main:x_12|3>=...|3]}
    \\
    \tilde{x}_{12}|3\rbrack &= \frac{1}{2m}\left(p_2-p_1\right)|3\rangle
    \label{eq:main:xt_12|3]=...|3>}
    \ .
\end{align}
By multiplying on the left by $(p_2-p_1)/2m$, we can show that $\tilde{x}=1/x$.
Further, a reference spinor $\langle\xi\rvert$ or $\lbrack\xi\rvert$ can be multiplied on the left to obtain the symmetrized
\begin{align}
    x_{12} &= \frac{\langle\xi\lvert\left(p_2-p_1\right)\rvert3\rbrack}{2m\langle\xi3\rangle}
    \label{eq:main:x_{12} = }
    \\
    \tilde{x}_{12} &=  \frac{\lbrack\xi\lvert\left(p_2-p_1\right)\rvert3\rangle}{2m\lbrack\xi3\rbrack} \ .
\end{align}
However, if the reference spinor is not physical, it must cancel before the end of the calculation.  Sometimes this is straight forward, but not always.  In other cases, a physicsal helicity spinor can be used as the references spinor.  This turns out to be useful when the photon is an external particle.  In any case, we would like to find identities that remove $x$ and $\tilde{x}$ that bypass the reference spinor and replace them directly with physical spinors, masses and momentum invariants.  

When the photon is on an internal line, we obtain expressions of the form $x_{ij}\tilde{x}_{kl}\langle\mathbf{ij}\rangle\lbrack\mathbf{kl}\rbrack+\tilde{x}_{ij}x_{kl}\lbrack\mathbf{ij}\rbrack\langle\mathbf{kl}\rangle$, where $i$ and $j$ are the particles on one side of the propagator and $k$ and $l$ are on the other side.  The first term is for the negative helicity photon and includes the vertices from both sides of the propagator while the second term is similarly for the positive helicity photon.  This can be simplified by using the identities
\begin{align}
    x_{ij}\tilde{x}_{kl}\left(
        \lbrack\mathbf{kl}\rbrack - 
        \langle\mathbf{kl}\rangle 
    \right) &=
    \frac{1}{m_jm_l}\langle\mathbf{k}|p_jp_i|\mathbf{l}\rangle+
    \frac{m_j}{m_l}\langle\mathbf{kl}\rangle
    \label{eq:main:xxt<ij>=xxt[ij]}
    \\
    x_{ij}\tilde{x}_{kl}\left(
        \langle\mathbf{ij}\rangle -
        \lbrack\mathbf{ij}\rbrack 
    \right) &=
    \frac{1}{m_jm_l}\lbrack\mathbf{i}|p_lp_k|\mathbf{j}\rbrack +
    \frac{m_l}{m_j}\lbrack\mathbf{ij}\rbrack
    \label{eq:main:xxt<kl>=xxt[kl]}
\end{align}
to turn angle brackets into square brackets or vice versa.  We give greater detail in Apps.~\ref{app:x_ij<ij>=x_ij[ij]+...} and \ref{app:xxtilde<>[]}.  Using these identities, the spinor products multiplying $x\tilde{x}$ in each term can be transformed into a common form and factored out of the expression.  Once this is done, we need to deal with $\left(x_{ij}\tilde{x}_{kl}+\tilde{x}_{ij}x_{kl}\right)$.  For this, we use the fully symmetrized identity
\begin{equation}
    x_{12}\tilde{x}_{34}+\tilde{x}_{12}x_{34} = 
    -\frac{(p_2-p_1)\cdot (p_4-p_3)}{2m_2m_4}\ .
    \label{eq:main:xxt+xtx}
\end{equation}
We review the derivation of this identity in App.~\ref{app:xxt+xtx} that uses the reference spinor.  We show alternate derivations of the same identity which do not use a reference spinor in App.~\ref{app:xxt+xtx revisited}.  In this appendix, we further show the importance of using the fully symmetrized version of these identities as inconsistent results can be obtained if the fully symmetrized versions are not used.  Unfortunately, although we have these identities for internal photons, we do not have an amplitude using them that agrees with Feynman diagrams, as we show in App.~\ref{sec:Constructive Massless eemm}.  It appears that further ingredients are required for internal massless particles.  

When external particles are massless, the amplitude expression contains an $x$ for every positive-helicity photon and $\tilde{x}$ for every negative-helicity photon.  Although we could use a reference spinor, we have found that $x$ and $\tilde{x}$ on external lines can be immediately replaced with physical expressions in the following way.  
We multiply $x$ or $\tilde{x}$ by the propagator denominator of the other diagram divided by itself.  In the numerator, we write the propagator denominator in terms of a spinor product and use Eqs.~(\ref{eq:main:x_12|3>=...|3]}) and (\ref{eq:main:xt_12|3]=...|3>}) on it.
For example, let's consider the case of one of the diagrams in this article, the s-channel diagram for $e\gamma^+\bar{e}h$.  The expression for this amplitude contains $x_{1,34}$, where the notation signifies that the particle entering the vertex has momentum $p_1$ and the antiparticle has momentum $p_3+p_4$.  The other diagram is in the u channel, therefore, we multiply by $\lbrack2\lvert p_3\rvert2\rangle/(u-m^2)$.  This is unity because $\lbrack2\lvert p_3\rvert2\rangle=2p_2\cdot p_3=u-m^2$.  However, we can now use $x_{1,34}\lvert2\rangle=\frac{1}{2m}\left(p_3+p_4-p_1\right)\lvert2\rbrack$.  Putting this together and simplifying, we obtain
\begin{equation}
    x_{1,34} 
    = x_{1,34}\frac{\lbrack2\lvert p_3\rvert2\rangle}{(u-m^2)}
    = \frac{\lbrack2\lvert p_3p_4\rvert2\rbrack}{m(u-m^2)}
    \ .
\end{equation}
However, we can also obtain the same result by using the reference spinor $\lbrack\xi\rvert = \lbrack2\rvert p_3$ in Eq.~(\ref{eq:main:x_{12} = }).  Greater detail and more examples can be found in App.~\ref{sec:x = alternate}.  We successfully use these identities to obtain the amplitudes for $e\gamma\bar{e}h$ in App.~\ref{sec:Constructive Massless heAe} and for $e\bar{e}\gamma\gamma$ in Apps.~\ref{app:eeA+A+} and \ref{app:eeA+A-}.  Our expressions agree with Feynman diagrams.

\subsection{\label{sec:main:constructive no Shift}Constructive Calculations}
Now that we have the ingredients for the constructive amplitude calculations, we will calculate the QED amplitudes containing photons.  We begin with the process $e\bar{e}\mu\bar{\mu}$ which has an internal photon in the s channel.  The vertices are $e x_{12}\langle\mathbf{12}\rangle$ on the left and $e \tilde{x}_{34}\lbrack\mathbf{34}\rbrack$ on the right for a negative helicity photon and
$e \tilde{x}_{12}\lbrack\mathbf{12}\rbrack$ and $e x_{34}\langle\mathbf{34}\rangle$ for a positive helicity photon.  Multiplying these and dividing by the propagator denominator gives us
\begin{equation}
    \mathcal{M} =
    \frac{e^2}{s}  \left(
    x_{34} \tilde{x}_{12} \lbrack\mathbf{1}\mathbf{2}\rbrack 
    \langle\mathbf{3}\mathbf{4}\rangle 
    +x_{12} \tilde{x}_{34}
    \langle\mathbf{1}\mathbf{2}\rangle \lbrack\mathbf{3}\mathbf{4}\rbrack  \right)\ .
\end{equation}
We next use the identities in Eqs.~(\ref{eq:main:xxt<ij>=xxt[ij]}) and (\ref{eq:main:xxt<kl>=xxt[kl]}) to convert the spinors in each term into a common spinor term that multiplies the $x$ factors and can be factored out.  After applying the identity in Eq.~(\ref{eq:main:xxt+xtx}), we are left with either
\begin{align}
    \mathcal{M} =
    &\ \frac{e^2}{2m_em_\mu s}  \Big[
    (u-t+2m_e^2+2m_\mu^2)
    \lbrack\mathbf{1}\mathbf{2}\rbrack 
    \lbrack\mathbf{3}\mathbf{4}\rbrack
    \nonumber\\
    &+2\left(
    \lbrack\mathbf{12}\rbrack \lbrack\mathbf{3}\lvert p_2p_1\rvert\mathbf{4}\rbrack+
    \lbrack\mathbf{1}\lvert p_4p_3\rvert\mathbf{2}\rbrack \lbrack\mathbf{34}\rbrack 
    \right)
    \Big]\ ,
    \label{eq:main:M_eemm massless no shift}
\end{align}
or its conjugate.  This amplitude does not agree with Feynman diagrams, so we think further constructive ingredients are necessary.  We will consider momentum and spinor shifts in Secs.~\ref{sec:main:spinor shifts}, \ref{sec:main:large z} and \ref{sec:main:constructive}.  Unfortunately, we will see that spinor shifts do not help.  

The main issue we see with this calculation is that the $x$ factor vertices do not expose the helicity of the photon in the propagator and, therefore, do not connect the helicity of the photon at opposite ends of the propagator.  This results in spinor products that contain fermions from the same side of the propagator (e.g. $\lbrack\mathbf{12}\rbrack\lbrack\mathbf{34}\rbrack$) rather than from the opposite ends of the propagator, as seen in the correct amplitude (e.g. $\langle\mathbf{13}\rangle\lbrack\mathbf{24}\rbrack$).  Secondly, we see that the $x$ factor in these vertices results in the amplitude involving division by the fermion masses, causing a bad high-energy growth, violating perturbative unitarity.  Speculatively, an improvement to this vertex would involve exposure of the photon helicity, allowing it to be connected on opposite ends of the propagator and no division by the fermion mass.  Further details for this calculation can be found in App.~\ref{sec:Constructive Massless eemm no Spinor Shift}.  

We next turn to the process $e\gamma^+\bar{e}h$, with one external photon.  For this process, there are potentially two diagrams that could be used.  It turns out that, when using a massless photon and the $x$ factor, we only need one of the diagrams and not both.  They both give the identical result.  
For the u-channel diagram, we have the Higgs-electron vertex $\frac{m_e}{v}\left(\langle\mathbf{1P}_{23\mathrm{I}}\rangle+\lbrack\mathbf{1P}_{23\mathrm{I}}\rbrack\right)$ and the photon-electron vertex $e x_{14,3}\langle\mathbf{P}^{\mathrm{I}}_{14}\mathbf{3}\rangle$.  Multiplying and dividing by the propagator denominator, we have
\begin{equation}
    \mathcal{M}_u^+ =
    \frac{em_e}{v}\frac{x_{14,3}
    (\langle\mathbf{1P}_{23\mathrm{I}}\rangle+\lbrack\mathbf{1P}_{23\mathrm{I}}\rbrack)
    \langle\mathbf{P}_{14}^{\mathrm{I}}\mathbf{3}\rangle}
    {u-m_e^2}
    \ .
\end{equation}
After using the spin-contraction, mass and momentum conservation identities, we find
\begin{equation}
    \mathcal{M}^+_u 
    = \frac{em_e}{v}\frac{x_{14,3}
    (2m_e\langle\mathbf{13}\rangle+\lbrack\mathbf{1}\lvert p_4\rvert\mathbf{3}\rangle)
    }
    {u-m_e^2}\ .
\end{equation}
Our next task is to remove the $x_{14,3}$.  Following the rules discussed in App.~\ref{sec:x = alternate}, we find
\begin{equation}
    x_{14,3} = \frac{\lbrack2\lvert p_1p_3\rvert2\rbrack}{m_e(s-m_e^2)}\ ,
\end{equation}
where we have used the form that contains the propagator denominator from the other diagram that we are not analyzing.  This makes obtaining the correct result easier.  Plugging this in, we find
\begin{equation}
    \mathcal{M}^+_u 
    = \frac{e}{v}\frac{\lbrack2\lvert p_1p_3\rvert2\rbrack
    (2m_e\langle\mathbf{13}\rangle+\lbrack\mathbf{1}\lvert p_4\rvert\mathbf{3}\rangle)
    }
    {\left(s-m_e^2\right)\left(u-m_e^2\right)}\ .
\end{equation}
At this point, we perform a series of Schouten, mass and momentum conservation identities to simplify this expression to
\begin{align}
    \mathcal{M}_u^+ &=
    -\frac{e m_e}{v\left(s-m_e^2\right)\left(u-m_e^2\right)}
      \Big( m_h ^{2} \lbrack\mathbf{1}2\rbrack\lbrack2\mathbf{3}\rbrack 
    - m_e \lbrack\mathbf{1}2\rbrack\lbrack2\lvert p_{4} \rvert \mathbf{3}\rangle 
    \nonumber\\
    &+ m_e \lbrack2\mathbf{3}\rbrack \langle\mathbf{1}\lvert p_{4} \rvert 2\rbrack
    - \langle\mathbf{1}\mathbf{3}\rangle \lbrack2\lvert p_{3} p_{4} \rvert 2\rbrack
    \Big),
\end{align}
which agrees with Feynman diagrams.  
Further details for this process can be found in App.~\ref{sec:Constructive Massless heAe no Shift}.

Turning to processes with two external photons, we begin with the process $e\bar{e}\gamma^+\gamma^+$.  Once again, there are two diagrams, this time in the t and u channels.  Once again, we can use either diagram and do not need both since they both give identical final results.  We will demonstrate the u channel, but have also done the t channel with full agreement.  The vertices are $e x_{1,23}\langle\mathbf{1P}_{23\mathrm{I}}\rangle$ and $e x_{14,2}\langle\mathbf{P}_{14}^{\mathrm{I}}\mathbf{2}\rangle$, giving us the amplitude
\begin{equation}
    \mathcal{M}_u^{++} =
    -\frac{e^2 m_e x_{1,23}x_{14,2}\langle\mathbf{12}\rangle}{u-m_e^2}
    \ ,
\end{equation}
where we have already simplified the contracted spinor index.  As in the previous process, we must now replace the $x$ factors.  There are multiple ways we could do this, including the use of Eqs.~(\ref{eq:x_1,23}) and (\ref{eq:x_14,2}), similar to our calculation of the previous process.  However, since we have two photons in this process, we can also use forms for $x_{1,23}$ and $x_{14,2}$ that involve the helicity spinor of the other photon as the reference spinor.  In particular, we can take 
\begin{align}
    x_{1,23} = -\frac{\langle3\lvert p_1\rvert4\rbrack}{m_e\langle34\rangle}
    \quad \mbox{and}\quad 
    x_{14,2} = \frac{\langle4\lvert p_2\rvert3\rbrack}{m_e\langle43\rangle}
    \ ,
\end{align}
where, we can show that these forms are equivalent to Eqs.~(\ref{eq:x_1,23}) and (\ref{eq:x_14,2}) on shell by use of the usual identities.  However, in some cases, a more efficient route to the final amplitude is possible by use of these simpler expressions.  In order to get the amplitude into the standard form where there are no spinors in the denominator, we multiply the numerator and denominator of these identies with $\lbrack34\rbrack$ to obtain
\begin{align}
    x_{1,23} = \frac{\langle3\lvert p_1\rvert4\rbrack\lbrack34\rbrack}{m_e s}
    \quad \mbox{and}\quad 
    x_{14,2} = \frac{\langle4\lvert p_2\rvert3\rbrack\lbrack34\rbrack}{m_e s}
    \ .
\end{align}
We don't simplify the numerator of these because we know we want the $\lbrack34\rbrack^2$ in the final form of the amplitude.  Plugging these identities into the amplitude, we have
\begin{equation}
\mathcal{M}_u^{++} =
    \frac{e^2 \langle4\lvert p_2\rvert3\rbrack \langle3\lvert p_1\rvert4\rbrack
    \lbrack34\rbrack^2\langle\mathbf{12}\rangle}{m_e s^2\left(u-m_e^2\right)}
    \ .
\end{equation}
In the denominator, we can use momentum conservation and the on-shell condition to modify $s = -t-u+2m_e^2 = -(t-m_e^2)$.  However, we only do this with one of them because we use the standard identities to simplify $\langle4\lvert p_2\rvert3\rbrack \langle3\lvert p_1\rvert4\rbrack = m_e^2 s$ and one $s$ cancels.  Our final result is
\begin{equation}
   \mathcal{M}_u^{++} =
    -\frac{e^2 m_e \langle\mathbf{12}\rangle \lbrack34\rbrack^2 }
    {\left(t-m_e^2\right)\left(u-m_e^2\right)} \ ,
\end{equation}
which agrees with Feynman diagrams.  Further details for this process can be found in App.~\ref{app:eeA+A+ no shift}.

For our final process, we consider $e\bar{e}\gamma^+\gamma^-$.  
Of the two diagrams, in the t and u channels, this time we will demonstrate the t-channel diagram.  As before, we have done the calculation with both diagrams and find exactly the same final amplitude with either one.  The vertices are $e x_{1,24}\langle\mathbf{1P}_{24\mathrm{I}}\rangle$ and $e \tilde{x}_{13,2}\lbrack\mathbf{P}_{13}^{\mathrm{I}}\mathbf{2}\rbrack$, giving the initial amplitude
\begin{equation}
    \mathcal{M}_t^{+-} =
    \frac{e^2 x_{1,24}\tilde{x}_{13,2}
    \left(m_e\lbrack\mathbf{12}\rbrack
    +\langle\mathbf{1}\lvert p_3\rvert\mathbf{2}\rbrack
    \right)}
    {t-m_e^2}
    \ ,
\end{equation}
where we have used the spin-index contraction and mass identities.  Similar to the $++$-helicity case, we must replace the $x$.  This time, since we have two photons \textit{and} we have one $x$ and one $\tilde{x}$, we get a particularly simple form if use $x_{1,24}=-\langle4\lvert p_1\rvert3\rbrack/m_e\langle43\rangle$ and $\tilde{x}_{13,2}=\lbrack3\lvert p_2\rvert4\rangle/m_e\lbrack34\rbrack$ with the product
\begin{equation}
    x_{1,24}\tilde{x}_{13,2}
    =
    \frac{\lbrack3\lvert p_1\rvert4\rangle^2}{m_e^2 s}
    \ ,
\end{equation}
after using momentum conservation.  Since, on shell, we have $s=-(u-m_e^2)$, we can plug this into the amplitude to obtain
\begin{equation}
    \mathcal{M}_t^{+-} =
    -\frac{e^2 \lbrack3\lvert p_1\rvert4\rangle^2
    \left(m_e\lbrack\mathbf{12}\rbrack
    +\langle\mathbf{1}\lvert p_3\rvert\mathbf{2}\rbrack
    \right)}
    {m_e^2\left(t-m_e^2\right)\left(u-m_e^2\right)}
    \ .
\end{equation}
We then perform a series of simplifications that use both the usual identities and the on-shell condition to end with
\begin{equation}
    \mathcal{M}_t^{+-} =
    \frac{e^2 \lbrack3\lvert p_1\rvert4\rangle
    \left(
    \langle\mathbf{2}4\rangle\lbrack\mathbf{1}3\rbrack 
    + \langle\mathbf{1}4\rangle\lbrack\mathbf{2}3\rbrack
    \right)
    }
    {\left(t-m_e^2\right)\left(u-m_e^2\right)}
    \ ,
\end{equation}
in agreement with Feynman diagrams.  
Further details can be found in App.~\ref{app:eeA+A- no shift}.

\subsection{\label{sec:main:spinor shifts}Spinor Shifts}
In order to extend the constructive methods from purely massless theories to theories with mass, we should generalize the spinor shifts.  They might be required to obtain the correct amplitudes in some cases.  We include much greater detail in App.~\ref{app:analytic continuation}.

In purely massless theories, we choose two of the external momenta that lie on opposite sides of the propagator and analytically continue them as in
\begin{align}
    \hat{p}_i &= p_i + z q\\
    \label{eq:main:phat = p+zq}
    \hat{p}_j &= p_j - z q,
\end{align}
where $z$ is a complex number and $q$ is a complex momentum.  It can be shown \cite{Britto:2005fq} that if the amplitude vanishes in the large $z$ limit, it can be built up recursively by stitching together smaller on-shell amplitudes and dividing by propagator denominators.  The on-shell condition for the internal line is accomplished by choosing a value for $z$ where the internal line goes on shell and relaxing this property at the end of the calculation.  

When the momenta are analytically continued in this way, the spinors are also impacted as they are functions of the momenta.  Therefore, we must also find the accompanying shifts of the spinors in order to work out the expressions for the amplitudes.  In purely massless theories, this has already been done.  
Given two momenta $p_i$ and $p_j$, there are two ways we can shift the spinors.  We can either do an $\lbrack i,j\rangle$ shift, where
\begin{align}
    \lbrack\hat{i}\rvert &= \lbrack i\rvert + z \lbrack j\rvert\\
    |\hat{j}\rangle &= |j\rangle - z |i\rangle\ ,
    \label{eq:main:|jhat> = |j>-z|i>}
\end{align}
and $\lvert j\rbrack$ and $\lvert i\rangle$ are not shifted, 
or the reverse.  It can be shown that this shift preserves momentum conservation as well as the on-shell property for both particles $i$ and $j$.  Additionally, if the special value of $z= -\lbrack ik\rbrack/\lbrack jk\rbrack$,
where $k$ represents all the other momenta on the same side of the propagator as $i$ is chosen, then the internal line also goes on shell as well.  For convenience of our calculations, we can insert this value of $z$ directly into our spinor shifts, to obtain
\begin{eqnarray}
    \lbrack\hat{i}\rvert &=& \lbrack i\rvert 
    -\frac{s_{ik}}{\langle i\lvert p_k\rvert j\rbrack} \lbrack j\rvert
    \\
    \lvert\hat{j}\rangle &=& \lvert j\rangle 
    +\lvert i\rangle\frac{s_{ik}}{\langle i\lvert p_k\rvert j\rbrack} \ ,
\end{eqnarray}
where $s_{ik}=(p_i+p_k)^2$, and we have used $z=-\lbrack ik\rbrack\langle ki\rangle/\lbrack jk\rbrack\langle ki\rangle = s_{ik}/\langle i\lvert p_k\rvert j\rbrack$ for convenient comparison with the massive expressions.

The simplest way to generalize this spinor shift is when the internal line becomes massive but the shifted spinors remain massless helicity spinors.  In this case, only the internal-line on-shell condition changes.  This results in the special value of $z=\left(M^2-s_{ik}\right)/\langle i\lvert p_k\rvert j\rbrack$, where $M$ is the mass of the internal particle, and the final shifts become
\begin{align}
    \lbrack\hat{i}\rvert &= \lbrack i\rvert 
    -\frac{\left(s_{ik}-M^2\right)}{\langle i\lvert p_k\rvert j\rbrack} \lbrack j\rvert
    \label{eq:main:[ihat|: 0,0,M}
    \\
    \lvert\hat{j}\rangle &= \lvert j\rangle 
    +\lvert i\rangle\frac{\left(s_{ik}-M^2\right)}{\langle i\lvert p_k\rvert j\rbrack} \ .
    \label{eq:main:|jhat>: 0,0,M}
\end{align}
We can see that this trivially reduces to the all-massless case in the massless limit.  Further details can be found in App.~\ref{app:[i,j> massless massless}

We must next consider spinor shifts where one of the shifted spinors is a massive spin spinor.  Let's begin with a $\lbrack\mathbf{i},j\rangle$ shift where particle $i$ is massive but particle $j$ remains massless.  In this case, the shifted spinor has an additional spin index $\lbrack \mathbf{i}_{\mathrm{I}}\rvert$ and, in order to shift it using the helicity spinor $\lbrack j\rvert$, this requires the complex number to generalize to a complex number with its own spin index $z^{\mathrm{I}}$.  In other words, the shift becomes
\begin{eqnarray}
    \lbrack\mathbf{\hat{i}}_\mathrm{I}| &=& \lbrack\mathbf{i}_\mathrm{I}| + z_\mathrm{I} \lbrack j| 
    \\
    |\hat{j}\rangle &=& |j\rangle - z_\mathrm{I} |\mathbf{i}^\mathrm{I}\rangle
    \ .
\end{eqnarray}
It is straight forward to see that momentum is still conserved and that particle $j$ remains massless, but the on-shell condition for particle $j$ requires the new constraint
\begin{equation}
    z_{\mathrm{I}}\lbrack\mathbf{i}^{\mathrm{I}}j\rbrack = 0
    \ .
\end{equation}
Additionally, requiring the internal line to be on shell, gives the constraint
\begin{equation}
    z_{\mathrm{I}}\langle \mathbf{i}^{\mathrm{I}} \lvert p_k\rvert j\rbrack =
    M^2 - s_{ik} \ ,
\end{equation}
where, once again, $M$ is the mass of the internal particle.  Solving these constraints simultaneously gives
\begin{equation}
    z_I = \frac{\lbrack j\mathbf{i}_I\rbrack}{
    \lbrack j\lvert p_i  p_k\rvert j\rbrack}
    \left(M^2 - s_{ik}\right)
    \ ,
\end{equation}
and this reduces to the all-massless result in the massless limit.  This can be plugged back into the spinor shifts to give our final shifts
\begin{eqnarray}
    \lbrack\mathbf{\hat{i}}| &=& \lbrack\mathbf{i}| + \frac{\lbrack \mathbf{i} j\rbrack\lbrack j\rvert}{
    \lbrack j\lvert p_i  p_k\rvert j\rbrack}
    \left(s_{ik}-M^2\right)  
    \label{eq:main:[ihat|: mi,0,M}
    \\
    |\hat{j}\rangle &=& |j\rangle - \frac{p_i\lvert j\rbrack}{
    \lbrack j\lvert p_i  p_k\rvert j\rbrack}
    \left(s_{ik}-M^2\right) \ .
    \label{eq:main:|jhat>: mi,0,M}
\end{eqnarray}
Further details can be found in App.~\ref{app:[i,j> massive massless}.

The opposite case, the shift $\lbrack i,\mathbf{j}\rangle$, where particle $i$ is massless and particle $j$ is massive, is analogous.  The final shifts are given by
\begin{eqnarray}
    \lbrack\hat{i}| &=& \lbrack\mathrm{i}| - \frac{\langle i \rvert p_j}{
    \langle i \lvert p_jp_k\rvert i\rangle}
    \left(s_{ik}-M^2\right) 
    \label{eq:main:[ihat|: 0,mj,M}
    \\
    |\mathbf{\hat{j}}\rangle &=& |\mathbf{j}\rangle + \frac{\lvert i\rangle\langle i \mathbf{j}\rangle}{
    \langle i \lvert p_jp_k\rvert i\rangle}
    \left(s_{ik}-M^2\right) \ .
    \label{eq:main:|jhat>: 0,mj,M}
\end{eqnarray}  
Further details can be found in App.~\ref{app:[i,j> massless massive}.

Our next case is the shift $\lbrack\mathbf{i,j}\rangle$, where both particles $i$ and $j$ are massive.  The most general case where all external particles and the internal line are massive turns out to be quite complicated.  On the other hand, if every particle in an amplitude, including the internal particles, were massive, we appear not to need the shift of the spinors, nor of the momenta.  So, the all-massive case does not appear to be of interest and a lack of simple formulas for spinors in this case does not seem important.  On the other hand, there are still cases where both shifted spinors are massive, even if other particles in the amplitude are not.  If the internal line is massive, but one or more of the unshifted external particles are massless, we might in principle need the formulas for the shifted spinors.  This could be important, for example, for the processes $e\gamma\bar{e}h$ and $e\bar{e}\gamma\gamma$.  However, it turns out that these shifts are also extremely complicated.  Moreover, because we can still calculate these processes using one of the previous shifts where at least one of the shifted spinors is massless, the lack of simple shifts in this case also does not appear important.  Therefore, we also skip it, but further details can be found in App.~\ref{app:[i,j> both massive massive internal line}.

We have one further case that we must consider, the shift $\lbrack\mathbf{i,j}\rangle$, where both spinors are massive, but the internal line is massless.  This case is important, for example, for the process $e\bar{e}\mu\bar{\mu}$, where we cannot use one of the previous spinor shifts, since all the external particles are massive.  In this case, both spinors have a spin index and this requires the complex number to have two spin indices.  We begin with the spinor shift
\begin{eqnarray}
    \lbrack\mathbf{\hat{i}}_\mathrm{I}| &=& \lbrack\mathbf{i}_\mathrm{I}| + z^{\ \mathrm{J}}_\mathrm{I} \lbrack\mathbf{j}_\mathrm{J}| 
    \\
    |\mathbf{\hat{j}}^\mathrm{J}\rangle &=& |\mathbf{j}^\mathrm{J}\rangle - z^{\ \mathrm{J}}_\mathrm{I} |\mathbf{i}^\mathrm{I}\rangle\ .
\end{eqnarray}
Momentum conservation is still straight forward, but now the on-shell condition for both external particles are non-trivial and are given by
\begin{align}
    z_\mathrm{I}^{\ \mathrm{J}}\left(
    2\lbrack\mathbf{i}^{\mathrm{I}}\mathbf{j}_{\mathrm{J}}\rbrack 
    - z_{\ \mathrm{J}}^{\mathrm{I}}m_j
    \right) &= 0
    \\
    z^{\ \mathrm{J}}_{\mathrm{I}}\left(
    2\langle\mathbf{i}^{\mathbf{I}}\mathbf{j}_\mathrm{J}\rangle
    - z^{\mathrm{I}}_{\ \mathrm{J}}m_i
    \right) &= 0\ .
\end{align}
Finally, we must combine this with the on-shell condition for the internal line, which is
\begin{equation}
    z_{\mathrm{I}}^{\ \mathrm{J}}\langle \mathbf{i}^{\mathrm{I}} \lvert p_k\rvert\mathbf{j}_{\mathrm{J}}\rbrack =
     -s_{ik}\ .
\end{equation}
The simplest solution to these three constraints is
\begin{equation}
    z_I^{\ J} = -\frac{\left(m_i\lbrack \mathbf{i}_{\mathrm{I}}\mathbf{j}^{\mathrm{J}} \rbrack + \langle \mathbf{i}_{\mathrm{I}}\lvert p_k\rvert \mathbf{j}^{\mathrm{J}}\rbrack\right)s_{ik}}{\left(m_i\lbrack \mathbf{i}_{\mathrm{K}}\mathbf{j}^{\mathrm{L}} \rbrack + \langle \mathbf{i}_{\mathrm{K}}\lvert p_k\rvert \mathbf{j}^{\mathrm{L}}\rbrack\right)\langle \mathbf{i}^{\mathrm{K}} \lvert p_k\rvert\mathbf{j}_{\mathrm{L}}\rbrack}\ ,
\end{equation}
which, it turns out, can be simplified to
\begin{equation}
    z_I^{\ J} = \frac{m_i\lbrack \mathbf{i}_{\mathrm{I}}\mathbf{j}^{\mathrm{J}} \rbrack + \langle \mathbf{i}_{\mathrm{I}}\lvert p_k\rvert \mathbf{j}^{\mathrm{J}}\rbrack}{m_im_j}\ .
\end{equation}
If we plug this back into our spinor shifts, we get a particularly simple form for the spinor shifts, namely
\begin{align}
    \lbrack\mathbf{\hat{i}}| &=
    - \frac{1}{m_i}\langle \mathbf{i}\lvert p_k
    \label{eq:main:[ihat|: mi,mj,0}
    \\
    |\mathbf{\hat{j}}\rangle
    &=
    \frac{1}{m_j}p_l\lvert\mathbf{j} \rbrack\ .    
    \label{eq:main:|jhat>: mi,mj,0}
\end{align}
The momentum shifts turn out to be even simpler.  They are
\begin{align}
    \hat{p}_i &= -p_k
    \\
    \hat{p}_j &= -p_l
    \ ,
\end{align}
where $p_k$ represents all the other momenta on the same side of the propagator as $p_i$ and $p_l$ represents all the other momenta on the same side as $p_j$.  In a 4-point amplitude, they are just the other two momenta.  In a higher-point amplitude, they are the sum of the other momenta. 
Further details can be found in App.~\ref{app:[i,j> both massive}.
Although we were successful finding spinor shifts for this case, the only process in this paper to which they apply, in principle, is to the process $e\bar{e}\mu\bar{\mu}$.  Unfortunately, we do not obtain the correct amplitude in this case.  However, we do not believe this failure is due to the spinor shifts being incorrect.  Rather, we think it is due to the amplitude not satisfying the large-$z$ condition for this method to work.  We will discuss the large-$z$ behavior next.

\subsection{\label{sec:main:large z}Large $\mathbf{z}$ Limit}
The theorem proving that the amplitude can be written as a product of smaller amplitudes divided by the propagator denominator depends on the amplitude vanishing in the large $z$ limit after the momentum shift $\hat{p}_i = p_i + z q$ and $\hat{p}_j = p_j - z q$ \cite{Britto:2005fq}.  This was proven for a purely massless theory, but it should be true for a partly massive theory as well.  On the other hand, the theorem only states that it is possible in this limit.  It does not state that the amplitude does not split up this way in the absence of an asymptotically vanishing amplitude.  Therefore, we might wonder whether we can still write a massive amplitude in some cases as a product of smaller amplitudes divided by a propagator denominator even when the amplitude does not vanish in the large $z$ limit.   We find that the answer is yes, sometimes, but not always, we can.  We present examples of each case in this article.  

We begin by analyzing the large $z$ behavior of the amplitude for $e\bar{e}\mu\bar{\mu}$.  We obtained its amplitude in Eq.~(\ref{eq:main:eemm massive})
\begin{equation}
    \mathcal{M} = \frac{e^2
    \left(\langle\mathbf{13}\rangle\lbrack\mathbf{24}\rbrack
    +\langle\mathbf{14}\rangle\lbrack\mathbf{23}\rbrack
    +\lbrack\mathbf{13}\rbrack\langle\mathbf{24}\rangle
    +\lbrack\mathbf{14}\rbrack\langle\mathbf{23}\rangle
    \right)}{(p_1+p_2)^2}\ .
\end{equation}
and it has been explicitly checked with Feynman diagrams (this large $z$ analysis could also be performed on squared Feynman diagrams with the same result.)  As we described in the last subsection, the two momenta that are shifted must come from opposite sides of the propagator.  Therefore, there are eight different spinor shifts that are allowed.  Each of them shifts either $p_1$ or $p_2$, but not both.  Therefore, the denominator grows as $z^2$ for every choice.  On the other hand, no matter what choice of particle $i$ and $j$ in the shift $\lbrack\mathbf{i,j}\rangle$ we make, it will always shift two spinors in at least one term in the numerator.  Therefore, the numerator also grows as $z^2$.  Altogether, this amplitude asymptotically approaches a constant in the large $z$ limit.
\begin{equation}
    \lim_{z\to\infty}\mathcal{M} = \mathcal{O}(z^0)
    \ .
\end{equation}
Since this amplitude does not vanish in the limit for any choice of momentum shift, the theorem does not apply.  In fact, as we show in App.~\ref{sec:Constructive Massless eemm}, we are unable to obtain this amplitude using a massless photon with the present set of tools.  However, on the other hand, we are able to find this amplitude beginning with a massive photon and taking the massless limit.  In fact, if we look at the amplitude before taking the massless limit,
\begin{equation}
  \mathcal{M}%^{ee\mu\mu} 
    = 
    e^2\frac{
    \langle\mathbf{13}\rangle\lbrack\mathbf{24}\rbrack
    +\lbrack\mathbf{13}\rbrack\langle\mathbf{24}\rangle
    +\lbrack\mathbf{14}\rbrack\langle\mathbf{23}\rangle
    +\langle\mathbf{14}\rangle\lbrack\mathbf{23}\rbrack
    }{\left(s-M_A^2\right)}
    \ ,
\end{equation}
we can see that it still does not vanish in the large $z$ limit of a momentum shift.  Nevertheless, we are able to find this amplitude as a product of smaller amplitudes divided by a propagator denominator.  Further details for this amplitude can be found in App.~\ref{app:eemm large z}.

For the process $e\gamma^+\bar{e}h$, the amplitude is given by Eq.~(\ref{eq:main:M_eAeh}).  In this case, we have two propagator denominators.  Because of this, some shifts will straddle both propagators and lead to $z^4$ growth in the denominator and others will only straddle one of the propagators and lead to $z^2$ growth in the denominator.  Additionally, there is great variety in the structure of the spinors in the numerator including a complete lack of spinors for the Higgs boson and only square spinors for the photon.  This suggests that shifts including the photon or the Higgs might have lower $z$ growth in the numerator.  All of this results in a variety of large-$z$ behavior for this process depending on the choice of spinor shift.  Of all the shifts, we find that $\lbrack\mathbf{4},2\rangle, \lbrack\mathbf{1,3}\rangle, \lbrack\mathbf{3,1}\rangle, \lbrack2,\mathbf{4}\rangle, \lbrack\mathbf{3},2\rangle$ and $\lbrack\mathbf{1},2\rangle$ lead to a vanishing amplitude in the large $z$ limit.  Our calculation of this amplitude was successful using either a massive photon or using a massless photon as we show in App.~\ref{sec:Constructive Massless heAe}.  We include further details for the large-$z$ behavior for this process in App.~\ref{app:large z:eAeh}. 

Turning to $e\bar{e}\gamma^+\gamma^+$, the amplitude is very simple
\begin{equation}
    \mathcal{M}^{++} = 
    \frac{e^2  m_e \lbrack34\rbrack ^{2} \langle\mathbf{1}\mathbf{2}\rangle }
    {\left(t-m_e^2\right)\left(u-m_e^2\right)} \ .
\end{equation}
Once again, we have a choice of shifts that straddle both propagators leading to $z^4$ denominator growth and others that only straddle one of the propagators leading to $z^2$ growth in the denominator.  The numerator is equally easy to analyze.  Any spinor shift that only has angle brackets for the photons and ony square brackets for the electron and positron does not lead to any numerator $z$ growth at all.  Therefore, we can choose many spinor shifts leading to a vanishing asymptotic limit.  They are the shifts $\lbrack\mathbf{1,2}\rangle, \lbrack\mathbf{2,1}\rangle, \lbrack3,4\rangle, \lbrack4,3\rangle, \lbrack\mathbf{1},3\rangle, \lbrack\mathbf{1},4\rangle, \lbrack\mathbf{2},3\rangle$ and $\lbrack\mathbf{2},4\rangle$.  We were able to find an amplitude that agreed with Feynman diagram for this process using both a massless photon and a massive photon.  
The amplitude for the process $e\bar{e}\gamma^+\gamma^-$ is nearly as simple and is
\begin{equation}
\mathcal{M}^{+-} = \frac{e^2\left(\lbrack\mathbf{1}3\rbrack\langle\mathbf{2}4\rangle+\langle\mathbf{1}4\rangle\lbrack\mathbf{2}3\rbrack\right)\lbrack3\lvert p_1\rvert4\rangle}{(t-M_e^2)(u-M_e^2)} \ .
\end{equation}
In the same way, we can see that this amplitude vanishes in the asymptotic limit for the shifts $\lbrack4,3\rangle, \lbrack\mathbf{2},3\rangle, \lbrack4,\mathbf{2}\rangle, \lbrack\mathbf{1,2}\rangle$ and $\lbrack\mathbf{2,1}\rangle$.
Greater detail for these processes can be found in App.~\ref{app:large z:eeAA}.

\subsection{\label{sec:main:constructive}Constructive Calculations with a Momentum and Spinor Shift}
We have already done the initial steps of the calculation for $e\bar{e}\mu\bar{\mu}$ and begin with the amplitude in the form
\begin{align}
    \mathcal{M} =
    &\ \frac{e^2}{2m_em_\mu s}  \Big[
    (u-t+2m_e^2+2m_\mu^2)
    \lbrack\mathbf{1}\mathbf{2}\rbrack 
    \lbrack\mathbf{3}\mathbf{4}\rbrack
    \nonumber\\
    &+2\left(
    \lbrack\mathbf{12}\rbrack \lbrack\mathbf{3}\lvert p_2p_1\rvert\mathbf{4}\rbrack+
    \lbrack\mathbf{1}\lvert p_4p_3\rvert\mathbf{2}\rbrack \lbrack\mathbf{34}\rbrack 
    \right)
    \Big]\ ,
\end{align}
from Eq.~(\ref{eq:main:M_eemm massless no shift}).   

None of the momentum shifts cause the correct amplitude to vanish in the limit of large $z$ (see App.~\ref{app:eemm large z}).  Therefore, we should not expect that a momentum shift must work.  Nevertheless, we attempt it anyway.   There are eight possible momentum shifts that straddle the s-channel propagator.  For example, a $\lbrack\mathbf{1,3}\rangle$ shift has the consequence that $\hat{u}=u$ and the following spinor relations
\begin{align}
    \langle\mathbf{12}\rangle\langle\hat{\mathbf{3}}\mathbf{4}\rangle
    &= \langle\mathbf{12}\rangle\lbrack\mathbf{34}\rbrack    
    \\
    \lbrack\hat{\mathbf{1}}\mathbf{2}\rbrack\langle\hat{\mathbf{3}}\mathbf{4}\rangle
    &= \langle\mathbf{12}\rangle\lbrack\mathbf{34}\rbrack
    \\
    \langle\mathbf{12}\rangle\lbrack\mathbf{3}\mathbf{4}\rbrack
    &=  \langle\mathbf{12}\rangle\lbrack\mathbf{34}\rbrack
    \\
    \lbrack\hat{\mathbf{1}}\mathbf{2}\rbrack\lbrack\mathbf{3}\mathbf{4}\rbrack
    &= \langle\mathbf{12}\rangle\lbrack\mathbf{34}\rbrack\ .
\end{align}
Interestingly, all the choices of spinor brackets are transformed into the same $\langle\mathbf{12}\rangle\lbrack\mathbf{34}\rbrack$ by this momentum shift.  The other terms are similarly transformed and the final result is
\begin{align}
    \mathcal{M}_{\lbrack\mathbf{1,3}\rangle,\lbrack\mathbf{2,4}\rangle,\lbrack\mathbf{1,4}\rangle,\lbrack\mathbf{2,3}\rangle} =
    &\frac{e^2(u-t)}{2m_em_\mu s}  
    \langle\mathbf{1}\mathbf{2}\rangle
    \lbrack\mathbf{3}\mathbf{4}\rbrack
    \ .
    \label{eq:M_eemm shifted A}
\end{align}
We get the same result if we do the shifts $\lbrack\mathbf{2,4}\rangle,\lbrack\mathbf{1,4}\rangle$ or $\lbrack\mathbf{2,3}\rangle$.  On the other hand, if we try the other momentum shifts, we obtain the conjugate
\begin{align}
    \mathcal{M}_{\lbrack\mathbf{3,1}\rangle,\lbrack\mathbf{4,2}\rangle,\lbrack\mathbf{4,1}\rangle,\lbrack\mathbf{3,2}\rangle} =
    &\frac{e^2(u-t)}{2m_em_\mu s}  
    \lbrack\mathbf{12}\rbrack
    \langle\mathbf{34}\rangle
    \ .
    \label{eq:M_eemm shifted B}
\end{align}
After squaring, there is only one unique form, since $\lvert\langle\mathbf{12}\rangle\lbrack\mathbf{34}\rbrack \rvert^2=\lvert\lbrack\mathbf{12}\rbrack\langle\mathbf{34}\rangle \rvert^2$.  Unfortunately, this result is wrong.  In fact, this is easily seen in the high-energy behavior where these amplitudes grow for all energies, due to the $u-t$ in the numerator, and would violate perturbative unitarity.  Further details for this process can be found in App.~\ref{sec:Constructive Massless eemm}.

We next turn to the process $e\gamma^+\bar{e}h$, with one external photon.  For this process, there are potentially two diagrams that could be used, we only need one of them.  Since we demonstrated the u-channel diagram in Sec.~\ref{sec:main:constructive no Shift}, we will perform the s-channel diagram here.  We have the Higgs-electron vertex  $\frac{m_e}{v}(\langle\mathbf{ P}_{12}^{\mathrm{I}}\mathbf{3}\rangle+\lbrack\mathbf{P}_{12}^{\mathrm{I}}\mathbf{3}\rbrack)$ and the photon-electron vertex $e(x_{1,34}\langle\mathbf{1P}_{34\mathrm{I}}\rangle)$.  Multiplying and dividing by the propagator denominator, we have
\begin{equation}
    \mathcal{M}_s^{+} =
    \frac{em_e}{v}\frac{x_{1,34}
    (2m_e\langle\mathbf{13}\rangle
    -\langle\mathbf{1}\lvert p_4\rvert\mathbf{3}\rbrack)
    }
    {s-m_e^2}\ ,
\end{equation}
after using the spin-contraction, mass and momentum conservation identities.

Our next task is to remove the $x_{1,34}$.  Following the rules discussed in App.~\ref{sec:x = alternate}, we find
\begin{equation}
    x_{1,34} =         \frac{\lbrack2|p_1p_3|2\rbrack}{m_e\left(u-m_e^2\right)}\ ,
\end{equation}
where we have used the form that contains the propagator denominator from the other diagram that we are not analyzing.  
Plugging this in, we find
\begin{equation}
    \mathcal{M}_s^{+} =
    \frac{e}{v}\frac{\lbrack2\lvert p_1p_3\rvert2\rbrack
    (2m_e\langle\mathbf{13}\rangle
    -\langle\mathbf{1}\lvert p_4\rvert\mathbf{3}\rbrack)
    }
    {\left(s-m_e^2\right)\left(u-m_e^2\right)}\ .
\end{equation}
We can see that this has directly exposed the positive helicity spinors for the photon, as expected.  We show in App.~\ref{app:large z:eAeh} that the amplitude for this process vanishes in the large-$z$ limit for the shifts $\lbrack\mathbf{4},2\rangle, \lbrack\mathbf{1},\mathbf{3}\rangle, \lbrack\mathbf{3},\mathbf{1}\rangle, \lbrack2,\mathbf{4}\rangle, \lbrack\mathbf{3},2\rangle$ or $ \lbrack\mathbf{1},2\rangle$.  Some of these momentum shifts straddle both propagator denominators, not just the denominator of the u-channel diagram that we are working on.  For this reason, we might wonder whether we should hold $(u-m_e^2)$ fixed, and not shift it, even though it is not the original propagator denominator.  The answer, we find, is that we must hold it fixed, no matter which momentum shift we choose, since it is already correct.  Therefore, this suggests that holding the denominator coming from an external $x$ factor fixed during momentum shifts might be a further rule to be followed in constructive calculations, although further examples should be considered to be sure.  Of the shifts, we only have simple formulas for the shifts where at least one of the shifted momenta are for massless particles.  We have tried the remaining four momentum shifts and have found the same unique result that agrees with Feynman diagrams with all of them except $\lbrack2,\mathbf{4}\rangle$, which we were unable to simplify to the correct form.  We have also found the same success with the u-channel diagram.  For example, if we do the $\lbrack\mathbf{4},2\rangle$ shift, which has momenta on both sides of both the s- and u-channel diagrams, we obtain
\begin{equation}
    \hat{\mathcal{M}}_s^{+} =
    \frac{e}{v}\frac{\lbrack2\lvert p_1p_3\rvert2\rbrack
    (2m_e\langle\mathbf{13}\rangle
    -\langle\mathbf{1}\lvert \hat{p}_4\rvert\mathbf{3}\rbrack)
    }
    {\left(s-m_e^2\right)\left(u-m_e^2\right)}\ ,
\end{equation}
where only the momentum $p_4$ is shifted in the numerator.  We then plug in the shift
\begin{equation}
    \langle\mathbf{1}\lvert \hat{p}_4\rvert\mathbf{3}\rbrack = \langle\mathbf{1}\lvert p_4\rvert\mathbf{3}\rbrack
    + \langle\mathbf{1}\lvert p_4\rvert 2\rbrack\lbrack2\mathbf{3}\rbrack
    \frac{(s-m_e^2)}{\lbrack 2\lvert p_4  p_3\rvert 2\rbrack}
    \ .
\end{equation}
After simplifying by applying the usual identities multiple times, we finally obtain
\begin{eqnarray}
    \mathcal{M}^{+} &=& 
    \frac{e m_e}{v\left(s-m_e^2\right)\left(u-m_e^2\right)}
      \Big( m_h ^{2} \lbrack\mathbf{1}2\rbrack\lbrack2\mathbf{3}\rbrack 
      \nonumber\\
    &&- m_e \lbrack\mathbf{1}2\rbrack\lbrack2\lvert p_{4} \rvert \mathbf{3}\rangle 
    + m_e \lbrack2\mathbf{3}\rbrack \langle\mathbf{1}\lvert p_{4} \rvert 2\rbrack
    \nonumber\\
    &&- \langle\mathbf{1}\mathbf{3}\rangle \lbrack2\lvert p_{3} p_{4} \rvert 2\rbrack
    \Big)\ .
\end{eqnarray}
Further details for this process can be found in App.~\ref{sec:Constructive Massless heAe}.

Turning to processes with two external photons, we begin with the process $e\bar{e}\gamma^+\gamma^+$.  Once again, there are two diagrams, this time in the t and u channels.  Once again, we can use either diagram and do not need both since they both give identical final results.  We will demonstrate the t channel, but have also done the u channel with full agreement.  The vertices are $e x_{1,24}\langle\mathbf{1P}_{24\mathrm{I}}\rangle$ and $e x_{13,2}\langle\mathbf{P}_{13}^{\mathrm{I}}\mathbf{2}\rangle$, giving us the amplitude
\begin{equation}
    \mathcal{M}_t^{++} =
    \frac{e^2 m_e x_{1,24}x_{13,2}\langle\mathbf{12}\rangle}{t-m_e^2} 
    \ ,
\end{equation}
where we have already simplified the contracted spinor index.  As in the previous process, we must now replace the $x$ factors.  Now that we have two $x$ factors, we must decide which propagator denominators to use in each of them.  It is convenient to use the other propagator denominator for at least one of them, but each choice is related by momentum conservation on shell.  We will demonstrate the shift $\lbrack\mathbf{1},4\rangle$, and replace $x$ with
\begin{align}
    x_{1,24} 
    &= \frac{\lbrack3\lvert p_1p_2\rvert3\rbrack}{m(u-m^2)}
    \\
    x_{13,2} 
    &= \frac{\lbrack4\lvert p_1p_2\rvert4\rbrack}{m(u-m^2)}
    \ ,
\end{align}
giving us
\begin{equation}
    \hat{\mathcal{M}}_t^{++} = 
    \frac{e^2 \langle\mathbf{12}\rangle \lbrack3\lvert \hat{p}_1p_2\rvert3\rbrack\lbrack4\lvert \hat{p}_1p_2\rvert4\rbrack }{m_e\left(t-m_e^2\right)\left(u-m_e^2\right)^2} \ .
\end{equation}
As before, since the denominator is already what we want, we only shifted the numerator.  Furthermore, we can see again that the positive helicity spinors for the two positive-helicity photons are immediately resolved.  Our next step is to plug in the momentum shifts, which after use of the standard identities, is
\begin{align}
    \lbrack3\lvert \hat{p}_1p_2\rvert3\rbrack\lbrack4\lvert \hat{p}_1p_2\rvert4\rbrack
    &= -m_e^2\left(u-m_e^2\right)\lbrack34\rbrack^2.
\end{align}
We see that the extra propagator denominator $(u-m_e^2)$ cancels and we are left with
\begin{equation}
   \mathcal{M}_t^{++} = 
    -\frac{e^2 m_e \langle\mathbf{12}\rangle \lbrack34\rbrack^2 }
    {\left(t-m_e^2\right)\left(u-m_e^2\right)} \ ,
\end{equation}
which agrees with Feynman diagrams.  (Presumably, if we had been doing a QCD calculation with quarks and gluons, since all 3 diagrams contribute, we would need to include both the other propagator denominators and they would not cancel.)
Further details for this process can be found in App.~\ref{app:eeA+A+}.

For our final process, we consider $e\bar{e}\gamma^+\gamma^-$.  
Of the two diagrams, in the t and u channels, this time we will demonstrate the u-channel diagram.  As before, we have done the calculation with both diagrams and find exactly the same final amplitude with either one.  The vertices are $e \tilde{x}_{1,23}\lbrack\mathbf{1P}_{23\mathrm{I}}\rbrack$ and $e x_{14,2}\langle\mathbf{P}_{14}^{\mathrm{I}}\mathbf{2}\rangle$, giving the initial amplitude
\begin{equation}
    \mathcal{M}_u^{+-} =
    \frac{e^2 \tilde{x}_{1,23}x_{14,2}
    \left(
    m_e\langle\mathbf{12}\rangle
    +\lbrack\mathbf{1}\lvert p_4\rvert \mathbf{2}\rangle
    \right)}
    {u-m_e^2}
    \ ,
\end{equation}
where we have used the spin-index contraction and mass identities.  Similar to the $++$-helicity case, we must replace one $x$ with a form containing $(t-m_e^2)$, but momentum conservation and the on-shell condition relates the choices.  We see in App.~\ref{app:large z:eeAA} that there are five momentum shifts that cause the amplitude to vanish at large $z$.  Of these, we have simple formulas for $\lbrack4,3\rangle, \lbrack\mathbf{2},3\rangle$ and $\lbrack4,\mathbf{2}\rangle$.  This time we will do the shift $\lbrack4,3\rangle$, which will make the second shift including the propagator denominator $s$ convenient.  Therefore, we use 
\begin{align}
    \tilde{x}_{1,23} &=
    \frac{\langle4\lvert p_1p_3\rvert4\rangle}{m_e\ s}
    \\
    x_{14,2} &=
    \frac{\lbrack3\lvert p_1p_2\rvert3\rbrack}{m_e(t-m_e^2)}
    \ .
\end{align}
Making these replacements and shifting the momenta gives
\begin{equation}
    \hat{\mathcal{M}}_u^{+-} =
    \frac{e^2 
    \lbrack3\lvert p_1p_2\rvert3\rbrack
    \langle4\lvert p_1\hat{p}_3\rvert4\rangle
    \left(
    m_e\langle\mathbf{12}\rangle
    +\lbrack\mathbf{1}\lvert \hat{p}_4\rvert \mathbf{2}\rangle
    \right)}
    {m_e^2\ s \left(t-m_e^2\right)\left(u-m_e^2\right)}
    \ .
\end{equation}
Our momentum shift is
\begin{equation}
    \hat{p}_3 =
    p_3 + \lvert3\rbrack\langle4\rvert \frac{u-m_e^2}{\langle4\lvert p_1\rvert 3\rbrack}
    \ ,
\end{equation}
which gives a $\langle44\rangle$, causing $\langle4\lvert p_1\hat{p}_3\rvert4\rangle=\langle4\lvert p_1p_3\rvert4\rangle$, due to the masslessness of particle $4$.  We also need \begin{equation}
    \lbrack\mathbf{1}\lvert \hat{p}_4\rvert \mathbf{2}\rangle = \lbrack\mathbf{1}\lvert p_4\rvert \mathbf{2}\rangle - \lbrack\mathbf{1}3\rbrack\langle4\mathbf{2}\rangle 2p_1\cdot p_4/\langle4\lvert p_1\rvert3\rbrack
    \ .
\end{equation}
After a series of identities and simplifications, $s$ can be factored out and cancel the $s$ in the denominator.  The final form can be brought to
\begin{equation}
    \mathcal{M}_u^{+-} =
    \frac{e^2 \lbrack3\lvert p_1\rvert4\rangle
    \left(
    \langle\mathbf{2}4\rangle\lbrack\mathbf{1}3\rbrack 
    + \langle\mathbf{1}4\rangle\lbrack\mathbf{2}3\rbrack
    \right)
    }
    {\left(t-m_e^2\right)\left(u-m_e^2\right)}
    \ .
\end{equation}
This form agrees with Feynman diagrams.  We have also done this calculation with the other momentum shifts and for both diagrams and get the same result.
Further details can be found in App.~\ref{app:eeA+A-}.

\subsection{\label{sec:main:amplitudes}The Amplitudes}
In this section, we add further comments about the spinor amplitudes for each process.

\subsubsection{\label{sec:eemm}$\mathbf{e,\bar{e},\bar{\mu},\mu}$ and $\mathbf{e,\bar{e},\bar{e},e}$}

We begin with the process $e\bar{e}\to\mu\bar{\mu}$.  As we saw in Sec.~\ref{sec:main:MA} and in greater detail in App.~\ref{sec:eemm spinor amp calculation}, we can calculate this amplitude using a intermediate massive photon and take the massless limit.  Unfortunately, as we saw in Sec.~\ref{sec:main:constructive} and in greater detail in App.~\ref{sec:Constructive Massless eemm}, we do not yet have the ingredients necessary to calculate the correct spinor amplitude using the constructive technique with massless photons.  

Before giving the result, let us note that the general structure of the amplitude allows us to enumerate the possible spinor terms that could appear.  In particular, since all four external particles are massive spin-$1/2$ particles, each term must be composed of one spin spinor for each particle.  Each of these can be either angle or square spinors.  There are ten possible combinations with no momenta sandwiched between the spinors, namely     $\langle\mathbf{12}\rangle\lbrack\mathbf{34}\rbrack$, 
$\langle\mathbf{13}\rangle\lbrack\mathbf{24}\rbrack$, 
$\langle\mathbf{14}\rangle\lbrack\mathbf{23}\rbrack$, 
$\lbrack\mathbf{12}\rbrack\langle\mathbf{34}\rangle$, 
$\lbrack\mathbf{13}\rbrack\langle\mathbf{24}\rangle$,  
$\lbrack\mathbf{14}\rbrack\langle\mathbf{23}\rangle$, 
$\langle\mathbf{13}\rangle\langle\mathbf{24}\rangle$, 
$\langle\mathbf{14}\rangle\langle\mathbf{23}\rangle$, 
$\lbrack\mathbf{13}\rbrack\lbrack\mathbf{24}\rbrack$ and 
$\lbrack\mathbf{14}\rbrack\lbrack\mathbf{23}\rbrack$.  (We do not include
$\langle\mathbf{12}\rangle\langle\mathbf{34}\rangle$ and 
$\lbrack\mathbf{12}\rbrack\lbrack\mathbf{34}\rbrack$ because they can be obtained from the others by a Schouten transformation.)  We further note that, due to the vectorial nature of the photon, we expect a symmetry between angle and square brackets.  Therefore, we expect the amplitude to contain some combination of $\langle\mathbf{12}\rangle\lbrack\mathbf{34}\rbrack+\lbrack\mathbf{12}\rbrack\langle\mathbf{34}\rangle, \langle\mathbf{13}\rangle\lbrack\mathbf{24}\rbrack+\lbrack\mathbf{13}\rbrack\langle\mathbf{24}\rangle, \langle\mathbf{14}\rangle\lbrack\mathbf{23}\rbrack+\lbrack\mathbf{14}\rbrack\langle\mathbf{23}\rangle, \langle\mathbf{13}\rangle\langle\mathbf{24}\rangle+\lbrack\mathbf{13}\rbrack\lbrack\mathbf{24}\rbrack$ and $\langle\mathbf{14}\rangle\langle\mathbf{23}\rangle+\lbrack\mathbf{14}\rbrack\lbrack\mathbf{23}\rbrack$.  We could also consider spinor products with a momentum, however, it turns out not to be necessary in this case and we prefer to present the simplest possible form of the amplitude.  In the absence of the massless-limit approach, we could try different combinations of these terms until we find one that agrees with Feynman diagrams.  In fact, this is how we initially found this amplitude.  As we saw in Sec.~\ref{sec:main:MA}, the correct amplitude only contains two of these, both with the same coefficient.  The amplitude is
\begin{equation}
    \mathcal{M} = \frac{e^2}{s}
    \left(\langle\mathbf{13}\rangle\lbrack\mathbf{24}\rbrack \!+\!
          \langle\mathbf{14}\rangle\lbrack\mathbf{23}\rbrack \!+\!
          \lbrack\mathbf{13}\rbrack\langle\mathbf{24}\rangle \!+\!
          \lbrack\mathbf{14}\rbrack\langle\mathbf{23}\rangle
    \right).
    \label{eq:M_eemm}
\end{equation}
We have checked explicitly and our result agrees analytically and exactly with Feynman diagrams at all energies and all angles.  In particular, following the procedures outlined in \cite{Christensen:2019mch}, we have squared our amplitude and checked this expression against the Feynman diagram result output by CalcHEP\cite{Belyaev:2012qa}.

As we can see, this amplitude does not contain any momenta in the numerator.  Further, every term is composed of an electron or positron paired with a muon or anti-muon in all possible combinations where one spinor product is an angle product and the other is a square product. 

We also note that the published version of this amplitude \cite{Arkani-Hamed:2017jhn} is given by 
\begin{equation}
    \mathcal{M}^{AHH} = \frac{e^2}{s} 
    \frac{\langle\mathbf{12}\rangle\langle\mathbf{34}\rangle}
         {m}
    \left(p_1-p_2\right) \!\cdot\! p_3 \ .
    \label{eq:M_eemm-Arkani}
\end{equation}
One may initially suspect that these amplitudes are equivalent after some sort of transformation, such as some combination of momentum conservation and Schouten identities.   However, this is not the case and, as we show in App.~\ref{sec:Constructive Massless eemm}, momentum shifts do not correct this amplitude.  In fact, we can see that this expression grows for all energies, violating perturbative unitarity, unlike the correct amplitude given above.

Now that we have the amplitude for $e\bar{e}\mu\bar{\mu}$, we can construct the amplitude for $e\bar{e}e\bar{e}$.  There are now two diagrams.  The $s$-channel diagram from the previous muon case and a $t$-channel diagram.  The amplitude is 
\begin{align}
    \mathcal{M} &= \frac{e^2}{s}
    \left(\langle\mathbf{13}\rangle\lbrack\mathbf{24}\rbrack \!+\!
          \langle\mathbf{14}\rangle\lbrack\mathbf{23}\rbrack \!+\!
          \lbrack\mathbf{13}\rbrack\langle\mathbf{24}\rangle \!+\!
          \lbrack\mathbf{14}\rbrack\langle\mathbf{23}\rangle
    \right)
    \nonumber\\
    &- \frac{e^2}{t}
    \left(\langle\mathbf{12}\rangle\lbrack\mathbf{34}\rbrack \!-\!
          \langle\mathbf{14}\rangle\lbrack\mathbf{23}\rbrack \!+\!
          \lbrack\mathbf{12}\rbrack\langle\mathbf{34}\rangle \!-\!
          \lbrack\mathbf{14}\rbrack\langle\mathbf{23}\rangle
    \right),
    \label{eq:M_eeee}
\end{align}
where we have interchanged $2\leftrightarrow3$.  We then used the antisymmetry property $\langle\mathbf{32}\rangle=-\langle\mathbf{23}\rangle$ and $\lbrack\mathbf{32}\rbrack=-\lbrack\mathbf{23}\rbrack$.  The relative negative sign between diagrams comes from the exchanged identical fermions.  Indeed, we have computed the squared amplitude and compared with Feynman diagrams and found exact agreement.  

We note that, even with two diagrams, this amplitude was still reducible to a form with no momenta in the numerator.   Also, as far as we can tell, we do not gain anything by combining the two terms into an expression with a common denominator.

\subsubsection{\label{sec:M_eAeh}$\mathbf{e,\gamma^{\pm},\bar{e},h}$}
We next turn to a process with only a single external photon.  We add the Higgs boson and construct $e\gamma^{\pm}\bar{e}h$.  The structure of this amplitude tells us we must have two helicity spinors for the photon, both square if the photon has positive helicity or both angle if the photon has negative helicity, and one spin spinor each for the electron and positron that can be either angle or square.  For example, let's first consider the process $e\gamma^+\bar{e}h$.  In this case, there is only one possible term with no momenta.  It is $\lbrack\mathbf{1}2\rbrack\lbrack2\mathbf{3}\rbrack$.  This turns out not to be enough and we need to consider terms with a momenta.  With one momentum, we could construct $\langle\mathbf{1}\lvert p_4\rvert2\rbrack\lbrack2\mathbf{3}\rbrack$ or $\lbrack\mathbf{1}2\rbrack\lbrack2\lvert p_4\rvert\mathbf{3}\rangle$.  The only other possibility would be to use $p_3$ or $p_1$ instead of $p_4$, but that would be equivalent to these by momentum conservation and the mass identities.  Finally, it turns out that we also need terms with two momenta.  There are only two unique terms possible.  They are $\langle\mathbf{13}\rangle\lbrack2\lvert p_3p_4\rvert2\rbrack$ and $\lbrack\mathbf{13}\rbrack\lbrack2\lvert p_3p_4\rvert2\rbrack$.  Other forms, such as $\langle\mathbf{1}\lvert p_4\rvert2\rbrack\lbrack2\lvert p_4\rvert\mathbf{3}\rangle$, can be converted to these by use of Schouten and other identities.  Once again, we could try different combinations of these terms until we find agreement with Feynman diagrams and this is how we initially found this amplitude.  Of these five possible terms, the correct amplitude uses four of them and each has a mass coefficient to bring them to the same mass dimension.  It is
\begin{align} \nonumber
\mathcal{M}^+ &=
    \frac{e m_e}
         {v\left(s-m_e^2\right)\left(u-m_e^2\right)} \\ \nonumber
     & \times
      \Big( m_h ^{2} 
            \lbrack\mathbf{1}2\rbrack
            \lbrack2\mathbf{3}\rbrack 
      - m_e \lbrack\mathbf{1}2\rbrack
            \lbrack 2\lvert p_{4} \rvert \mathbf{3}\rangle \nonumber \\
     &+ m_e \lbrack2\mathbf{3}\rbrack 
            \langle\mathbf{1}\lvert p_{4} \rvert 2\rbrack
          - \langle\mathbf{1}\mathbf{3}\rangle 
            \lbrack2\lvert p_{3} p_{4} \rvert 2\rbrack \Big).
\label{eq:M_eA+eh}         
\end{align}
For the negative-helicity photon case, $e\gamma^-\bar{e}h$, we find
\begin{align} \nonumber
    \mathcal{M}^- = &
    \frac{e m_e}{v\left(s-m_e^2\right)\left(u-m_e^2\right)} \times \qquad \\ \nonumber
    & \Big( 
      m_h ^{2} \langle \mathbf{1} 2 \rangle
               \langle 2 \mathbf{3} \rangle  
      - m_e \langle \mathbf{1} 2 \rangle
            \langle 2 \lvert p_{4} \rvert \mathbf{3} \rbrack \\
    & + m_e \langle 2 \mathbf{3} \rangle 
            \lbrack \mathbf{1} \lvert p_{4} \rvert 2 \rangle 
      - \lbrack \mathbf{1} \mathbf{3} \rbrack 
        \langle 2 \lvert p_{3} p_{4} \rvert 2 \rangle
   \Big)\ ,
   \label{eq:M_eAeh}
\end{align}
with all angle brackets replaced with square brackets and vice versa.
In addition to trying multiple different forms and testing, we found this form both as the massless limit of a massive photon theory in Sec.~\ref{sec:main:MA} and App.~\ref{sec:eAeh spinor amp calculation} and by a purely massless-photon theory using the $x$ factor in Sec.~\ref{sec:main:constructive no Shift} and ~\ref{sec:Constructive Massless heAe no Shift} and also using momentum shifts in Sec.~\ref{sec:main:constructive} and App.~\ref{sec:Constructive Massless heAe}.

In order to compare this with Feynman diagrams, we calculated the same process using Feynman diagrams and multiplied by the positive-helicity (negative-helicity) polarization vector before squaring for the positive- (negative-)helicity case.  We found exact analytic agreement with our result.  We also summed the squared amplitude over helicities and compared with the output of CalcHEP and found exact analytic agreement.

An interesting feature of the propagator denominators in the cases where there are external photons is that they can be written more compactly and suggestively in spinor form.  For example, $(s-m_e^2)=2p_1\cdot p_2 = \lbrack2\lvert p_1\rvert2\rangle$ and $(u-m_e^2)=2p_2\cdot p_3=\lbrack2\lvert p_3\rvert2\rangle$, where both have the square and angle helicity spinors for the photon and have the electron and positron momenta in the middle.  With this, we could write the positive-helicity amplitude as
\begin{align} \nonumber
\mathcal{M}^+ &=
    \frac{e m_e}
         {v\lbrack2\lvert p_1\rvert2\rangle\lbrack2\lvert p_3\rvert2\rangle} \\ \nonumber
    & \times
    \Big( m_h ^{2} 
          \lbrack\mathbf{1}2\rbrack
          \lbrack2\mathbf{3}\rbrack 
    - m_e \lbrack\mathbf{1}2\rbrack
          \lbrack2\lvert p_{4} \rvert \mathbf{3}\rangle \nonumber \\
   &+ m_e \lbrack2\mathbf{3}\rbrack 
          \langle\mathbf{1}\lvert p_{4} \rvert 2\rbrack
        - \langle\mathbf{1}\mathbf{3}\rangle 
          \lbrack2\lvert p_{3} p_{4} \rvert 2\rbrack \Big),
\end{align}
and similarly for the negative-helicity photon.  This denominator likely generalizes in an interesting way to amplitudes with more photons and more electron lines.

\subsubsection{\label{sec:M_eeA+A+}$\mathbf{e,\bar{e},\gamma^{\pm},\gamma^{\pm}}$}
Turning to the process with two same-helicity photons, we consider $e,\bar{e},\gamma^{\pm},\gamma^{\pm}$.  In order to obtain the correct transformation properties, each term of the amplitude must contain the right spinors.  For example, for the $e,\bar{e},\gamma^+,\gamma^+$ case, each term must have two square helicity spinors for particle 3, two square helicity spinors for particle 4, and one spin spinor of either type for particle 1 and also for particle 2.  If we do not need a term with a momentum, there are only a few possibilities.  Particles 3 and 4 can be together in terms like $\lbrack34\rbrack^2\langle\mathbf{12}\rangle$ and $\lbrack34\rbrack^2\lbrack\mathbf{12}\rbrack$.  We could also have particles 1 and 2 together with particles 3 and 4 in terms such as $\lbrack34\rbrack\lbrack\mathbf{1}3\rbrack\lbrack\mathbf{2}4\rbrack$ and $\lbrack34\rbrack\lbrack\mathbf{1}4\rbrack\lbrack\mathbf{2}3\rbrack$, but one of these last two are equivalent to the other three by Schouten identities.  We could, and originally did, try different combinations of these terms until we found the amplitude that agrees with Feynman diagrams.  It is
\begin{align}
\mathcal{M}^{++} &= 
   \frac{e^2 m_e \lbrack 34 \rbrack^2 \langle \mathbf{12} \rangle}
        {(t-m_e^2)(u-m_e^2)} .
 \label{eq:M_eeA+A+}
\end{align}
For the double-negative-helicity case, we have
\begin{align}
\mathcal{M}^{--} &= 
   \frac{e^2 m_e \langle 34 \rangle^2 \lbrack \mathbf{12} \rbrack}
        {(t-m_e^2)(u-m_e^2)} .
\end{align}
We have also found these amplitudes as the massless limit of a massive-photon theory in Sec.~\ref{sec:main:MA} and App.~\ref{sec:Massive eeAA} and using a purely massless theory in Sec.~\ref{sec:main:constructive no Shift} and App.~\ref{app:eeA+A+ no shift} without the momentum shift and Sec.~\ref{sec:main:constructive} and App.~\ref{app:eeA+A+} with the momentum shift.  We have checked our results with Feynman diagrams and found exact analytic agreement.  We did this by calculating the Feynman diagrams and multiplying by polarization vectors for the photons before squaring.  We also summed over all the helicity combinations after squaring and comparing with the output of CalcHEP.

As for the single-photon amplitude, we could write the denominator in spinor form to obtain
\begin{align}
\mathcal{M}^{++} 
 &= \frac{e^2 m_e \lbrack 34 \rbrack^2 \langle \mathbf{12} \rangle}
        {\lbrack 3 \lvert p_1 \rvert3 \rangle \lbrack4 \lvert p_1 \rvert 4 \rangle} 
        \label{eq:main:MeeA+A+ alternate den}
        \\
\mathcal{M}^{--} 
 &= \frac{e^2 m_e \langle 34 \rangle^2 \lbrack \mathbf{12} \rbrack}
        {\lbrack 3 \lvert p_1 \rvert 3 \rangle \lbrack 4 \lvert p_1 \rvert 4 \rangle} .
        \label{eq:main:MeeA-A- alternate den}
\end{align}
The denominator of both of these could also be written as $\lbrack3\lvert p_1\rvert3\rangle\lbrack3\lvert p_2\rvert3\rangle = \lbrack4\lvert p_2\rvert4\rangle\lbrack4\lvert p_1\rvert4\rangle$, each related by momentum conservation.  There may be some further insight from these forms.

\subsubsection{\label{sec:M_eeA+A-} $\mathbf{e,\bar{e},\gamma^{\pm},\gamma^{\mp}}$}
Our final process has two photons of opposite helicity.  If we think about the process $e\bar{e}\gamma^+\gamma^-$, each term must have two square helicity spinors for particle 3, two angle helicity spinors for particle 4, and one spin spinor of either type for particle 1 and particle 2.  It turns out that there is no way to write a term with no momenta.  We can't have a spinor product between the square bracket of particle 3 and the angle bracket of particle 4 without any momenta in between and there aren't enough spinors from particles 1 and 2 to contract with all the helicity spinors from particles 3 and 4.  So, this amplitude only has terms with one or more momenta.  If we consider the case where each term has one momentum, we must put the momentum in between the helicity spinors for particle 3 and 4.  Therefore, we have $\lbrack\mathbf{1}3\rbrack\langle\mathbf{2}4\rangle\lbrack3\lvert p_1\rvert4\rangle$ and $\lbrack\mathbf{2}3\rbrack\langle\mathbf{1}4\rangle\lbrack3\lvert p_1\rvert4\rangle$.  Other possibilities are equivalent by application of the identities.  A simple combination of these two terms gives the correct amplitude as
\begin{align}
\mathcal{M}^{+-} &= 
  e^2 
   \frac{ 
     \left( \lbrack\mathbf{1}3\rbrack\langle\mathbf{2}4\rangle
           +\langle\mathbf{1}4\rangle\lbrack\mathbf{2}3\rbrack 
     \right)
   \lbrack 3 \lvert p_1 \rvert 4 \rangle}
        {(t-m_e^2)(u-m_e^2)} .
\label{eq:M_eeA+A-}
\end{align}
Similarly,
\begin{align}
\mathcal{M}^{-+} 
&= e^2 
  \frac{
    \left(
      \langle\mathbf{1}3\rangle\lbrack\mathbf{2}4\rbrack
      +\lbrack\mathbf{1}4\rbrack\langle\mathbf{2}3\rangle
    \right)
    \lbrack4\lvert p_1\rvert3\rangle}
       {(t-m_e^2)(u-m_e^2)} .
       \label{eq:M_-+ 109}
\end{align}
We also find these amplitudes as the massless limit of the massive theory in Sec.~\ref{sec:main:MA} and App.~\ref{sec:Massive eeAA} and with the massless photon theory without a momentum shift in Sec.~\ref{sec:main:constructive no Shift} and App.~\ref{app:eeA+A- no shift} and with a momentum shift in Sec.~\ref{sec:main:constructive} and App.~\ref{app:eeA+A-}.  As in past cases, we have compared this with Feynman diagrams and obtain exact analytic agreement for the squared amplitude, both for the individual helicity-amplitudes squared and for their sum over helicities.  As before, the denominator can be written in multiple ways, such as $(t-m_e^2)(u-m_e^2)=\lbrack 3 \lvert p_1 \rvert 3 \rangle \lbrack 4 \lvert p_1 \rvert 4 \rangle = \lbrack3\lvert p_1\rvert3\rangle\lbrack3\lvert p_2\rvert3\rangle = \lbrack4\lvert p_2\rvert4\rangle\lbrack4\lvert p_1\rvert4\rangle$.  It would be interesting to find the amplitude with more than two photons with a variety of helicities.

The authors of \cite{Arkani-Hamed:2017jhn} also calculated this process.  From their Eq.~(5.21), we copy
\begin{equation}
    \mathcal{M}^{-+} = \frac{e^2\langle3|p_1-p_2|4\rbrack}{2(t-m_e^2)(u-m_e^2)}\left(\langle\mathbf{2}3\rangle\lbrack\mathbf{1}4\rbrack+\langle\mathbf{1}3\rangle\lbrack\mathbf{2}4\rbrack\right)\ ,
\end{equation}
where we have interchanged particles $2$ and $4$ to bring them into the same order as our calculation.  We can simplify this by using momentum conservation in the first spinor product.  If we make the replacement $p_2=-p_1-p_3-p_4$, the $p_3$ and $p_4$ terms vanish due to the masslessness of the photons and we are left with Eq.~(\ref{eq:M_-+ 109}), above.

\section{\label{sec:conclusions}Summary and Conclusions} 
In this paper, we considered the constructive approach to calculating the 4-point scattering amplitudes of QED, namely the processes $e\bar{e}\mu\bar{\mu}$, $e\bar{e}e\bar{e}$, $e\gamma^{\pm}\bar{e}h$, $e\bar{e}\gamma^{\pm}\gamma^{\pm}$ and $e\bar{e}\gamma^{\pm}\gamma^{\mp}$.  We found that, if we begin with a massless photon, the current tools are insufficient for correctly calculating the processes $e\bar{e}\mu\bar{\mu}$ and $e\bar{e}e\bar{e}$, with an internal photon, but that they are sufficient for the processes with an external photon,  $e\gamma^{\pm}\bar{e}h$, $e\bar{e}\gamma^{\pm}\gamma^{\pm}$ and $e\bar{e}\gamma^{\pm}\gamma^{\mp}$.  This, moreover, continues to be true after analytically continuing the momenta and generalizing the spinor shifts to massive spin spinors.  Further refinement of the constructive approach is still necessary.  On the other hand, all of these processes can be obtained if we begin with a massive photon and take the massless limit at the end of the calculation.  These processes are also simple and sufficiently constrained that the amplitudes can be obtained by trial and error.

In greater detail, in Sec.~\ref{sec:main:MA} and App.~\ref{sec:Massive Photon}, we gave the photon a mass $M_A$, updated its vertex in Eqs.~(\ref{eq:M_eeA massive vertex}) and described Taylor expansions of its massive spinors in Eqs.~(\ref{eq:<ij><ik> expansion}) through (\ref{eq:[ij][ik] expansion}).  We then calculated each of the scattering amplitudes of this paper using this massive theory and showed that the correct physical amplitudes were obtained in the massless $M_A\to0$ limit.  For the process $e\bar{e}\mu\bar{\mu}$, the amplitude was given in Eq.~(\ref{eq:main:eemm massive}), for $e\gamma^+\bar{e}h$ in Eq.~(\ref{eq:main:M_eAeh}), for $e\bar{e}\gamma^+\gamma^+$ in Eq.~(\ref{eq:main:M_eeA+A+}) and for $e\bar{e}\gamma^+\gamma^-$ in Eq.~(\ref{eq:main:M_eeA+A- massive}).  

Although using a massive photon was successful in all the amplitudes calculated here, it would be preferable to have a fully consistent massless photon theory.  Giving the photon a mass requires that its two physical helicities be combined into a a spin-1 object that also contains an unphysical helicity-0 object.  The helicity-0 state must fall out of all physical amplitudes in the massless limit, and does in our calculations, but having an unphysical state in intermediate calculations is a shortcoming that we would like to overcome in constructive calculations.  Furthermore, when using a massive photon, every diagram that would be included in a Feynman-diagram calculation must also be included in the massive spinor calculation.  There is no improvement in the efficiency of the calculation.  Whereas, in a massless-photon theory, where successful, when the photon is external, only one of the diagrams is required to give the full result.  This leads to an improved efficiency of the calculation and suggests that the multiple diagrams are redundant and a shortcoming of Feymman diagrams and of the massive-photon spinor calculation.  For these reasons, although success was achieved with the massive photon, we further explore the massless-photon theory.

In Sec.~\ref{sec:main:x} and App.~\ref{app:x}, we review the x factor.  We derive new useful formulas for $x$ and $\tilde{x}$ when the massless particle is external.  The details can be found App.~\ref{sec:x = alternate}.  In App.~\ref{app:xxt+xtx revisited}, we discuss the importance of using the fully symmetrized identities involving $x$.

In Sec.~\ref{sec:main:constructive no Shift} and App.~\ref{app:Calc AHH Spinor Amps No Shifts}, we calculate the amplitudes using massless photons and the $x$ factor.  We begin with $e\bar{e}\mu\bar{\mu}$ and show that the result [in Eq.~(\ref{eq:main:M_eemm massless no shift})] is not in agreement with Feynman diagrams.  On the other hand, we also calculate the amplitude for $e\gamma^+\bar{e}h$ and $e\bar{e}\gamma^+\gamma^+$ and $e\bar{e}\gamma^+\gamma^-$ using the standard constructive technique and find agreement with Feynman diagrams as well as with the results already obtained using the massless limit of a massive photon.

In Sec.~\ref{sec:main:spinor shifts} and App.~\ref{app:analytic continuation}, we describe the analytical continuation of two external momenta on opposite sides of the propagator and the associated shift in the spinors.  We begin with a review of the massless case and then generalize this shift to the massive case.  We begin by finding the shift $\lbrack i,j\rangle$ when the two spinors are massless but the internal line is massive in Eqs.~(\ref{eq:main:[ihat|: 0,0,M}) and (\ref{eq:main:|jhat>: 0,0,M}).  We then find the shift $\lbrack\mathbf{i},j\rangle$ when particle $i$ is massive, particle $j$ is massless and the internal line is massive in Eqs.~(\ref{eq:main:[ihat|: mi,0,M}) and (\ref{eq:main:|jhat>: mi,0,M}).  We find the opposite case $\lbrack i,\mathbf{j}\rangle$ when particle $i$ is massless, particle $j$ is massive and the internal line is massive in Eqs.~(\ref{eq:main:[ihat|: 0,mj,M}) and (\ref{eq:main:|jhat>: 0,mj,M}).  Finally, we turn to the case $\lbrack\mathbf{i,j}\rangle$ where both external particles are massive.  We note that the formulas are excessively complicated and not necessary when the internal line is also massive, but work out the expressions when the internal line is massless and give it in Eqs.~(\ref{eq:main:[ihat|: mi,mj,0}) and (\ref{eq:main:|jhat>: mi,mj,0}).

We consider the large $z$ limit of the analytically continued amplitudes in Sec.~\ref{sec:main:large z} and App.~\ref{app:large z behavior}.   This is important because the proof that amplitudes can be split up into smaller amplitudes in this way relies on the amplitude vanishing for asymptotically large $z$ \cite{Britto:2005fq}.  We find that the amplitude for $e\bar{e}\mu\bar{\mu}$ does not vanish for any choice of momentum shift and suggest that this may be part of the reason that the constructive techniques with a massless photon fail to obtain the correct result.  However, we note that not vanishing for large $z$ does not imply that the amplitude cannot be built up in this way.  In fact, we point out that the amplitude for $e\bar{e}\mu\bar{\mu}$ using a massive photon also does not asymptotically vanish.  Nevertheless, as we showed in Sec.~\ref{sec:main:MA}, the constructive method does succeed in finding this amplitude.  On the other hand, we find that for the processes $e\gamma^{\pm}\bar{e}h$, $e\bar{e}\gamma^{\pm}\gamma^{\pm}$ and $e\bar{e}\gamma^{\pm}\gamma^{\mp}$, the amplitude does vanish for multiple choices of momentum shifts.  We take this as confirmation that the techniques should work for these processes and they do.  

In Sec.~\ref{sec:main:constructive} and App.~\ref{app:Calc AHH Spinor Amps}, we recalculate the amplitudes where we explicitly shift the momenta and the spinors.  Although the large $z$ limit of the process $e\bar{e}\mu\bar{\mu}$ does not vanish, we considered it instructive to see what occurs when the shift is applied to the constructive amplitude.  Unfortuantely, perhaps unsurprisingly, the resulting amplitude in Eqs.~(\ref{eq:M_eemm shifted A}) and (\ref{eq:M_eemm shifted B}) still do not agree with Feynman diagrams.  Using the shifts on the processes $e\gamma^{\pm}\bar{e}h$, $e\bar{e}\gamma^{\pm}\gamma^{\pm}$ and $e\bar{e}\gamma^{\pm}\gamma^{\mp}$ on the other hand, produces the correct final amplitudes, suggesting that the shifts are correct and not the source of the problem with the process $e\bar{e}\mu\bar{\mu}$.  

In Sec.~\ref{sec:main:amplitudes}, we further discuss the correct amplitudes for all the processes discussed in this paper.  For each of them, we note that most of the structure can be determined by considering the transformation properties of the amplitude and that the remaining ambiugiuty can be determined by trial and error.  We also note that all the amplitudes have been compared analytically with Feynman diagrams and found to be in agreement for all energies and all angles.  We extend the result for the process $e\bar{e}\mu\bar{\mu}$ to the process $e\bar{e}e\bar{e}$ by exchanging spinor states $2\leftrightarrow3$ and making appropriate sign changes.  Again, we checked this process against Feynman diagrams.  Although this process with an internal photon has two diagrams, we do not see any simplification occurring by combining with a common denominator, unlike the processes with an external photon and two diagrams where the amplitude is much simpler with a common denominator.  For the processes with an external photon, we also note that the propagator denominator can be written in an alternate way, more reminiscent of purely massless amplitudes.  For example, for the process $e\bar{e}\gamma^{\pm}\gamma^{\pm}$,  $(t-m_e^2)= \lbrack3\lvert p_1\rvert3\rangle$ and $(u-m_e^2)=\lbrack4\lvert p_1\rvert4\rangle$, giving us the amplitude in the form of Eqs.~(\ref{eq:main:MeeA+A+ alternate den}) and (\ref{eq:main:MeeA-A- alternate den}).  This amplitude is suggestive for a generalization for the amplitude with an arbitrary number of same-helicity photons.

In the future, the open question is how to calculate scattering amplitudes constructively with an internal massless photon (or gluon).  New ingredients appear to be needed.  Beyond that, we would like to calculate all the 4-point amplitudes of the Standard Model, followed by higher-point and higher-loop amplitudes and compare with Feynman diagrams to determine whether these spinor amplitudes would potentially improve the efficiency of matrix element generators used to compare theory with experiment at the colliders.  We also hope that these structures will lead to improved understanding of the SM.

\section{Acknowledgements}
We would like to thank Yu-tin Huang for suggesting adding a spin index to $z$ in the massive spinor shifts.

\appendix

\section{\label{sec:Massive Photon}Spinor Amplitudes Using a Massive Photon}

\subsection{\label{sec:eemm spinor amp calculation}$\mathbf{e,\bar{e},\mu,\bar{\mu}}$}
Following the usual rules for massive propagators, as outlined in \cite{Christensen:2019mch}, the amplitude is given by
\begin{align}
    \mathcal{M}%^{ee\mu\mu} 
    &=
    \frac{-e^2}{2 M_A^2\left(s\!-\!M_A^2\right)}
    \left(  \langle\mathbf{1P_{12}}^{I}\rangle
            \lbrack\mathbf{2P_{12}}^{J}\rbrack \!+\! 
            \lbrack\mathbf{1P_{12}}^{I}\rbrack
            \langle\mathbf{2P_{12}}^{J}\rangle
    \right) \nonumber \\
    &\times
     \Big( 
       \langle\mathbf{3P_{12}}_{I}\rangle
       \lbrack\mathbf{4P_{12}}_{J}\rbrack + 
       \lbrack\mathbf{3P_{12}}_{I}\rbrack
       \langle\mathbf{4P_{12}}_{J}\rangle \nonumber \\ 
    & \qquad
      +\langle\mathbf{3P_{12}}_{J}\rangle
       \lbrack\mathbf{4P_{12}}_{I}\rbrack + 
       \lbrack\mathbf{3P_{12}}_{J}\rbrack
       \langle\mathbf{4P_{12}}_{I}\rangle
     \Big) ,
\end{align}
where the minus sign is because the incoming momentum of the first vertex is $P_{34}=-P_{12}$ and the indices $I$ and $J$ are symmetrized since the massive photon is spin 1.  Our next step is to expand this expression and use the contraction rules from \cite{Christensen:2019mch} to obtain
\begin{align}
    \mathcal{M}%^{ee\mu\mu} 
    &= 
    \frac{e^2}{2M_A^2\left(s\!-\!M_A^2\right)}
    \Big(
    M_A^2 \langle\mathbf{13}\rangle
          \lbrack\mathbf{24}\rbrack 
      \!+\!\lbrack\mathbf{1}\lvert P_{12}\rvert\mathbf{3}\rangle
           \langle\mathbf{2}\lvert P_{12}\rvert\mathbf{4}\rbrack \nonumber \\
    & \qquad
      +\langle\mathbf{1}\lvert P_{12}\rvert\mathbf{3}\rbrack
       \lbrack\mathbf{2}\lvert P_{12}\rvert\mathbf{4}\rangle 
      + M_A^2
       \lbrack\mathbf{13}\rbrack
       \langle\mathbf{24}\rangle \nonumber \\
    & \qquad
     +\langle\mathbf{1}\lvert P_{12}\rvert\mathbf{4}\rbrack
      \lbrack\mathbf{2}\lvert P_{12}\rvert\mathbf{3}\rangle
     + M_A^2
      \lbrack\mathbf{14}\rbrack\langle\mathbf{23}\rangle \nonumber \\
    & \qquad
     + M_A^2
      \langle\mathbf{14}\rangle\lbrack\mathbf{23}\rbrack
     +\lbrack\mathbf{1}\lvert P_{12}\rvert\mathbf{4}\rangle
      \langle\mathbf{2}\lvert P_{12}\rvert\mathbf{3}\rbrack
    \Big)\ .
\end{align}
At this point, we use Schouten identities on the terms that do not have a factor of $M_A$.  The second term becomes $\lbrack\mathbf{1}\lvert P_{12}\rvert\mathbf{3}\rangle\langle\mathbf{2}\lvert P_{12}\rvert\mathbf{4}\rbrack = \lbrack\mathbf{4}\lvert P_{12}\rvert\mathbf{3}\rangle\langle\mathbf{2}\lvert P_{12}\rvert\mathbf{1}\rbrack - \lbrack\mathbf{4}\lvert P_{12}^2\rvert\mathbf{1}\rbrack\langle\mathbf{23}\rangle = -\lbrack\mathbf{4}\lvert (p_3+p_4)\rvert\mathbf{3}\rangle\langle\mathbf{2}\lvert (p_1+p_2)\rvert\mathbf{1}\rbrack + M_A^2\lbrack\mathbf{14}\rbrack\langle\mathbf{23}\rangle = m_em_\mu( - \lbrack\mathbf{34}\rbrack\langle\mathbf{12}\rangle + \lbrack\mathbf{34}\rbrack\lbrack\mathbf{12}\rbrack + \langle\mathbf{34}\rangle\langle\mathbf{12}\rangle - \langle\mathbf{34}\rangle\lbrack\mathbf{12}\rbrack ) + M_A^2\lbrack\mathbf{14}\rbrack\langle\mathbf{23}\rangle$, 
and similarly for the third term, $\langle\mathbf{1}\lvert P_{12}\rvert\mathbf{3}\rbrack\lbrack\mathbf{2}\lvert P_{12}\rvert\mathbf{4}\rangle = m_em_\mu( - \langle\mathbf{34}\rangle\lbrack\mathbf{12}\rbrack + \langle\mathbf{34}\rangle\langle\mathbf{12}\rangle + \lbrack\mathbf{34}\rbrack\lbrack\mathbf{12}\rbrack - \lbrack\mathbf{34}\rbrack\langle\mathbf{12}\rangle ) + M_A^2\langle\mathbf{14}\rangle\lbrack\mathbf{23}\rbrack$, 
the fifth term, $\langle\mathbf{1}\lvert P_{12}\rvert\mathbf{4}\rbrack\lbrack\mathbf{2}\lvert P_{12}\rvert\mathbf{3}\rangle = m_em_\mu (  \langle\mathbf{34}\rangle\lbrack\mathbf{12}\rbrack - \langle\mathbf{34}\rangle\langle\mathbf{12}\rangle - \lbrack\mathbf{34}\rbrack\lbrack\mathbf{12}\rbrack + \lbrack\mathbf{34}\rbrack\langle\mathbf{12}\rangle ) + M_A^2\langle\mathbf{13}\rangle\lbrack\mathbf{24}\rbrack$,
and eigth term, 
$\lbrack\mathbf{1}\lvert P_{12}\rvert\mathbf{4}\rangle\langle\mathbf{2}\lvert P_{12}\rvert\mathbf{3}\rbrack = m_em_\mu(  \lbrack\mathbf{34}\rbrack\langle\mathbf{12}\rangle - \lbrack\mathbf{34}\rbrack\lbrack\mathbf{12}\rbrack - \langle\mathbf{34}\rangle\langle\mathbf{12}\rangle + \langle\mathbf{34}\rangle\lbrack\mathbf{12}\rbrack ) + M_A^2\lbrack\mathbf{13}\rbrack\langle\mathbf{24}\rangle$.  Between these, we see that the $m_em_\mu$ terms cancel and we are left with,
\begin{equation}
  \mathcal{M}%^{ee\mu\mu} 
    = 
    e^2\frac{
    \langle\mathbf{13}\rangle\lbrack\mathbf{24}\rbrack
    +\lbrack\mathbf{13}\rbrack\langle\mathbf{24}\rangle
    +\lbrack\mathbf{14}\rbrack\langle\mathbf{23}\rangle
    +\langle\mathbf{14}\rangle\lbrack\mathbf{23}\rbrack
    }{\left(s-M_A^2\right)}
    \ .
\end{equation}
As we can see, division by $M_A$ has been canceled and we are now in a position to take the limit as $M_A\to0$, giving
\begin{equation}
    \mathcal{M}%^{ee\mu\mu} 
    = 
    \frac{e^2}{s}\left(
    \langle\mathbf{13}\rangle\lbrack\mathbf{24}\rbrack
    +\lbrack\mathbf{13}\rbrack\langle\mathbf{24}\rangle
    +\lbrack\mathbf{14}\rbrack\langle\mathbf{23}\rangle
    +\langle\mathbf{14}\rangle\lbrack\mathbf{23}\rbrack
    \right)\ ,
\end{equation}
in agreement with Eq.~(\ref{eq:M_eemm}).

\subsection{\label{sec:eAeh spinor amp calculation}$\mathbf{e,\gamma,\bar{e},h}$}

For this amplitude, we will need the Higgs vertex, which is \cite{Christensen:2018zcq},
\begin{equation}
    \mathcal{M}_{eeh} = -\frac{m_e}{v}\left(\langle\mathbf{12}\rangle+\lbrack\mathbf{12}\rbrack\right)\ ,
\end{equation}
where $v=2M_Ws_W/e$, $M_W$ is the mass of the W boson and $s_W$ is the sin of the Weinberg angle.

There are two diagrams.  The first is an s-channel diagram given by
\begin{align}
  \mathcal{M}_{eAeh}^{(s)} &= -\frac{em_e}{vM_A\left(s-m_e^2\right)}
     \left(
       \lbrack\mathbf{12}\rbrack \langle\mathbf{2P}_{12I}\rangle -
       \langle\mathbf{12}\rangle \lbrack\mathbf{2P}_{12I}\rbrack 
     \right) \nonumber \\
    & \qquad \times 
      \left(
       \langle\mathbf{P}_{12}^{I}\mathbf{3}\rangle +
       \lbrack\mathbf{P}_{12}^{I}\mathbf{3}\rbrack 
     \right) . 
\end{align}
From \cite{Christensen:2019mch},  $\langle\mathbf{2P}_{12I}\rangle\langle\mathbf{P}_{12}^I\mathbf{3}\rangle=-m_e\langle\mathbf{23}\rangle$ and $\langle\mathbf{2P}_{12I}\rangle\lbrack\mathbf{P}_{12}^I\mathbf{3}\rbrack=-\langle\mathbf{2}\lvert (p_1+p_2)\rvert\mathbf{3}\rbrack = -\langle\mathbf{2}\lvert 
p_1\rvert\mathbf{3}\rbrack-M_A\lbrack\mathbf{23}\rbrack$, and similarly for the other two products giving us
\begin{align}
  \mathcal{M}_{eAeh}^{(s)} = \frac{e \, m_e}{v(s-m_e^2)} %\times \nonumber \\
   &\Biggl(
    \frac{m_e}{M_A}
          \left(  \lbrack\mathbf{12}\rbrack\langle\mathbf{23}\rangle +
                  \langle\mathbf{12}\rangle\lbrack\mathbf{23}\rbrack
          \right) \nonumber \\
    &+\frac{
         \left(   \lbrack\mathbf{12}\rbrack
                  \langle\mathbf{2}\lvert p_1\rvert\mathbf{3}\rbrack +
                  \langle\mathbf{12}\rangle
                  \lbrack\mathbf{2}\lvert p_1\rvert\mathbf{3}\rangle
         \right)
         }{M_A} \nonumber \\
    &+ 
         \left(   \lbrack\mathbf{12}\rbrack
                  \lbrack\mathbf{23}\rbrack +
                  \langle\mathbf{12}\rangle
                  \langle\mathbf{23}\rangle \right) 
    \Biggr) \ . %\nonumber \\
\label{eq:M_eAeh^s}
\end{align}
The u-channel diagram gives
\begin{align}
    \mathcal{M}_{eAeh}^{(u)} = 
   -&\frac{em_e}{vM_A\left(u-m_e^2\right)} \times \nonumber\\
    &\left( \langle\mathbf{1}\mathbf{P}_{14I}\rangle -
            \lbrack\mathbf{1}\mathbf{P}_{14I}\rbrack 
     \right) \times \nonumber \\
    &\left( \langle\mathbf{23}\rangle 
            \lbrack\mathbf{P}_{14}^I\mathbf{2}\rbrack +
            \langle\mathbf{P}_{14}^I\mathbf{2}\rangle 
            \lbrack\mathbf{23}\rbrack 
     \right)  ,
\end{align}
giving us,
\begin{align}
    \mathcal{M}_{eAeh}^{(u)} &= \frac{e \, m_e}{v M_A (u-m_e^2)} \Biggl(
    2 m_e\left(
      \langle\mathbf{23}\rangle\lbrack\mathbf{12}\rbrack
     +\langle\mathbf{12}\rangle \lbrack\mathbf{23}\rbrack \right) \nonumber\\
    &+ \lbrack\mathbf{1}\lvert p_4\rvert\mathbf{2}\rangle 
       \lbrack\mathbf{23}\rbrack +
       \langle\mathbf{1}\lvert p_4\rvert\mathbf{2}\rbrack 
       \langle\mathbf{23}\rangle
    \Biggr) \ .
\label{eq:M_eAeh^u}
\end{align}

Each of these amplitudes potentially has three helicities for the massive photon.  We will just focus on the physical process where the photon is helicity $\pm1$ in the massless limit.  Before we consider specific helicity cases, let's note that, since the denominator has $M_A$ to the first power, we need only expand the numerator to linear order as well.  We expect the zeroth-order terms in the numerator will cancel and that we will be left with a numerator term that is first order in $M_A$ (for $\pm$ helicity).  Since the third term of the s channel is already first order, we only need to expand the photon spinors to leading order, while for the first and second terms, we need to expand them  to linear order.  As we Taylor expand the massive spinors, we will also multiply and divide by the propagator denominator of the other diagram.  In other words, we will multiply the s-channel diagram by $\lbrack2\lvert p_3\rvert2\rangle/\left(u-m_e^2\right)$ and the u-channel diagram by $\lbrack2\lvert p_1\rvert2\rangle/(s-m_e^2)$, where $\lbrack2\lvert$ and $\rvert2\rangle$ are understood to be in their final massless forms.  This will allow us to combine the two diagrams and cancel the singularity between them.  It also gives us greater freedom to combine $\lvert2\rangle$ with $\lvert\zeta_2\rangle$ and use the identity, $\langle2\zeta_2\rangle=\sqrt{2E_2}$ in order to write the final expressions in terms of invariants.

\subsubsection{$+$ Helicity}
Upon Taylor expansion and combination, we obtain
\begin{align}
    \mathcal{M}^+_{eAeh} &= \frac{e m_e}{v(s-m_e^2)(u-m_e^2)} \times \nonumber \\
    &\Biggl[ \lbrack2\lvert p_{3} \rvert 2\rangle
             \lbrack2\mathbf{3}\rbrack \lbrack\mathbf{1}2\rbrack 
            -\frac{m_e}{ \sqrt{2E_2}} \Bigl(      
             \nonumber\\
    &+ \lbrack2\lvert p_{3} \rvert 2\rangle
       \bigl(
         \lbrack2\mathbf{3}\rbrack \langle\mathbf{1}\zeta_{2}\rangle+ 
         \lbrack\mathbf{1}2\rbrack \langle\zeta_{2}\mathbf{3}\rangle 
       \bigr) \nonumber \\
    &+2 \lbrack2\lvert p_{1} \rvert 2\rangle
       \bigl(
          \lbrack2\mathbf{3}\rbrack 
          \langle\mathbf{1}\zeta_{2}\rangle
         +\lbrack\mathbf{1}2\rbrack 
          \langle\zeta_{2}\mathbf{3}\rangle
       \bigr) \Bigr) \nonumber \\
    &-\frac{\lbrack2\lvert p_{1} \rvert 2\rangle\left(
    \langle\zeta_{2}\mathbf{3}\rangle \lbrack2\lvert p_{4} \rvert \mathbf{1}\rangle +
    \lbrack2\mathbf{3}\rbrack \lbrack\mathbf{1}\lvert p_{4} \rvert \zeta_{2}\rangle
    \right)}{\sqrt{2E_2}} \nonumber \\
    &-\frac{\lbrack2\lvert p_{3} \rvert 2\rangle\left(
    \langle\mathbf{1}\zeta_{2}\rangle\lbrack2\lvert p_{1} \rvert \mathbf{3}\rangle +
    \lbrack\mathbf{1}2\rbrack \lbrack\mathbf{3}\lvert p_{1} \rvert \zeta_{2}\rangle
    \right)}{ \sqrt{2E_2}} \Biggr] \ ,
\end{align}
where we can see that $M_A$ has already dropped out and we will take $M_A=0$ for the rest of the calculation.  Our next step is to use a series of identities including Schouten identities, anticommutation rules, mass relations and momentum conservation to simplify this expression until the $\zeta_2$ and the $E_2$ drop out.  Notice that in the terms with $\lvert\zeta_2\rangle$, we don't want to use $\lvert2\rangle\lbrack2\rvert=p_2$ right away because we want to use $\lbrack2\lvert p_1\rvert2\rangle$ and $\lbrack2\lvert p_3\rvert2\rangle$ in the Schouten identities with the $\lvert\zeta_2\rangle$.  The sequence of steps could vary, of course, and there are enough steps to be tedious and not be enlightening.  Although this can be done by hand, we find this calculation to require enough steps to not make the full details enlightening.  We complete a series of identities including Schouten identities, mass identities and momentum conservation.  We also bring the spinor products into a standard form with the spinors in ascending order.  After this series of steps, we arrive at,
\begin{eqnarray}
    \mathcal{M}^+_{eAeh} 
    &=&
    \frac{e m_e}{v\left(s-m_e^2\right)\left(u-m_e^2\right)}
    \Big( m_h ^{2} \lbrack\mathbf{1}2\rbrack\lbrack2\mathbf{3}\rbrack 
    \nonumber\\
    &&- m_e \lbrack\mathbf{1}2\rbrack\lbrack2\lvert p_{4} \rvert \mathbf{3}\rangle 
    + m_e \lbrack2\mathbf{3}\rbrack \lbrack2\lvert p_4 \rvert\mathbf{1}\rangle
    \nonumber\\
    &&- \langle\mathbf{1}\mathbf{3}\rangle \lbrack2\lvert p_{3} p_{4} \rvert 2\rbrack\Big)\ ,
\end{eqnarray}
where the denominator could be replaced with $\left(s-m_e^2\right)\left(u-m_e^2\right)=\lbrack2\lvert p_1\rvert2\rangle\lbrack2\lvert p_3\rvert2\rangle$.

The steps for the negative helicity case $\mathcal{M}^-_{eAeh}$ are much the same and we will not describe it here, although we have performed the calculation.

\subsection{\label{sec:Massive eeAA}$\mathbf{e,\bar{e},\gamma,\gamma}$}
For our final process, with two photon legs, we have a $t$- and a $u$-channel diagram with the mathematical forms
\begin{eqnarray}
    \mathcal{M}_{eeAA}^{(t)} &=&
    \frac{e^2}{M_A^2 \left(t-m_e^2 \right)} \left(
    \langle\mathbf{2}\mathbf{4}\rangle \lbrack\mathbf{4}\mathbf{P_{13} }^{ I}\rbrack 
    +\lbrack\mathbf{2}\mathbf{4}\rbrack \langle\mathbf{4}\mathbf{P_{13} }^{ I}\rangle  
    \right)
    \nonumber\\
    &&\hspace{1in}\left(
    \langle\mathbf{3}\mathbf{1}\rangle \lbrack\mathbf{P_{13} }_{ I}\mathbf{3}\rbrack 
    -\lbrack\mathbf{3}\mathbf{1}\rbrack \langle\mathbf{P_{13} }_{ I}\mathbf{3}\rangle
    \right) 
    \nonumber\\
    &=& 
    -\frac{ e^2 M_A \left(
    \langle\mathbf{1}\mathbf{3}\rangle \lbrack\mathbf{2}\mathbf{4}\rbrack \langle\mathbf{3}\mathbf{4}\rangle 
    + \lbrack\mathbf{1}\mathbf{3}\rbrack\langle\mathbf{2}\mathbf{4}\rangle  \lbrack\mathbf{3}\mathbf{4}\rbrack 
    \right)}{M_A^2 \left(t-m_e^2 \right)} 
    \nonumber\\
    &&-\frac{e^2 m_e \left( 
    \langle\mathbf{1}\mathbf{3}\rangle \langle\mathbf{2}\mathbf{4}\rangle \lbrack\mathbf{3}\mathbf{4}\rbrack 
    +\lbrack\mathbf{1}\mathbf{3}\rbrack \lbrack\mathbf{2}\mathbf{4}\rbrack \langle\mathbf{3}\mathbf{4}\rangle
    \right)}{M_A^2 \left(t-m_e^2 \right)} 
    \nonumber\\
    &&-\frac{ e^2 \left(
    \langle\mathbf{1}\mathbf{3}\rangle \lbrack\mathbf{2}\mathbf{4}\rbrack \lbrack\mathbf{3}\lvert p_{1} \rvert \mathbf{4}\rangle 
    +\lbrack\mathbf{1}\mathbf{3}\rbrack\langle\mathbf{2}\mathbf{4}\rangle  \lbrack\mathbf{4}\lvert p_{1} \rvert \mathbf{3}\rangle 
    \right)}{M_A^2 \left(t-m_e^2 \right)}
    \nonumber\\
    \label{eq:M_eeAA t-channel}
\end{eqnarray}
and
\begin{eqnarray}
    \mathcal{M}_{eeAA}^{(u)} &=& 
    \frac{e^2}{M_A^2 \left(u-m_e^2 \right)}
    \left(
    \langle\mathbf{2}\mathbf{3}\rangle \lbrack\mathbf{3}\mathbf{P_{14} }^{I}\rbrack 
    +\lbrack\mathbf{2}\mathbf{3}\rbrack \langle\mathbf{3}\mathbf{P_{14} }^{I}\rangle 
    \right)
    \nonumber\\
    &&\hspace{1in}\left( 
    \langle\mathbf{4}\mathbf{1}\rangle \lbrack\mathbf{P_{14} }_{I}\mathbf{4}\rbrack 
    -\lbrack\mathbf{4}\mathbf{1}\rbrack \langle\mathbf{P_{14} }_{I}\mathbf{4}\rangle
    \right)
    \nonumber\\
    &=& 
    \frac{e^2 
    M_A \left( 
    \langle\mathbf{1}\mathbf{4}\rangle \lbrack\mathbf{2}\mathbf{3}\rbrack\langle\mathbf{3}\mathbf{4}\rangle  
    + \lbrack\mathbf{1}\mathbf{4}\rbrack\langle\mathbf{2}\mathbf{3}\rangle  \lbrack\mathbf{3}\mathbf{4}\rbrack 
    \right)}{M_A^2 \left(u-m_e^2 \right)} 
    \nonumber\\
    &&+ \frac{
    e^2 m_e \left(
    \langle\mathbf{1}\mathbf{4}\rangle \langle\mathbf{2}\mathbf{3}\rangle \lbrack\mathbf{3}\mathbf{4}\rbrack 
    +  \lbrack\mathbf{1}\mathbf{4}\rbrack \lbrack\mathbf{2}\mathbf{3}\rbrack \langle\mathbf{3}\mathbf{4}\rangle
    \right)
    }{M_A^2 \left(u-m_e^2 \right)} 
    \nonumber\\
    &&-\frac{e^2 \left(
    \langle\mathbf{1}\mathbf{4}\rangle \lbrack\mathbf{2}\mathbf{3}\rbrack \lbrack\mathbf{4}\lvert p_{1} \rvert \mathbf{3}\rangle 
    +\lbrack\mathbf{1}\mathbf{4}\rbrack\langle\mathbf{2}\mathbf{3}\rangle  \lbrack\mathbf{3}\lvert p_{1} \rvert \mathbf{4}\rangle 
    \right)}{M_A^2 \left(u-m_e^2 \right)} .
    \nonumber\\
    \label{eq:M_eeAA u-channel}
\end{eqnarray}
As we can see, we must expand the photon spinors to second order in the mass in this case in order to cancel the $M_A^2$ in the denominator.  As before, we also combine the two diagrams with a common denominator.  There are two nonzero helicity combinations and their parity partner.  We begin with both helicities the same and then do the opposite-helicity case.  

\subsubsection{++ Helicity}
After Taylor expansion and combination, we have,
\begin{align}
    \mathcal{M}_{eeAA}^{++} &= 
    -\frac{e^2  \lbrack4\lvert p_{1} \rvert 4\rangle}{\sqrt{2E_3}\sqrt{2E_4}\left(t-m_e^2\right)\left(u-m_e^2\right)}
    \Big(
    \nonumber\\
    &
    \hspace{0.5in}m_e \lbrack34\rbrack \langle\mathbf{1}\zeta_{3}\rangle\langle\mathbf{2}\zeta_{4}\rangle
    + m_e \lbrack\mathbf{1}3\rbrack \lbrack\mathbf{2}4\rbrack \langle\zeta_{3}\zeta_{4}\rangle
    \nonumber\\
    &
    \hspace{0.5in}+\langle\mathbf{1}\zeta_{3}\rangle\lbrack\mathbf{2}4\rbrack \lbrack3\lvert p_{1} \rvert \zeta_{4}\rangle
    + \langle\mathbf{2}\zeta_{4}\rangle\lbrack\mathbf{1}3\rbrack \lbrack4\lvert p_{1} \rvert \zeta_{3}\rangle
    \nonumber\\
    &
    \hspace{0.5in}-\sqrt{2E_3}\lbrack34\rbrack \langle\mathbf{2}\zeta_{4}\rangle\lbrack\mathbf{1}3\rbrack 
    \Big)
    \nonumber\\
    &- \frac{e^2  \lbrack3\lvert p_{1} \rvert 3\rangle}{\sqrt{2E_3}\sqrt{2E_4}\left(t-m_e^2\right)\left(u-m_e^2\right)}
    \Big(
    \nonumber\\
    &\hspace{0.5in}
    - m_e \lbrack34\rbrack \langle\mathbf{1}\zeta_{4}\rangle\langle\mathbf{2}\zeta_{3}\rangle
    - m_e \lbrack\mathbf{1}4\rbrack \lbrack\mathbf{2}3\rbrack \langle\zeta_{3}\zeta_{4}\rangle
    \nonumber\\
    &\hspace{0.5in}
    + \langle\mathbf{2}\zeta_{3}\rangle\lbrack\mathbf{1}4\rbrack \lbrack3\lvert p_{1} \rvert \zeta_{4}\rangle
    + \langle\mathbf{1}\zeta_{4}\rangle\lbrack\mathbf{2}3\rbrack \lbrack4\lvert p_{1} \rvert \zeta_{3}\rangle
    \nonumber\\
    &\hspace{0.5in}
    +\sqrt{2E_4}\lbrack34\rbrack \langle\mathbf{2}\zeta_{3}\rangle\lbrack\mathbf{1}4\rbrack 
    \Big)
    \ .
\end{align}
After applying the usual identities multiple times, we bring it into the form
\begin{align}
    \mathcal{M}_{eeAA}^{++} &=
    \frac{e^2  m_e \lbrack34\rbrack ^{2} \langle\mathbf{1}\mathbf{2}\rangle }
    {\left(t-m_e^2\right)\left(u-m_e^2\right)} 
    \nonumber\\
    &\hspace{-0.25in}
    -\frac{e^2  \lbrack\mathbf{1}3\rbrack   \lbrack4\lvert p_{1} \rvert \zeta_{4}\rangle}
    {\sqrt{2E_4}\left(t-m_e^2\right)\left(u-m_e^2\right)} 
    \left(
    m_e\lbrack\mathbf{2}3\rbrack +
    \lbrack3\lvert (p_{1}+p_4) \rvert\mathbf{2}\rangle
    \right)
    \nonumber\\
    &\hspace{-0.25in}
    -\frac{e^2 \lbrack\mathbf{1}4\rbrack \lbrack3\lvert p_{1} \rvert \zeta_{3}\rangle  }
    {\sqrt{2E_3}\left(t-m_e^2\right)\left(u-m_e^2\right)} 
    \left(
    m_e \lbrack\mathbf{2}4\rbrack 
    +\lbrack4\lvert (p_{1}+p_3) \rvert \mathbf{2}\rangle
    \right)
    \nonumber\\
    &\hspace{-0.25in}
    +\frac{e^2  \lbrack3\lvert p_{1} \rvert \zeta_{3}\rangle\lbrack4\lvert p_{1} \rvert \zeta_{4}\rangle}
    {\sqrt{2E_3}\sqrt{2E_4}\left(t-m_e^2\right)\left(u-m_e^2\right)}
    \Big(
    \lbrack\mathbf{1}\lvert (p_{3}+p_4) \rvert \mathbf{2}\rangle
    \nonumber\\
    &\hspace{0.25in}
    +\lbrack\mathbf{2}\lvert (p_{3}+p_4) \rvert \mathbf{1}\rangle \Big) \ ,
\end{align}
where, at the end, we also used $\lvert3\rangle\lbrack3\rvert=p_3$ and $\lvert4\rangle\lbrack4\rvert=p_4$ and grouped the momenta inside the spinor products.  (We found that the benefit of using these contractions outweighed the benefit of keeping the spinors separate at this point.)  We again use momentum conservation, $\lbrack3\lvert(p_1+p_4)\rvert\mathbf{2}\rangle = - m_e\lbrack\mathbf{2}3\rbrack$, $\lbrack4\lvert(p_1+p_3)\rvert\mathbf{2}\rangle = - m_e\lbrack\mathbf{2}4\rbrack$ and $\lbrack\mathbf{1}\lvert(p_3+p_4)\rvert\mathbf{2}\rangle + \lbrack\mathbf{2}\lvert(p_3+p_4)\rvert\mathbf{1}\rangle =  2m_e\left(\lbrack\mathbf{12}\rbrack - \langle\mathbf{12}\rangle\right)$.  Finally, we collect what is left and apply a few more standard identities to obtain
\begin{equation}
    \mathcal{M}_{eeAA}^{++} = 
    \frac{e^2  m_e \lbrack34\rbrack ^{2} \langle\mathbf{1}\mathbf{2}\rangle }
    {\left(t-m_e^2\right)\left(u-m_e^2\right)} \ .
\end{equation}
Both negative helicity follows the same series of steps.

\subsubsection{+- Helicity}
Taylor expansion and combination gives us
\begin{align}
    \mathcal{M}_{eeAA}^{+-} &=
    -\frac{e^2 \lbrack4\lvert p_{1} \rvert 4\rangle}
    {\sqrt{2E_3}\sqrt{2E_4}\left(t-m_e^2\right)\left(u-m_e^2\right)}
    \Big(
    \nonumber\\
    &\hspace{0.5in}
    m_e \lbrack3\tilde{\zeta}_{4}\rbrack\langle\mathbf{1}\zeta_{3}\rangle\langle\mathbf{2}4\rangle
    +m_e \lbrack\mathbf{1}3\rbrack \lbrack\mathbf{2}\tilde{\zeta}_{4}\rbrack\langle\zeta_{3}4\rangle
    \nonumber\\
    &\hspace{0.5in}
    +\langle\mathbf{1}\zeta_{3}\rangle\lbrack\mathbf{2}\tilde{\zeta}_{4}\rbrack\lbrack3\lvert p_{1} \rvert 4\rangle
    +\langle\mathbf{2}4\rangle\lbrack\mathbf{1}3\rbrack \lbrack\tilde{\zeta}_{4}\lvert p_{1} \rvert \zeta_{3}\rangle
    \nonumber\\
    &\hspace{0.5in}
    -\sqrt{2E_3}\lbrack3\tilde{\zeta}_{4}\rbrack\langle\mathbf{2}4\rangle\lbrack\mathbf{1}3\rbrack 
    \Big)
    \nonumber\\
    &
    -\frac{e^2 \lbrack3\lvert p_{1} \rvert 3\rangle}
    {\sqrt{2E_3}\sqrt{2E_4}\left(t-m_e^2\right)\left(u-m_e^2\right)}
    \Big(
    \nonumber\\
    &\hspace{0.5in}
    -m_e \lbrack3\tilde{\zeta}_{4}\rbrack\langle\mathbf{1}4\rangle\langle\mathbf{2}\zeta_{3}\rangle
    -m_e \lbrack\mathbf{1}\tilde{\zeta}_{4}\rbrack\lbrack\mathbf{2}3\rbrack \langle\zeta_{3}4\rangle
    \nonumber\\
    &\hspace{0.5in}
    +\langle\mathbf{2}\zeta_{3}\rangle\lbrack\mathbf{1}\tilde{\zeta}_{4}\rbrack\lbrack3\lvert p_{1} \rvert 4\rangle
    +\langle\mathbf{1}4\rangle\lbrack\mathbf{2}3\rbrack \lbrack\tilde{\zeta}_{4}\lvert p_{1} \rvert \zeta_{3}
    \rangle
    \nonumber\\
    &\hspace{0.5in}
    +\sqrt{2E_4}\langle\mathbf{1}4\rangle\lbrack\mathbf{2}3\rbrack \langle\zeta_{3}4\rangle
    \Big) 
    \ .
\end{align}
Although the exact sequence of identities used differs from one helicity combination to the next, the broad outline is the same.  We begin by performing a series of identities, but leave out $\lvert3\rangle\lbrack3\rvert=p_3$ and $\lvert4\rangle\lbrack4\rvert=p_4$ until the end, to obtain,
\begin{align}
    \mathcal{M}_{eeAA}^{+-} &=
    \frac{e^2  \left(\langle\mathbf{2}4\rangle\lbrack\mathbf{1}3\rbrack +\langle\mathbf{1}4\rangle\lbrack\mathbf{2}3\rbrack \right)\lbrack3\lvert p_{1} \rvert 4\rangle}{\left(t-m_e^2\right)\left(u-m_e^2\right)} 
    \nonumber\\
    &\hspace{-0.3in}
    -\frac{e^2  \lbrack\mathbf{1}3\rbrack  \lbrack\tilde{\zeta}_{4}\lvert p_{1} \rvert 4\rangle}{\sqrt{2E_4}\left(t-m_e^2\right)\left(u-m_e^2\right)}
    \left(
    m_e \lbrack\mathbf{2}3\rbrack 
    + \lbrack3\lvert (p_{1}+p_4) \rvert \mathbf{2}\rangle  
    \right)
    \nonumber\\
    &\hspace{-0.3in}
    -\frac{e^2  \langle\mathbf{1}4\rangle\lbrack3\lvert p_{1} \rvert \zeta_{3}\rangle }{\sqrt{2E_3}\left(t-m_e^2\right)\left(u-m_e^2\right)} 
    \left(
    m_e\langle\mathbf{2}4\rangle 
    +\lbrack\mathbf{2}\lvert (p_{1}+p_3) \rvert 4\rangle
    \right)
    \nonumber\\
    &\hspace{-0.3in}
    +\frac{e^2  \lbrack3\lvert p_{1} \rvert \zeta_{3}\rangle\lbrack\tilde{\zeta}_{4}\lvert p_{1} \rvert 4\rangle}{\sqrt{2E_3}\sqrt{2E_4}\left(t-m_e^2\right)\left(u-m_e^2\right)}
    \Big(
    \lbrack\mathbf{1}\lvert (p_{3}+p_4) \rvert \mathbf{2}\rangle
    \nonumber\\
    &\hspace{0.25in}
    +\lbrack\mathbf{2}\lvert (p_{3}+p_4) \rvert \mathbf{1}\rangle  
    \Big)
    \ .
\end{align}
Next, we use a similar set of momentum conservation identities followed by a few more identities to finally obtain,
\begin{eqnarray}
    \mathcal{M}_{eeAA}^{+-} &=& 
    \frac{e^2 \left(\langle\mathbf{2}4\rangle\lbrack\mathbf{1}3\rbrack +\langle\mathbf{1}4\rangle\lbrack\mathbf{2}3\rbrack \right)\lbrack3\lvert p_{1} \rvert 4\rangle}{\left(t-m_e^2\right)\left(u-m_e^2\right)}\ . 
\end{eqnarray}
The negative-positive-helicity case is obtained by basically the same set of steps.

\section{\label{app:x}The $\mathbf{x}$ Factor}
In this appendix, we review the $x$ factor.  As we will see, these methods, with or without a momentum shift described in App.~\ref{app:analytic continuation}, are sufficient to obtain agreement with Feynman diagrams for the processes with an external photon but not the ones with an internal photon.
Indeed, the only method to agree with Feynman diagrams for the process $e\bar{e}\mu\bar{\mu}$ was to begin with a massive photon and take the massless limit at the end of the calculation as described in App.~\ref{sec:Massive Photon}.  Nevertheless, we fully describe our $x$-factor techniques to be used where they are successful and to hopefully inspire a generalization that works for all diagrams containing massless particles.

%\subsection{\label{sec:x-factor AHH}A Brief Review of the x Factor}
The $x$ factor only appears in 3-point vertices with one massless particle and two massive particles of the same mass and is important in the photon processes discussed here.  It's definition and properties are described in \cite{Arkani-Hamed:2017jhn,Christensen:2018zcq} and we review them here.  Suppose particle 3 is massless.  Then, the two helicity spinors $|3\rangle$ and $\frac{1}{m}p_2|3\rbrack$ are parallel in helicity-spinor space as can be seen by considering their inner product.
\begin{eqnarray}
    \langle3|p_2|3\rbrack &=&
    2p_2\cdot p_3 =
    \left(p_2+p_3\right)^2 - m^2 
    \nonumber\\
    &=&
    p_1^2 - m^2 = 
    0\ .
\end{eqnarray}
Therefore, we can write one as the other times some coefficient, a constant of proportionality, that we call $x$, as in
\begin{equation}
    x_{12}|3\rangle_\alpha = \frac{p_{2\alpha\dot{\beta}}}{m}|3\rbrack^{\dot{\beta}} =
    -\frac{p_{1\alpha\dot{\beta}}}{m}|3\rbrack^{\dot{\beta}}\ ,
    \label{eq:x|3> unsymmetrized}
\end{equation}
where the subscript $12$ represents the identical-mass particles in the vertex with the massless particle $3$, where the first index is the particle and the second is the antiparticle.  We also used $p_2=-p_1-p_3$ and $p_3|3\rangle=p_3|3\rbrack=0$.  As we can see, there is an ambiguity about whether to include $p_2$ or $-p_1$.  They are related by momentum conservation here, but in the final identities below, that will not always be the case.  In fact, choosing one or the other sometimes gives different, inequivalent, results.  Therefore, following \cite{Arkani-Hamed:2017jhn}, we will symmetrize the formula by taking the average of the two results.  We will use
\begin{equation}
    x_{12}|3\rangle_\alpha = \frac{1}{2m}\left(p_{2\alpha\dot{\beta}}-p_{1\alpha\dot{\beta}}\right)|3\rbrack^{\dot{\beta}}\ .
    \label{eq:x|3>=}
\end{equation}
Similarly, we define $\tilde{x}$ in
\begin{equation}
    \tilde{x}_{12}|3\rbrack^{\dot{\alpha}} = \frac{p_2^{\dot{\alpha}\beta}}{m}|3\rangle_{\beta} = -\frac{p_1^{\dot{\alpha}\beta}}{m}|3\rangle_{\beta}\ ,
    \label{eq:xt|3] unsymmetrized}
\end{equation}
and after symmetrizing,
\begin{equation}
    \tilde{x}_{12}|3\rbrack^{\dot{\alpha}} = \frac{1}{2m}\left(p_2^{\dot{\alpha}\beta}-p_1^{\dot{\alpha}\beta}\right)|3\rangle_{\beta} \ .
    \label{eq:xtilde|]=}
\end{equation}
Moreover, we note that $\tilde{x}$ is the reciprocal of $x$.  To see this, multiply both sides of this expression by $(p_2-p_1)/m$ to obtain,
\begin{eqnarray}
    \tilde{x}_{12}\frac{\left(p_{2\beta\dot{\alpha}}-p_{1\beta\dot{\alpha}}\right)}{2m}\lvert3\rbrack^{\dot{\alpha}} = \frac{2m^2-2p_1\cdot p_2}{4m^2}\lvert3\rangle_{\beta}\ ,
    \nonumber\\
\end{eqnarray}
where we have used $p_1^2=p_2^2=m^2$ and $p_1p_2+p_2p_1 = 2p_1\cdot p_2$.  However, $2p_1\cdot p_2 = (p_1+p_2)^2-2m^2 = p_3^2-2m^2 = -2m^2$, giving us
\begin{eqnarray}
    \frac{\left(p_{2\beta\dot{\alpha}}-p_{1\beta\dot{\alpha}}\right)}{2m}\lvert3\rbrack^{\dot{\alpha}} = \frac{1}{\tilde{x}_{12}}\lvert3\rangle_{\beta}\ ,
    \nonumber\\
\end{eqnarray}
where we have moved $\tilde{x}_{12}$ to the other side.
Therefore,
\begin{equation}
    \tilde{x} = \frac{1}{x}\ .
    \label{eq:xtilde = 1/x}
\end{equation}
These identities will be useful in our calculations.  We can multiply these expressions by a linearly independent reference helicity-spinor to obtain expressions for $x$ and $\tilde{x}$ 
\begin{equation}
    x_{12} = \frac{\langle\xi\lvert\left(p_2-p_1\right)\rvert3\rbrack}{2m\langle\xi3\rangle}\ ,
    \label{eq:x_12 = xi...}
\end{equation}
and
\begin{equation}
    \tilde{x}_{12} =  \frac{\lbrack\xi\lvert\left(p_2-p_1\right)\rvert3\rangle}{2m\lbrack\xi3\rbrack} \ .
    \label{eq:xtilde_12 = xi...}
\end{equation}
In principle, we have the freedom to choose any linearly independent spinor for $\lvert\xi\rangle$ and $\vert\xi\rbrack$ and, if it is not physical, it should cancel at the end of the calculation.  It is not a gauge, there is no gauge in constructive amplitude theory, but it is similar in the sense that it must cancel out and not impact the calculation.  However, we usually find it more straight forward in our calculations to bypass this reference spinner and directly replace $x$ and $\tilde{x}$ with formulas involving only momenta and external spinors.

\subsection{\label{app:x_ij<ij>=x_ij[ij]+...}$x_{ij}\langle\mathbf{ij}\rangle=x_{ij}\lbrack\mathbf{ij}\rbrack +\cdots$}
Although we can always refer back to Eqs.~(\ref{eq:x|3>=}) and (\ref{eq:xtilde|]=}), there are certain products involving $x$ factors that come up frequently and it is useful to create specialized identities for them.  For example, we can write
\begin{equation}
    x_{ij}\langle\mathbf{i}|(p_i+p_j)|\mathbf{j}\rbrack =
    m x_{ij}\left( \lbrack\mathbf{ij}\rbrack - \langle\mathbf{ij}\rangle \right)
    \ ,
    \label{eq:B11}
\end{equation}
where we have used $\langle\mathbf{i}\lvert p_i=m\lbrack\mathbf{i}\lvert$, $p_j\rvert\mathbf{j}\rbrack=-m\rvert\mathbf{j}\rangle$ and $p_i+p_j$ is on shell and massless, and the $x_{ij}$ on both sides is in preparation for the next step.  The left side can be written as $x_{ij}\langle\mathbf{i}p_{ij}\rangle\lbrack p_{ij}\mathbf{j}\rbrack$ and we can use Eq.~(\ref{eq:x|3>=}), where $\rvert3\rangle$ has become $\rvert p_{ij}\rangle$, to write this as
\begin{equation}
    \frac{1}{2m}\langle\mathbf{i}|\left(p_j-p_i\right)|p_{ij}\rbrack\lbrack p_{ij}\mathbf{j}\rbrack = m x_{ij} \left( \lbrack\mathbf{ij}\rbrack - \langle\mathbf{ij}\rangle \right)
    \ .
    \label{eq:-[][]=mx([]-<>)}
\end{equation}
Similarly, we will sometimes use 
\begin{equation}
    \tilde{x}_{ij}\lbrack\mathbf{i}|(p_i+p_j)|\mathbf{j}\rangle = 
    m \tilde{x}_{ij}\left(\langle\mathbf{ij}\rangle - \lbrack\mathbf{ij}\rbrack\right)
    \ ,
\end{equation}
which, through a similar set of steps, gives us
\begin{equation}
    \frac{1}{2m}\lbrack\mathbf{i}\lvert(p_j-p_i)\rvert p_{ij}\rangle\langle p_{ij}\mathbf{j}\rangle = 
    m \tilde{x}_{ij}\left(\langle\mathbf{ij}\rangle - \lbrack\mathbf{ij}\rbrack\right)
    \ .
    \label{eq:-<><>=mx(<>-[])}
\end{equation}

We can then use Eqs.~(\ref{eq:-[][]=mx([]-<>)}) and (\ref{eq:-<><>=mx(<>-[])}) to switch from angle brackets to square brackets and vice versa when they are multiplied by appropriate factors of $x$ or $\tilde{x}$.  

It seems appropriate here to comment on an issue with all the identities involving the $x$ factor on an internal line.  Note that between Eq.~(\ref{eq:B11}) and (\ref{eq:-[][]=mx([]-<>)}), we assume that the internal particle with momentum $p_i+p_j$ is on shell and massless.  This is necessary so that we can use the identity $p_i+p_j = \lvert p_{ij}\rangle\lbrack p_{ij}\rvert$, where these are helicity spinors, which is required by the identity in Eq.~(\ref{eq:x|3>=}).  In order to place the internal line on shell, we must complexify the momenta and spinors as we describe in App.~\ref{app:analytic continuation}.  However, we show in App.~\ref{app:eemm large z} that these amplitudes with internal photons do not vanish in the large $z$ limit, throwing this procedure, and these identities, into doubt for these processes.  These same comments apply to the identities in Apps.~\ref{app:xxtilde<>[]}, \ref{app:xxt+xtx} and \ref{app:xxt+xtx revisited}, and may partly explain why the constructive approach does not agree for these processes.  On the other hand, when the photon is an external particle, it is on shell, and there are no issues applying the identities in Eqs.~(\ref{eq:x|3>=}) and (\ref{eq:xtilde|]=}).  Consequently, the identities in App.~\ref{sec:x = alternate} do not appear to suffer from this problem.

\subsection{\label{app:xxtilde<>[]}$x_{ij}\tilde{x}_{kl}\langle\mathbf{ij}\rangle=x_{ij}\tilde{x}_{kl}\lbrack\mathbf{ij}\rbrack+\cdots$}

When we have a massless particle along an internal line, we will get a product of the form $x_{ij}\tilde{x}_{kl}$ times an angle or a square bracket that matches the indices of $x_{ij}$ or $\tilde{x}_{kl}$.  We will often need to switch angle for square brackets or vice versa.  As an example, consider an s-channel massless particle.  In this case we typically obtain $\tilde{x}_{12}x_{34}\lbrack\mathbf{12}\rbrack$ and need to switch to angle brackets to match the other term with a $x_{12}\tilde{x}_{34}\langle\mathbf{12}\rangle$ (or vice versa).  In order to achieve this, we can multiply Eq.~(\ref{eq:-<><>=mx(<>-[])}), where $i=1$ and $j=2$, by $
x_{34}$ to obtain
\begin{equation}
    \tilde{x}_{12}x_{34}\left(\langle\mathbf{12}\rangle - \lbrack\mathbf{12}\rbrack\right) =
    \frac{1}{2m_2^2}x_{34}\lbrack\mathbf{1}\lvert(p_2-p_1)\rvert p_{12}\rangle\langle p_{12}\mathbf{2}\rangle
    \ .
\end{equation}
However, since $p_{34}=-p_{12}$, we have  $|p_{12}\rangle\langle p_{12}|=|p_{34}\rangle\langle p_{34}|$ (any sign change to the spinors cancel in the pair).  Now we can write this as
\begin{equation}
    \tilde{x}_{12}x_{34}\left(\langle\mathbf{12}\rangle - \lbrack\mathbf{12}\rbrack\right) =
    \frac{1}{2m_2^2}x_{34}\lbrack\mathbf{1}\lvert(p_2-p_1)\rvert p_{34}\rangle\langle p_{34}\mathbf{2}\rangle
    \ .
\end{equation}
We can then use Eq.~(\ref{eq:x|3>=}) to obtain
\begin{eqnarray}
    &&\tilde{x}_{12}x_{34}\left(\langle\mathbf{12}\rangle - \lbrack\mathbf{12}\rbrack\right) 
    =
    \nonumber\\
    &&\hspace{0.75in}
    \frac{1}{4m_2^2m_4}\lbrack\mathbf{1}\lvert(p_2-p_1)(p_4-p_3)\rvert p_{34}\rbrack\langle p_{34}\mathbf{2}\rangle 
    \ .
    \nonumber\\
\end{eqnarray}
However, now we can use $p_2-p_1=-2p_1-p_3-p_4$ and $|p_{34}\rbrack\langle p_{34}|=(p_3+p_4)$ leading us to
\begin{eqnarray}
    &&\tilde{x}_{12}x_{34}\left(\langle\mathbf{12}\rangle - \lbrack\mathbf{12}\rbrack\right) =
    -\frac{1}{2m_2^2m_4}\lbrack\mathbf{1}\lvert p_1(p_4-p_3)(p_3+p_4)\rvert\mathbf{2}\rangle 
    \nonumber\\
    &&\hspace{0.75in}
    -\frac{1}{4m_2^2m_4}\lbrack\mathbf{1}\lvert(p_3+p_4)(p_4-p_3)(p_3+p_4)\rvert\mathbf{2}\rangle 
    \ .
    \nonumber\\
\end{eqnarray}
In the second term, we use $\lbrack\mathbf{1}\lvert(p_3+p_4)(p_4-p_3)(p_3+p_4)\rvert\mathbf{2}\rangle = 2(p_4-p_3)\cdot(p_3+p_4)\lbrack\mathbf{1}\lvert(p_3+p_4)\rvert\mathbf{2}\rangle - (p_3+p_4)^2\lbrack\mathbf{1}\lvert(p_4-p_3)\rvert\mathbf{2}\rangle = 0$, since $(p_4-p_3)\cdot(p_3+p_4) = m_4^2-m_3^2=0$ and $(p_3+p_4)^2=0$ is the on-shell condition.  We next use $\lbrack\mathbf{1}\lvert p_1=m_1\langle\mathbf{1}\lvert$ and $(p_4-p_3)(p_3+p_4) = m_4^2-m_3^2 + p_4p_3-p_3p_4 = -2p_3\cdot p_4 + 2p_4p_3$ in the first term to obtain
\begin{equation}
    \tilde{x}_{12}x_{34}\left(\langle\mathbf{12}\rangle - \lbrack\mathbf{12}\rbrack\right) =
    -\frac{1}{m_2m_4}\langle\mathbf{1}\lvert p_4p_3\rvert\mathbf{2}\rangle 
    +\frac{2p_3\cdot p_4}{2m_2m_4}\langle\mathbf{12}\rangle .
\end{equation}
Finally, $2p_3\cdot p_4 = (p_3+p_4)^2-2m_4^2 = -2m_4^2$ for
\begin{equation}
    \tilde{x}_{12}x_{34}\left(\lbrack\mathbf{12}\rbrack - \langle\mathbf{12}\rangle\right) =
    \frac{1}{m_2m_4}\langle\mathbf{1}\lvert p_4p_3\rvert\mathbf{2}\rangle 
    +\frac{m_4}{m_2}\langle\mathbf{12}\rangle 
    \ ,
    \label{eq:xtildex([12]-<12>)=...}
\end{equation}
where we also flipped the sign.
Similarly, we find
\begin{equation}
    \tilde{x}_{12}x_{34}\left(
        \langle\mathbf{34}\rangle -
        \lbrack\mathbf{34}\rbrack 
    \right) =
    \frac{1}{m_2m_4}\lbrack\mathbf{3}|p_2p_1|\mathbf{4}\rbrack +
    \frac{m_2}{m_4}\lbrack\mathbf{34}\rbrack\ ,
    \label{eq:xtildex(<34>-[34])=...}
\end{equation}
as well as
\begin{eqnarray}
    x_{12}\tilde{x}_{34}\left(
        \lbrack\mathbf{34}\rbrack - 
        \langle\mathbf{34}\rangle 
    \right) &=&
    \frac{1}{m_2m_4}\langle\mathbf{3}|p_2p_1|\mathbf{4}\rangle+
    \frac{m_2}{m_4}\langle\mathbf{34}\rangle\ ,
    \label{eq:xtildex([34]-<34>)=...}
    \nonumber\\\\
    x_{12}\tilde{x}_{34}\left(
        \langle\mathbf{12}\rangle -
        \lbrack\mathbf{12}\rbrack 
    \right) &=&
    \frac{1}{m_2m_4}\lbrack\mathbf{1}|p_4p_3|\mathbf{2}\rbrack +
    \frac{m_4}{m_2}\lbrack\mathbf{12}\rbrack\ .
    \label{eq:xtildex(<12>-[12])=...}
    \nonumber\\
\end{eqnarray}
We see that these last two differ from the previous identities by the change $x\leftrightarrow\tilde{x}$ and $\langle\cdots\rangle\leftrightarrow\lbrack\cdots\rbrack$.  Or, in other words,
\begin{eqnarray}
    x_{ij}\tilde{x}_{kl}\left(
        \lbrack\mathbf{kl}\rbrack - 
        \langle\mathbf{kl}\rangle 
    \right) &=&
    \frac{1}{m_jm_l}\langle\mathbf{k}|p_jp_i|\mathbf{l}\rangle+
    \frac{m_j}{m_l}\langle\mathbf{kl}\rangle\ ,
    \label{eq:xtildex([kl]-<kl>)=...}
    \nonumber\\\\
    x_{ij}\tilde{x}_{kl}\left(
        \langle\mathbf{ij}\rangle -
        \lbrack\mathbf{ij}\rbrack 
    \right) &=&
    \frac{1}{m_jm_l}\lbrack\mathbf{i}|p_lp_k|\mathbf{j}\rbrack +
    \frac{m_l}{m_j}\lbrack\mathbf{ij}\rbrack\ ,
    \label{eq:xtildex(<ij>-[ij])=...}
    \nonumber\\
\end{eqnarray}
where $p_i+p_j+p_k+p_l=0$.
%We will see these identities used in App.~\ref{sec:Constructive Massless eemm no Spinor Shift}.

If we look at the transition between Eqs.~(5.42) and (5.44) in \cite{Arkani-Hamed:2017jhn}, it appears that the authors use a truncated version of this identity, namely $x_{ij}\tilde{x}_{kl}\langle\mathbf{ij}\rangle = x_{ij}\tilde{x}_{kl}\lbrack\mathbf{ij}\rbrack$ and $   x_{ij}\tilde{x}_{kl}\langle\mathbf{kl}\rangle = x_{ij}\tilde{x}_{kl}\lbrack\mathbf{kl}\rbrack$.  But, this can not be correct since $\langle\mathbf{ij}\rangle\neq\lbrack\mathbf{ij}\rbrack$ and the final amplitude will be different whether we change all the spinor products into angle spinor products or square spinor products.  Unfortunately, even the full identities in Eqs.~(\ref{eq:xtildex([kl]-<kl>)=...}) and (\ref{eq:xtildex(<ij>-[ij])=...}) will not bring the amplitude into agreement with Feynman diagrams, as we will see in App.~\ref{sec:Constructive Massless eemm no Spinor Shift}.

\subsection{\label{app:xxt+xtx}Replacing $\mathbf{x_{12}\tilde{x}_{34}+\tilde{x}_{12}x_{34}}$}
As mentioned in the previous subsection, we frequently have a massless particle on an internal line.  After using the identities in that subsection, we always end with the $x_{12}$ and $x_{34}$ from the vertex factoring into the form  $x_{12}\tilde{x}_{34}+\tilde{x}_{12}x_{34}$, assuming the massless particle is in the s-channel.  In this section, we work out an identity to replace this factor with mass and momentum products, thereby eliminating the x factors from the expression.  

If we plug in the explicit expressions for $x$ and $\tilde{x}$ from Eqs.~(\ref{eq:x_12 = xi...}) and (\ref{eq:xtilde_12 = xi...}), we have
\begin{eqnarray}
    4m_2m_4x_{12}\tilde{x}_{34} &=&
    -\frac{\langle\xi_{12}|(p_2-p_1)|p_{12}\rbrack\lbrack\xi_{34}|(p_4-p_3)|p_{12}\rangle}{\langle\xi_{12}p_{12}\rangle\lbrack\xi_{34}p_{12}\rbrack} 
    \nonumber\\
    4m_2m_4\tilde{x}_{12}x_{34} &=&
    -\frac{\lbrack\xi_{12}|(p_2-p_1)|p_{12}\rangle\langle\xi_{34}|(p_4-p_3)|p_{12}\rbrack}{\lbrack\xi_{12}p_{12}\rbrack\langle\xi_{34}p_{12}\rangle}\ ,
    \nonumber\\
    \label{eq:m1m3(xxt+xtx) =}
\end{eqnarray}
where we have used $p_{34}=-p_{12}$, which gives $\lvert p_{34}\rangle = -\lvert p_{12}\rangle$ and $\lvert p_{34}\rbrack = \lvert p_{12}\rbrack$, so that $\lvert p_{34}\rangle\lbrack p_{34}\rvert = -\lvert p_{12}\rangle\lbrack p_{12}\rvert$.  (Whether the angle or the square bracket flips sign is a convention.)

Our next step is to apply the Schouten identity [see App.~B from \cite{Christensen:2019mch}, especially Eqs.~(B19) and (B20)] to the numerators.  We obtain
\begin{eqnarray}
    4m_2m_4x_{12}\tilde{x}_{34} &=& 
    -\frac{\lbrack\xi_{34}|(p_4-p_3)(p_2-p_1)|p_{12}\rbrack}{\lbrack\xi_{34}p_{12}\rbrack}
    \nonumber\\
    4m_2m_4\tilde{x}_{12}x_{34} &=& 
    -\frac{\langle\xi_{34}|(p_4-p_3)(p_2-p_1)|p_{12}\rangle}{\langle\xi_{34}p_{12}\rangle}\ ,
    \nonumber\\
    \label{eq:4mmxxtilde intermediate 1}
\end{eqnarray}
where we have dropped vanishing terms of the form $\langle p_{12}|p_i|p_{12}\rbrack$.  We can see this by first noting that $\langle p_{12}|p_i|p_{12}\rbrack=2p_i\cdot p_{12}=2\left(p_1\cdot p_i+p_2\cdot p_i\right)$.  We can consider each case separately.  If $p_i=p_1$, we find $2p_1\cdot p_1 + 2p_1\cdot p_2 = 2m_1^2+(p_1+p_2)^2-m_1^2-m_2^2$.  However, $(p_1+p_2)^2=0$ is the on-shell condition for the massless particle connecting the vertices.  Moreover, a massless particle only connects particles of the same mass, therefore, $m_2=m_1$ and we have $\langle p_{12}|p_1|p_{12}\rbrack=0$.  Similarly, $\langle p_{12}|p_2|p_{12}\rbrack=0$.  But, considering momenta from the other side of the propagator is no different since $|p_{12}\rbrack\langle p_{12}|=-|p_{34}\rbrack\langle p_{34}|$.  Therefore, $\langle p_{12}|p_3|p_{12}\rbrack=-\langle p_{34}|p_3|p_{34}\rbrack=0$ by the same reasoning.  Finally, in the same way, we find $\langle p_{12}|p_4|p_{12}\rbrack=0$.
Next, obtaining a common denominator gives us
\begin{eqnarray}
    4m_2m_4x_{12}\tilde{x}_{34} &=&
    \frac{\lbrack\xi_{34}|(p_4-p_3)(p_2-p_1)|p_{12}\rbrack\langle p_{12}\xi_{34}\rangle}{\lbrack\xi_{34}p_{12}\rbrack\langle\xi_{34}p_{12}\rangle}
    \nonumber\\
    4m_2m_4\tilde{x}_{12}x_{34} &=&
    \frac{\langle\xi_{34}|(p_4-p_3)(p_2-p_1)|p_{12}\rangle\lbrack p_{12}\xi_{34}\rbrack}{\lbrack\xi_{34}p_{12}\rbrack\langle\xi_{34}p_{12}\rangle}\ .
    \nonumber\\
\end{eqnarray}
Now, we use a Schouten identity on the second equation giving
\begin{eqnarray}
    4m_2m_4x_{12}\tilde{x}_{34} &=& 
    \frac{\lbrack\xi_{34}|(p_4-p_3)(p_2-p_1)|p_{12}\rbrack\langle p_{12}\xi_{34}\rangle}{\lbrack\xi_{34}p_{12}\rbrack\langle\xi_{34}p_{12}\rangle}
    \nonumber\\
    4m_2m_4\tilde{x}_{12}x_{34} &=&
    -\frac{\lbrack\xi_{34}|(p_2-p_1)|p_{12}\rangle\lbrack p_{12}|(p_4-p_3)|\xi_{34}\rangle}{\lbrack\xi_{34}p_{12}\rbrack\langle\xi_{34}p_{12}\rangle}\ ,
    \nonumber\\
\end{eqnarray}
where we have again dropped terms of the form $\langle p_{12}|p_i|p_{12}\rbrack$.  We perform one more Schouten identity on the right to obtain
\begin{eqnarray}
    4m_2m_4x_{12}\tilde{x}_{34} &=& 
    \frac{\lbrack\xi_{34}|(p_4-p_3)(p_2-p_1)|p_{12}\rbrack\langle p_{12}\xi_{34}\rangle}{\lbrack\xi_{34}p_{12}\rbrack\langle\xi_{34}p_{12}\rangle}
    \nonumber\\
    4m_2m_4\tilde{x}_{12}x_{34} &=&
    \frac{\lbrack p_{12}|(p_4-p_3)(p_2-p_1)|\xi_{34}\rbrack\langle\xi_{34}p_{12}\rangle}{\lbrack\xi_{34}p_{12}\rbrack\langle\xi_{34}p_{12}\rangle}\ ,
    \nonumber\\
\end{eqnarray}
which, after cancelling $\langle p_{12}\xi_{34}\rangle$ and reversing the spinor product in the second term, is equal to
\begin{eqnarray}
    4m_2m_4x_{12}\tilde{x}_{34} &=&
    -\frac{\lbrack\xi_{34}|(p_4-p_3)(p_2-p_1)|p_{12}\rbrack}{\lbrack\xi_{34}p_{12}\rbrack}
    \nonumber\\
    4m_2m_4\tilde{x}_{12}x_{34} &=&
    -\frac{\lbrack \xi_{34}|(p_2-p_1)(p_4-p_3)|p_{12}\rbrack}{\lbrack\xi_{34}p_{12}\rbrack}\ .
    \nonumber\\
\end{eqnarray}
Finally, adding these two terms together and using $(p_4-p_3)(p_2-p_1)+(p_2-p_1)(p_4-p_3)=2(p_2-p_1)\cdot (p_4-p_3)$, we obtain
\begin{equation}
    x_{12}\tilde{x}_{34}+\tilde{x}_{12}x_{34} = 
    -\frac{(p_2-p_1)\cdot (p_4-p_3)}{2m_2m_4}\ .
    \label{eq:xxt+xtx}
\end{equation}
Our results match the final result of Eq.~(5.40) from \cite{Arkani-Hamed:2017jhn}, up to a sign, although we have included the symmetrization in $p_3$ and $p_4$ as well as in $p_1$ and $p_2$ and we allow the masses on the two ends to be different.  We will see this identity used in App.~\ref{sec:Constructive Massless eemm no Spinor Shift}.

\subsection{\label{sec:x = alternate}$x$ on an External Line} 
In this subsection, we find identities involving the x factor on external lines.  Although we could use Eqs.~(\ref{eq:x_12 = xi...}) and (\ref{eq:xtilde_12 = xi...}) and cancel the reference spinor in the final result, these identities prove to be very useful.  In fact, using them, we immediately expose the helicity spinors of the external photons and quickly find very simple final results.  In fact, combined with the momentum shifts described in App.~\ref{app:analytic continuation}, we find simple expressions for the amplitudes that agree with Feynman diagrams.  

For concreteness, we consider a four-point amplitude where particle $3$ is massless and the propagator is in the s channel.  In this case, we obtain an amplitude with an $x_{12,4}$, where the subscript represents that the massive particle entering this vertex has momentum $p_{12}$ and the same-mass antiparticle has momentum $p_4$.  Our next step is to multiply the $x$ factor by a special form of $1$.  In particular, we multiply by the other propagator denominator divided by itself.  Assuming the other diagram is in the u channel, so that $m_2=m_4=m$, we obtain
\begin{eqnarray}
    x_{12,4} &=&
    x_{12,4}\frac{(p_2+p_3)^2-m^2}{(u-m^2)}
    \nonumber\\
    &=& x_{12,4}\frac{\lbrack3\lvert p_2\rvert3\rangle}{(u-m^2)}\ ,
    \label{eq:x12,4=x12,4[3|p2|3>/(u-m^2)}
\end{eqnarray}
where we have used $p_2^2=m^2$, $p_3^2=0$ and $2p_2\cdot p_3 = \lbrack3\lvert p_2\rvert3\rangle$.  We can now use Eq.~(\ref{eq:x|3>=}) to obtain
\begin{eqnarray}
    x_{12,4} &=&
    \frac{\lbrack3|p_2(p_4-p_1-p_2)|3\rbrack}{2m\left(u-m^2\right)}\ .
    \label{eq:x_12,4 = intermediate}
\end{eqnarray}
However, we can use $(p_1+p_2)\lvert3\rbrack = -(p_3+p_4)\lvert3\rbrack = -p_4\lvert3\rbrack$ to write it more simply as
\begin{eqnarray}
    x_{12,4} &=&
    \frac{\lbrack3|p_2p_4|3\rbrack}{m\left(u-m^2\right)}\ .
    \label{eq:x12,4...p1p2}
\end{eqnarray}
As we can see, this immediately exposes the helicity spinor of the external massless particle as desired.  We can also obtain the same result, more directly, by using the definition of $x$ from Eq.~(\ref{eq:x|3>=})
\begin{equation}
    x_{12,4}\lvert3\rangle = \frac{1}{2m}\left(p_4-p_2-p_1\right)\lvert3\rbrack\ ,
\end{equation}
and multiplying on the left by $\lbrack3\rvert p_2$, giving
\begin{equation}
    x_{12,4} = \frac{\lbrack3\rvert p_2\left(p_4-p_2-p_1\right)\lvert3\rbrack}{2m\lbrack3\rvert p_2\lvert3\rangle}
    \ ,
\end{equation}
where we have moved the coefficient of $x_{12,4}$ to the right side.  This is the same expression that we found in Eq.~(\ref{eq:x_12,4 = intermediate}) and the final result is the same.  Another (equivalent) way to see this identity is to use Eq.~(\ref{eq:x_12 = xi...}) and set the reference spinor to $\lbrack3\rvert p_2$.  In fact, we see that we could choose either of the propagator denominators depending on the momentum we choose in $\lbrack\xi\rvert=\lbrack3\rvert p$.

We can follow a similar procedure for the other diagram.  In the u channel, we would have $x_{2,14}$ to obtain
\begin{eqnarray}
    x_{2,14} 
    &=& \frac{\lbrack3\lvert p_4(-p_2+p_1+p_4)\rvert3\rbrack}{2m\lbrack3\lvert p_4\rvert3\rbrack}\ .
    \label{eq:x12,4=[3|p4(-p2+p1+p4)|3]/2m(s-m^2)}
\end{eqnarray}
Once again, we can use $(p_1+p_4)\rvert3\rbrack = -(p_2+p_3)\rvert3\rbrack = -p_2\rvert3\rbrack$, to write this as
\begin{eqnarray}
    x_{2,14} &=&
    \frac{\lbrack3\lvert p_2p_4\rvert3\rbrack}{m(s-m^2)}\ ,
    \label{x_{2,14}=}
\end{eqnarray}
where we used the antisymmetry of the spinor product.  Interestingly, this suggests $x_{12,4}$ and $x_{2,14}$ differ only by their final propagator denominator.  At this point, we can include these in the amplitude and, now that they have a common denominator, simplify the result considerably.  This method also works for $\tilde{x}$ where the final result has an angle-bracket spinor product rather than a square-bracket spinor product.
\begin{eqnarray}
    \tilde{x}_{12,4} &=& \frac{\langle3|p_2p_4|3\rangle}{m\left(u-m^2\right)}
    \\
    \tilde{x}_{2,14} &=& \frac{\langle3\lvert p_2p_4\rvert3\rangle}{m(s-m^2)}\ .
\end{eqnarray}
We could also use the $t$ channel and so on.  There are several useful combinations, depending on the process.  
The important thing to remember when using these identities is that the numerator should always include the massless particle's momentum.  That is, whether to use $(p_i+p_j)^2-m^2$ or $(p_k+p_l)^2-m^2$ in the numerator, we always choose the form with the massless momentum.
As another example, if the massless particle were instead particle 2, we would have
\begin{eqnarray}
    x_{1,34} &=&         \frac{\lbrack2|p_1p_3|2\rbrack}{m\left(u-m^2\right)}
    \label{eq:x_1,34=}
    \\
    x_{14,3} &=& \frac{\lbrack2\lvert p_1p_3\rvert2\rbrack}{m(s-m^2)}
    \label{eq:x_14,3=}
    \\
    \tilde{x}_{1,34} &=&         \frac{\langle2|p_1p_3|2\rangle}{m\left(u-m^2\right)}
    \label{eq:xt_1,34=}
    \\
    \tilde{x}_{14,3} &=& \frac{\langle2\lvert p_1p_3\rvert2\rangle}{m(s-m^2)}
    \label{eq:xt_14,3=}
    \ .
\end{eqnarray}
We will use these identities in Apps.~\ref{sec:Constructive Massless heAe no Shift}, \ref{app:eeA+A+ no shift} and \ref{app:eeA+A- no shift}.

\subsection{\label{app:xxt+xtx revisited}$x_{12}\tilde{x}_{34}+\tilde{x}_{12}x_{34}$ Revisited}
Now that we have useful identities when the x factors appear on internal and external lines, let's revisit $x_{12}\tilde{x}_{34}+\tilde{x}_{12}x_{34}$, and see whether we can achieve an identity by multiplying by a special form of $1$ as we did in the previous subsection.
Unlike in the previous cases, we do not want to use the propagator denominator of another diagram.  The reason is that we need the s-channel momentum in the numerator in order to apply the identities in Eqs.~(\ref{eq:x|3>=}) and (\ref{eq:xtilde|]=}).  The form we will choose is a product of $p_1+p_2$ with a reference momentum in the following way
\begin{equation}
    x_{12}\tilde{x}_{34} = 
    x_{12}\tilde{x}_{34}
    \frac{\lbrack p_{12}\lvert p_\xi \rvert p_{12}\rangle}
    {2p_{12}\cdot p_\xi}\ .
\end{equation}
We use $\lvert p_{12}\rbrack=\lvert p_{34}\rbrack$ to obtain
\begin{equation}
    x_{12}\tilde{x}_{34} = 
    x_{12}\tilde{x}_{34}
    \frac{\lbrack p_{34}\lvert p_\xi \rvert p_{12}\rangle}
    {2p_{12}\cdot p_\xi}\ .
\end{equation}
We can now apply Eqs.~(\ref{eq:x|3>=}) and (\ref{eq:xtilde|]=}) and find
\begin{equation}
    x_{12}\tilde{x}_{34} = 
    -\frac{\langle p_{34}\lvert (p_4-p_3)p_\xi(p_2-p_1) \rvert p_{12}\rbrack}
    {8m_2m_4p_{12}\cdot p_\xi}\ .
\end{equation}
Using $\lvert p_{34}\rangle = -\lvert p_{12}\rangle$ gives
\begin{eqnarray}
    x_{12}\tilde{x}_{34} &=& 
    \frac{\mbox{Tr}[ (p_4-p_3)p_\xi(p_2-p_1)(p_1+p_2)]}
    {8m_2m_4p_{12}\cdot p_\xi}
    \nonumber\\
    &=& \frac{ -(p_4-p_3)\cdot (p_2-p_1) p_{12}\cdot p_\xi 
    }{4m_2m_4p_{12}\cdot p_\xi}
    \ ,
\end{eqnarray}
since $(p_4-p_3)\cdot (p_1+p_2)=(p_2-p_1)\cdot (p_1+p_2)=0$.  We are now in a position to cancel the $p_{12}\cdot p_{\xi}$ to obtain
\begin{eqnarray}
    x_{12}\tilde{x}_{34} &=& 
    -\frac{(p_2-p_1)\cdot(p_4-p_3)}{4m_2m_4}
    \ .
\end{eqnarray}
This is interesting since, unlike in App.~\ref{app:xxt+xtx}, we didn't even need $\tilde{x}_{12}x_{34}$ this time to obtain this result.  Because of the symmetry, we can see that we get the same result from $\tilde{x}_{12}x_{34}$, giving us the same final result
\begin{eqnarray}
    x_{12}\tilde{x}_{34} + \tilde{x}_{12}x_{34} &=& 
    -\frac{(p_2-p_1)\cdot(p_4-p_3)}{2m_2m_4}
    \ .
\end{eqnarray}
It appears that this time we did not need a reference spinor, however, if $p_\xi$ is any linear combination of the external momenta, $p_{12}\cdot p_{\xi}=0$ [as we showed in the text below Eq.~(\ref{eq:4mmxxtilde intermediate 1})].  Therefore, $p_{\xi}$ has taken the place of a reference spinor in this calculation, but is removed before ever appearing in the amplitude.

There are also ways of deriving this identity without the use of either a reference spinor or a reference momentum.  Whenever we encounter an $x_{ij}\tilde{x}_{kl}$ on an internal line, it is always accompanied by $\langle\mathbf{ij}\rangle\lbrack\mathbf{kl}\rbrack$.  So, let's consider
\begin{eqnarray}
    x_{ij}\tilde{x}_{kl}\langle\mathbf{ij}\rangle &=& 
    \frac{1}{s}x_{ij}\tilde{x}_{kl}\langle\mathbf{i}\lvert (p_i+p_j)^2 \rvert\mathbf{j}\rangle
    \nonumber\\
    &=& \frac{1}{s}x_{ij}\tilde{x}_{kl}\langle\mathbf{i}p_{ij}\rangle\lbrack p_{ij}\lvert (p_i+p_j)\rvert\mathbf{j}\rangle
    \nonumber\\
    &=& \frac{1}{2m_js}\tilde{x}_{kl}\langle\mathbf{i}\lvert(p_j-p_i)\rvert p_{ij}\rbrack\lbrack p_{ij}\lvert (p_i+p_j)\rvert\mathbf{j}\rangle
    \nonumber\\
    &=& \frac{-\langle\mathbf{i}\lvert(p_j-p_i)(p_l-p_k)\rvert p_{ij}\rangle\lbrack p_{ij}\lvert (p_i+p_j)\rvert\mathbf{j}\rangle}{4m_jm_ls}
    \nonumber\\
    &=& \frac{-1}{4m_jm_l}\langle\mathbf{i}\lvert(p_j-p_i)(p_l-p_k)\rvert\mathbf{j}\rangle\ ,
\end{eqnarray}
where the $s/s=0/0$ has cancelled (the numerators of the diagrams are considered on shell during intermediate steps of the calculation).  Similarly, we find
\begin{eqnarray}
    \tilde{x}_{ij}x_{kl}\langle\mathbf{ij}\rangle &=& 
    \frac{1}{s}\tilde{x}_{ij}x_{kl}\langle\mathbf{i}\lvert (p_i+p_j)^2 \rvert\mathbf{j}\rangle
    \nonumber\\
    &=& \frac{1}{s}\tilde{x}_{ij}x_{kl}\langle\mathbf{i}p_{ij}\rangle\lbrack p_{ij}\lvert (p_i+p_j)\rvert\mathbf{j}\rangle
    \nonumber\\
    &=& \frac{-1}{2m_js}\tilde{x}_{ij}\langle\mathbf{i}\lvert(p_l-p_k)\rvert p_{ij}\rbrack\lbrack p_{ij}\lvert (p_i+p_j)\rvert\mathbf{j}\rangle
    \nonumber\\
    &=& \frac{-\langle\mathbf{i}\lvert(p_l-p_k)(p_j-p_i)\rvert p_{ij}\rangle\lbrack p_{ij}\lvert (p_i+p_j)\rvert\mathbf{j}\rangle}{4m_jm_ls}
    \nonumber\\
    &=& \frac{-1}{4m_jm_l}\langle\mathbf{i}\lvert(p_l-p_k)(p_j-p_i)\rvert\mathbf{j}\rangle\ .
\end{eqnarray}
Putting these together, we have,
\begin{equation}
    \left(x_{ij}\tilde{x}_{kl}+\tilde{x}_{ij}x_{kl}\right)\langle\mathbf{ij}\rangle =
    -\frac{(p_l-p_k)\cdot(p_j-p_i)}{2m_jm_l}\langle\mathbf{ij}\rangle
    \ ,
\end{equation}
where we have used $(p_j-p_i)(p_l-p_k)+(p_l-p_k)(p_j-p_i)=2(p_j-p_i)\cdot(p_l-p_k)$.
We could also have used the square bracket.  Presumably, there are other ways to achieve this result.

Before leaving this section, we note the importance of using the already symmetrized version of Eqs.~(\ref{eq:x|3>=}) and (\ref{eq:xtilde|]=}) and not the unsymmetrized versions found in Eqs.~(\ref{eq:x|3> unsymmetrized}) and (\ref{eq:xt|3] unsymmetrized}).  If we use the unsymmetrized identity, not only do we get a different result, but we get inconsistent results.  To see this, let's focus on the middle form of Eqs.~(\ref{eq:x|3> unsymmetrized}) and (\ref{eq:xt|3] unsymmetrized}).  For simplicity, let's assume we are in the s channel and consider
\begin{eqnarray}
    x_{1,2}\tilde{x}_{3,4} &=& 
    x_{1,2}\tilde{x}_{3,4}
    \frac{\lbrack p_{12}\lvert p_2\rvert p_{12}\rangle}
    {s}\ ,
\end{eqnarray}
where we used  $\lbrack p_{12}\lvert p_2\rvert p_{12}\rangle=\mbox{Tr}[p_2(p_1+p_2)]=2p_1\cdot p_2+2m^2 = s$ in the denominator.  Switching $\lbrack p_{12}\rvert=\lbrack p_{34}\rvert$, as usual, applying the identity and switching back, we obtain
\begin{eqnarray}
    x_{1,2}\tilde{x}_{3,4} &=& 
    \frac{\langle p_{12}\lvert p_4p_2p_2\rvert p_{12}\rbrack}
    {m_2m_4s}\ .
\end{eqnarray}
However, since $p_2p_2=p_2^2=m_2^2$ and $\langle p_{12}\lvert p_4\rvert p_{12}\rbrack = 2p_{12}\cdot p_4=s$, we end with
\begin{equation}
    x_{1,2}\tilde{x}_{3,4} = 
    \frac{m_2}{m_4}\ .
\end{equation}
This incredibly simply formula gives very compact results for four-point amplitudes which disagree with Feynman diagrams.  Not only that, but if we had instead sandwiched $p_4$ in the middle, we would have obtained
\begin{eqnarray}
    x_{1,2}\tilde{x}_{3,4} &=& 
    x_{1,2}\tilde{x}_{3,4}
    \frac{\lbrack p_{12}\lvert p_4\rvert p_{12}\rangle}
    {s}
    \nonumber\\
    &=& \frac{\langle p_{12}\lvert p_4p_4p_2\rvert p_{12}\rbrack}
    {m_2m_4s}
    \nonumber\\
    &=& \frac{m_4}{m_2} \ .
\end{eqnarray}
Unfortunately, we see we cannot trust this result.  In fact, we have only found consistency when we have used the fully symmetrized Eqs.~(\ref{eq:x|3>=}) and (\ref{eq:xtilde|]=}).

\section{\label{app:Calc AHH Spinor Amps No Shifts}Constructive Diagram Calculations}
In this appendix, we will describe the constructive calculation of these amplitudes using a massless photon from the beginning.  With the details of $x$ described in App.~\ref{app:x}, we achieve agreement with Feynman diagrams for the diagrams with external photons but not with the diagram with an internal photon.  This last case requires ingredients beyond what are currently published, as we will see.

\subsection{\label{sec:Constructive Massless eemm no Spinor Shift}$\mathbf{e,\bar{e},\mu,\bar{\mu}}$}

There is only one diagram for this process, involving a photon in the s channel.  The vertex is given on the left by
$e x_{12}\langle\mathbf{12}\rangle$ and on the right by $e \tilde{x}_{34}\lbrack\mathbf{34}\rbrack$ for a negative helicity photon and
$e \tilde{x}_{12}\lbrack\mathbf{12}\rbrack$ and $e x_{34}\langle\mathbf{34}\rangle$ for a positive helicity photon.  Combining these for each photon helicity and dividing by the propagator denominator gives us
\begin{equation}
    \mathcal{M} =
    \frac{e^2}{s}  \left(
    x_{34} \tilde{x}_{12} \lbrack\mathbf{1}\mathbf{2}\rbrack 
    \langle\mathbf{3}\mathbf{4}\rangle 
    +x_{12} \tilde{x}_{34}
    \langle\mathbf{1}\mathbf{2}\rangle \lbrack\mathbf{3}\mathbf{4}\rbrack  \right)\ .
    \label{eq:M_eemm-constructive Initial}
\end{equation}
We now need to use identities that eventually remove the $x$ and $\tilde{x}$.  As we can see in Eqs.~(\ref{eq:xtildex([12]-<12>)=...}) through (\ref{eq:xtildex(<12>-[12])=...}), we can convert square brackets to angle brackets or vice versa, but in doing so, we get extra terms.  There are two different final forms depending on whether we end with $\lbrack\mathbf{12}\rbrack$ or $\langle\mathbf{12}\rangle$ and $\lbrack\mathbf{34}\rbrack$ or $\langle\mathbf{34}\rangle$.  We begin by putting all the spinors multiplying $x\tilde{x}$ in the form of square brackets.  We use Eq.~(\ref{eq:xtildex(<34>-[34])=...}) on the left term and Eq.~(\ref{eq:xtildex(<12>-[12])=...}) on the right term to obtain
\begin{align}
    \mathcal{M} =
    &\ \frac{e^2}{s}  \Bigg[
    \left(
    x_{34} \tilde{x}_{12} + +x_{12} \tilde{x}_{34}
    +\frac{m_e}{m_\mu}
    +\frac{m_\mu}{m_e}
    \right)
    \lbrack\mathbf{1}\mathbf{2}\rbrack 
    \lbrack\mathbf{3}\mathbf{4}\rbrack
    \nonumber\\
    &+\frac{1}{m_em_\mu}\left(
    \lbrack\mathbf{12}\rbrack\lbrack\mathbf{3}\lvert p_2p_1\rvert\mathbf{4}\rbrack+
    \lbrack\mathbf{1}\lvert p_4p_3\rvert\mathbf{2}\rbrack\lbrack\mathbf{34}\rbrack
    \right)
    \Bigg]\ .
    \label{eq:M_eemm-constructive [][] with xxt+xtx}
\end{align}
To remove $\left(
    x_{34} \tilde{x}_{12} + +x_{12} \tilde{x}_{34}
    \right)$, we use the identity in Eq.~(\ref{eq:xxt+xtx}) to find
\begin{align}
    \mathcal{M} =
    &\ \frac{e^2}{2m_em_\mu s}  \Big[
    (u-t+2m_e^2+2m_\mu^2)
    \lbrack\mathbf{1}\mathbf{2}\rbrack 
    \lbrack\mathbf{3}\mathbf{4}\rbrack
    \nonumber\\
    &+2\left(
    \lbrack\mathbf{12}\rbrack \lbrack\mathbf{3}\lvert p_2p_1\rvert\mathbf{4}\rbrack+
    \lbrack\mathbf{1}\lvert p_4p_3\rvert\mathbf{2}\rbrack \lbrack\mathbf{34}\rbrack 
    \right)
    \Big]\ ,
    \label{eq:M_eemm-constructive [][]}
\end{align}
where we have used $(p_2-p_1)\cdot(p_4-p_3)=t-u$.  We could reduce the terms with two momenta in the spinor products, but we find that this would result in a more complex expression, therefore, we will stop here.  We find the same expression if we first apply Eq.~(\ref{eq:xtildex(<12>-[12])=...}) on the right of Eq.~(\ref{eq:M_eemm-constructive Initial}) followed by application of Eq.~(\ref{eq:xtildex([34]-<34>)=...}) on the resulting term.  We also find this same expression if we apply Eq.~(\ref{eq:xtildex(<34>-[34])=...}) to the left term of Eq.~(\ref{eq:M_eemm-constructive Initial}) followed by Eq.~(\ref{eq:xtildex([12]-<12>)=...}) on the resulting term.  In order to show their equality, application of the identities, including the on-shell property of the internal line are necessary.

On the other hand, if we apply Eq.~(\ref{eq:xtildex([12]-<12>)=...}) to the left and Eq.~(\ref{eq:xtildex([34]-<34>)=...}) to the right side of Eq.~(\ref{eq:M_eemm-constructive Initial}), we obtain
\begin{align}
    \mathcal{M} =
    &\ \frac{e^2}{2m_em_\mu s}  \Big[
    (u-t+2m_e^2+2m_\mu^2)
    \langle\mathbf{1}\mathbf{2}\rangle 
    \langle\mathbf{3}\mathbf{4}\rangle
    \nonumber\\
    &+2\left(
    \langle\mathbf{12}\rangle \langle\mathbf{3}\lvert p_2p_1\rvert\mathbf{4}\rangle+
    \langle\mathbf{1}\lvert p_4p_3\rvert\mathbf{2}\rangle \langle\mathbf{34}\rangle
    \right)
    \Big]\ .
    \label{eq:M_eemm-constructive <><>}
\end{align}
We can see that this is the complex conjugate of Eq.~(\ref{eq:M_eemm-constructive [][]}).  We obtain this same expression if we apply Eq.~(\ref{eq:xtildex([34]-<34>)=...}) on the right of Eq.~(\ref{eq:M_eemm-constructive Initial}) followed by application of Eq.~(\ref{eq:xtildex(<12>-[12])=...}) on the resulting term or if we apply Eq.~(\ref{eq:xtildex([12]-<12>)=...}) to the left term of Eq.~(\ref{eq:M_eemm-constructive Initial}) followed by Eq.~(\ref{eq:xtildex(<34>-[34])=...}) on the resulting term.  These last two are the opposite order of applying these identities compared to the order that obtains Eq.~(\ref{eq:M_eemm-constructive [][]}).

All together, we have two different forms for this amplitude, which are complex conjugates of one another, coming from the different ways we apply the identities in Eqs.~(\ref{eq:xtildex([12]-<12>)=...}) through (\ref{eq:xtildex(<12>-[12])=...}).  Unfortunately, none of these agree with the correct form given in Eq.~(\ref{eq:M_eemm}), not even with the use of identities.  In fact, we can see that Eq.~(\ref{eq:M_eemm}) is self conjugate, while these are not.
Since the two forms here are conjugates of each other, their squares are identical.  However, the square does not agree with Feynman diagrams.  In fact, in the high-energy limit, this form of the amplitude grows quadratically with energy and, therefore, cannot be correct.  We will consider a momentum and accompanying spinor shift for this process in App.~\ref{sec:Constructive Massless eemm}, but will find that this will not improve success in obtaining the correct amplitude either.

We note that the authors of \cite{Arkani-Hamed:2017jhn} found a simpler form as can be seen in their Eq.~(5.44), namely $\mathcal{M} = e^2/s*(u-t)/(2m_em_\mu)*   \langle\mathbf{12}\rangle\langle\mathbf{34}\rangle$.  It appears as if they got this form by dropping the right side of Eqs.~(\ref{eq:xtildex([kl]-<kl>)=...}) and (\ref{eq:xtildex(<ij>-[ij])=...}).  This form also grows quadratically at high energy, violating perturbative unitarity, and is not in agreement with Feynman diagrams.  Although this form is incorrect, it interestingly simplifies to the same form as our result after a momentum shift.  We will show this in App.~\ref{sec:Constructive Massless eemm}.

The issue with the $x$ factor is that it does not connect the helicities at the opposite ends of the photon propagator.  For example, the vertices $e x_{12}\langle\mathbf{12}\rangle$ and $e\tilde{x}_{34}\lbrack\mathbf{34}\rbrack$, for this process, do not expose the helicities of the photon and the product $x_{12}\tilde{x}_{34}$ does not connect the helicities of the two vertices at all.  This leads to all the spinor products containing fermions from the same side of the propagator (for example, terms like $\langle\mathbf{12}\rangle\langle\mathbf{34}\rangle$).  Contrast this with the massive photon calculation.  As seen in Eq.~(\ref{eq:M_eeA massive vertex}), the vertex has the form $e/M_A\left(\langle\mathbf{31}\rangle\lbrack\mathbf{23}\rbrack+\lbrack\mathbf{31}\rbrack\langle\mathbf{23}\rangle\right)$, where the photon is particle $3$ in this vertex and its spin is clearly exposed.  Then, when the diagram is formed, the spin of the photon from one vertex is transferred to the other vertex, as described in App.~\ref{sec:eemm spinor amp calculation}.  In this case, this results in all spinor products containing fermions from opposite ends of the propagator (for example, terms like $\langle\mathbf{13}\rangle\lbrack\mathbf{24}\rbrack$), the opposite of the constructive result with the $x$ factor.  In fact, the correct result cannot be put into a form with fermions from the same side of the propagator appearing in spinor products by use of any identities.  We might also note that Feynman diagrams similarly connect the helicities of the photon at the two vertices by inserting $\gamma^\mu$ and $\gamma^\nu$ in the two fermion lines and contracting them with $-g_{\mu\nu}$ (in Feynman gauge).  This leads us to suspect that an improvement to the $x$ factor vertex would expose and connect the helicities of the photon along its propagator.  An obvious naive candidate would be $e\langle\mathbf{1}3\rangle\langle\mathbf{2}3\rangle/m_e$ for the negative helicity vertex and $e\lbrack\mathbf{1}3\rbrack\lbrack\mathbf{2}3\rbrack/m_e$ for the positive helicity vertex.  However, we found that this did not agree with Feynman diagrams either.  Although it connected the helicities of the photon, it leads to a bad high-energy growth, once again violating perturbative unitarity.  Therefore, it appears to us that a second requirement for an improved vertex is that it does not involve a division by the fermion mass.  We leave the correct specification as an open problem.

\subsection{\label{sec:Constructive Massless heAe no Shift}$\mathbf{e,\gamma^+,\bar{e},h}$}
In this subsection, we consider electron-photon diagrams with only one photon.  Since this will require another boson, we must add the Higgs boson to this calculation.
The photon and Higgs can couple on either end of the electron propagator giving us s and u channels.  This means there will be a spin index contracted on the propagator line between the two ends.
In order to describe this calculation, we must make an important note about how we choose which vertex has an upper index and which has a lower index.  The overall choice does not matter since the amplitude is squared, but a relative choice between diagrams does matter as it can flip the relative sign between diagrams.  Therefore, it is important that we make a consistent recipe that treats all diagrams the same.  Since the internal line is a particle going into one vertex and an antiparticle into the opposite vertex, we can use this property.   We will choose the particle end of the propagator to have an upper index and the antiparticle end to have a lower index and note that this choice works in App.~\ref{sec:eAeh spinor amp calculation} as well.
As a result, in the present calculation, for the s channel, the higgs-electron vertex is given by $\frac{m_e}{v}(\langle\mathbf{ P}_{12}^{\mathrm{I}}\mathbf{3}\rangle+\lbrack\mathbf{P}_{12}^{\mathrm{I}}\mathbf{3}\rbrack)$ and the photon-electron vertex is given by $e(x_{1,34}\langle\mathbf{1P}_{34\mathrm{I}}\rangle)$.  On the other hand, for the u-channel, the higgs-electron vertex is given by $\frac{m_e}{v}(\langle\mathbf{1P}_{23\mathrm{I}}\rangle+\lbrack\mathbf{1P}_{23\mathrm{I}}\rbrack)$ and the photon-electron vertex by $e(x_{14,3}\langle\mathbf{P}_{14}^{\mathrm{I}}\mathbf{3}\rangle)$. 

Taking the product of these vertices and dividing by the propagator denominators, we obtain
\begin{align}
    \mathcal{M}_s^+ &=
    \frac{em_e}{v}\frac{x_{1,34}
    (\langle\mathbf{ P}_{12}^{\mathrm{I}}\mathbf{3}\rangle+\lbrack\mathbf{P}_{12}^{\mathrm{I}}\mathbf{3}\rbrack)
    \langle\mathbf{1P}_{34\mathrm{I}}\rangle}
    {s-m_e^2}
    \\
    \mathcal{M}_u^+ &=
    \frac{em_e}{v}\frac{x_{14,3}
    (\langle\mathbf{1P}_{23\mathrm{I}}\rangle+\lbrack\mathbf{1P}_{23\mathrm{I}}\rbrack)
    \langle\mathbf{P}_{14}^{\mathrm{I}}\mathbf{3}\rangle}
    {u-m_e^2}.
\end{align}
In order to simplify these expressions, we need to reverse the momentum of the spinors.  That is, for the s-channel diagram, we need to either (a) set $(\langle\mathbf{ P}_{12}^{\mathrm{I}}\mathbf{3}\rangle+\lbrack\mathbf{P}_{12}^{\mathrm{I}}\mathbf{3}\rbrack)\to(-\langle\mathbf{ P}_{34}^{\mathrm{I}}\mathbf{3}\rangle+\lbrack\mathbf{P}_{34}^{\mathrm{I}}\mathbf{3}\rbrack)$ or (b) set $\langle\mathbf{1P}_{34\mathrm{I}}\rangle\to-\langle\mathbf{1P}_{12\mathrm{I}}\rangle$.  When we reverse the momentum, the sign appears on angle spinors but not square spinors, by convention.  At first look, this appears to create two very different results,
\begin{align}
    \mathcal{M}_s^{(a)+} &=
    \frac{em_e}{v}\frac{x_{1,34}
    (-\langle\mathbf{ P}_{34}^{\mathrm{I}}\mathbf{3}\rangle+\lbrack\mathbf{P}_{34}^{\mathrm{I}}\mathbf{3}\rbrack)
    \langle\mathbf{1P}_{34\mathrm{I}}\rangle}
    {s-m_e^2}
    \nonumber\\
    \mathcal{M}_s^{(b)+} &=
    -\frac{em_e}{v}\frac{x_{1,34}
    (\langle\mathbf{ P}_{12}^{\mathrm{I}}\mathbf{3}\rangle+\lbrack\mathbf{P}_{12}^{\mathrm{I}}\mathbf{3}\rbrack)
    \langle\mathbf{1P}_{12\mathrm{I}}\rangle}
    {s-m_e^2}\ .
    \nonumber
\end{align}
Not only is the overall sign different, but choice (a) creates a relative sign between $-\langle\mathbf{ P}_{34}^{\mathrm{I}}\mathbf{3}\rangle$ and $\lbrack\mathbf{P}_{34}^{\mathrm{I}}\mathbf{3}\rbrack$.  However, after simplifying the contraction, we have
\begin{eqnarray}
    \mathcal{M}_s^{(a)+} &=&
    \frac{em_e}{v}\frac{x_{1,34}
    (m_e\langle\mathbf{13}\rangle
    -\langle\mathbf{1}\lvert P_{34}\rvert\mathbf{3}\rbrack)
    }
    {s-m_e^2}
    \nonumber\\
    \mathcal{M}_s^{(b)+} &=&
    -\frac{em_e}{v}\frac{x_{1,34}
    (-m_e\langle\mathbf{13}\rangle
    -\langle\mathbf{1}\lvert P_{12}\rvert\mathbf{3}\rbrack)
    }
    {s-m_e^2}\ .
    \nonumber
\end{eqnarray}
After switching $P_{34}=-P_{12}$ in the top formula, and factoring out the minus sign in the bottom formula, we get that $\mathcal{M}_s^{(a)+}=\mathcal{M}_s^{(b)+}$.  So, we need not worry which direction we switch the momenta whether $P_{12}\to-P_{34}$ or $P_{34}\to-P_{12}$.  The rules are consistent.  Our next step is to expand $P_{34}=p_3+p_4$ and use the mass identity $p_3\lvert\mathbf{3}\rbrack=-m_e\lvert\mathbf{3}\rangle$ to obtain
\begin{align}
    \mathcal{M}_s^{+} &=
    \frac{em_e}{v}\frac{x_{1,34}
    (2m_e\langle\mathbf{13}\rangle
    -\langle\mathbf{1}\lvert p_4\rvert\mathbf{3}\rbrack)
    }
    {s-m_e^2}
    \\
    \mathcal{M}^+_u 
    &= \frac{em_e}{v}\frac{x_{14,3}
    (2m_e\langle\mathbf{13}\rangle+\lbrack\mathbf{1}\lvert p_4\rvert\mathbf{3}\rangle)
    }
    {u-m_e^2}
    \ ,
\end{align}
where we have followed similar steps for the u-channel diagram.

Next, we remove the $x$ factors by use of Eqs.~(\ref{eq:x_1,34=}) and (\ref{eq:x_14,3=}) described in App.~\ref{sec:x = alternate}.  We include them here for convenience.
\begin{eqnarray}
    x_{1,34} &=&         \frac{\lbrack2|p_1p_3|2\rbrack}{m_e\left(u-m_e^2\right)}
    \\
    x_{14,3} &=& \frac{\lbrack2\lvert p_1p_3\rvert2\rbrack}{m_e(s-m_e^2)}\ .
\end{eqnarray}
As we can see, these forms of the $x$ factor immediately expose the positive helicity spinors of the photon.  Plugging these in, we obtain,
\begin{align}
    \mathcal{M}_s^{+} &=
    \frac{e}{v}\frac{\lbrack2\lvert p_1p_3\rvert2\rbrack
    (2m_e\langle\mathbf{13}\rangle
    -\langle\mathbf{1}\lvert p_4\rvert\mathbf{3}\rbrack)
    }
    {\left(s-m_e^2\right)\left(u-m_e^2\right)}
    \label{eq:M_s^+ after replacing x}
    \\
    \mathcal{M}^+_u 
    &= \frac{e}{v}\frac{\lbrack2\lvert p_1p_3\rvert2\rbrack
    (2m_e\langle\mathbf{13}\rangle+\lbrack\mathbf{1}\lvert p_4\rvert\mathbf{3}\rangle)
    }
    {\left(s-m_e^2\right)\left(u-m_e^2\right)}
    \label{eq:M_u^+ after replacing x}
    \ .
\end{align}

At this point, we must simplify these expressions.  We will show this for the u-channel diagram but have found the same final result using the s-channel diagram as well.  Focusing on the numerator after removing the electric charge, we have
\begin{equation}
    \frac{\mbox{num}}{e} = 2m_e\langle\mathbf{13}\rangle\lbrack2\lvert p_3p_4\rvert2\rbrack 
    + \lbrack\mathbf{1}\lvert p_4\rvert\mathbf{3}\rangle\lbrack2\lvert p_3p_4\rvert2\rbrack
    \ ,
\end{equation}
where we have used momentum conservation to replace $p_1$ with $p_4$.  We see that we have twice the first term compared to the result we obtained in Eq.~(\ref{eq:M_eA+eh}), so we Schouten transform half of it, $\langle\mathbf{13}\rangle\lbrack2\lvert p_3p_4\rvert2\rbrack = -\lbrack2\lvert p_4\rvert\mathbf{3}\rangle\lbrack2\lvert p_3\rvert\mathbf{1}\rangle+\lbrack2\lvert p_4\rvert\mathbf{1}\rangle\lbrack2\lvert p_3\rvert\mathbf{3}\rangle = -m_e\lbrack2\lvert p_4\rvert\mathbf{3}\rangle\lbrack2\mathbf{1}\rbrack+\lbrack2\lvert p_4\rvert\mathbf{3}\rangle\lbrack2\lvert p_4\rvert\mathbf{1}\rangle-m_e\lbrack2\lvert p_4\rvert\mathbf{1}\rangle\lbrack2\mathbf{3}\rbrack$, and we also Schouten transform the second term, $\lbrack\mathbf{1}\lvert p_4\rvert\mathbf{3}\rangle\lbrack2\lvert p_3p_4\rvert2\rbrack = \lbrack2\lvert p_4\rvert\mathbf{3}\rangle\lbrack2\lvert p_3p_4\rvert\mathbf{1}\rbrack - \lbrack2\mathbf{1}\rbrack\lbrack2\lvert p_3p_4p_4\rvert\mathbf{3}\rangle = \lbrack2\lvert p_4\rvert\mathbf{3}\rangle\lbrack2\lvert p_3p_4\rvert\mathbf{1}\rbrack +m_em_h^2 \lbrack2\mathbf{1}\rbrack\lbrack2\mathbf{3}\rbrack$, where we have also used momentum conservation and the mass identities in these manipulations.  Plugging these back in, we have
\begin{align}
    \frac{\mbox{num}}{e} &= m_e\langle\mathbf{13}\rangle\lbrack2\lvert p_3p_4\rvert2\rbrack 
    + m_e^2\lbrack\mathbf{1}2\rbrack\lbrack2\lvert p_4\rvert\mathbf{3}\rangle
    \nonumber\\
    &-m_e^2\lbrack2\mathbf{3}\rbrack\lbrack2\lvert p_4\rvert\mathbf{1}\rangle
    - m_em_h^2 \lbrack\mathbf{1}2\rbrack\lbrack2\mathbf{3}\rbrack
    \nonumber\\
    &+ m_e\lbrack2\lvert p_4\rvert\mathbf{3}\rangle\lbrack2\lvert p_4\rvert\mathbf{1}\rangle
    + \lbrack2\lvert p_4\rvert\mathbf{3}\rangle\lbrack2\lvert p_3p_4\rvert\mathbf{1}\rbrack 
    \ .
\end{align}
We can see that the first four terms are what we expect.  We just need to show that the last two terms cancel.  
In the second-to-last term, we push the $m_e$ back into the spinor product, $-m_e\lbrack2\lvert p_4\rvert\mathbf{1}\rangle = \lbrack2\lvert p_4p_1\rvert\mathbf{1}\rbrack$, followed by a reversal of the momenta, $-m_e\lbrack2\lvert p_4\rvert\mathbf{1}\rangle = -2p_1\cdot p_4 \lbrack\mathbf{1}2\rbrack - \lbrack2\lvert p_1p_4\rvert\mathbf{1}\rbrack$ and momentum conservation, $-m_e\lbrack2\lvert p_4\rvert\mathbf{1}\rangle = -2p_1\cdot p_4 \lbrack\mathbf{1}2\rbrack + \lbrack2\lvert p_3p_4\rvert\mathbf{1}\rbrack-m_h^2 \lbrack\mathbf{1}2\rbrack$, giving us
\begin{align}
    \frac{\mbox{num}}{e} &= m_e\langle\mathbf{13}\rangle\lbrack2\lvert p_3p_4\rvert2\rbrack 
    + m_e^2\lbrack\mathbf{1}2\rbrack\lbrack2\lvert p_4\rvert\mathbf{3}\rangle
    \nonumber\\
    &-m_e^2\lbrack2\mathbf{3}\rbrack\lbrack2\lvert p_4\rvert\mathbf{1}\rangle
    - m_em_h^2 \lbrack\mathbf{1}2\rbrack\lbrack2\mathbf{3}\rbrack
    \nonumber\\
    &+\left(2p_1\cdot p_4+m_h^2\right) \lbrack\mathbf{1}2\rbrack \lbrack2\lvert p_4\rvert\mathbf{3}\rangle
    \  .
\end{align}
However, $2p_1\cdot p_4+m_h^2 = u-m_e^2$, which vanishes due to this diagram being on shell.  Therefore, we are left with
\begin{align}
    \frac{\textrm{num}}{e} &=
    m_e\langle\mathbf{13}\rangle\lbrack2\lvert p_3p_4\rvert2\rbrack
    + m_e^2\lbrack\mathbf{1}2\rbrack \lbrack2\lvert p_4\rvert\mathbf{3}\rangle
    \nonumber\\
    &-m_e^2 \lbrack2\mathbf{3}\rbrack\lbrack2\lvert p_4\rvert\mathbf{1}\rangle
    -m_em_h^2\lbrack\mathbf{1}2\rbrack\lbrack2\mathbf{3}\rbrack   \ .
\end{align}
in agreement with Feynman diagrams as we discussed with Eq.~(\ref{eq:M_eA+eh}).

\subsection{\label{app:eeA+A+ no shift}$\mathbf{e,\bar{e},\gamma^{\pm},\gamma^{\pm}}$}
We next turn to the amplitude with two photons.  In this subsection, we consider the case where they both have the same helicity and demonstrate the calculation when both have positive helicity.  We will take the electron and positron to be particles 1 and 2, respectively.  There are two diagrams possible in the $t$ and $u$ channels.  As in the previous subsection, either one can be used and they result in the same final amplitude and agree with Feynman diagrams.  In the case of the t-channel diagram, we have the vertices $e x_{1,24}\langle\mathbf{1P}_{24\mathrm{I}}\rangle$ and $e x_{13,2}\langle\mathbf{P}_{13}^{\mathrm{I}}\mathbf{2}\rangle$, while in the case of the u-channel diagram, we have $e x_{1,23}\langle\mathbf{1P}_{23\mathrm{I}}\rangle$ and $e x_{14,2}\langle\mathbf{P}_{14}^{\mathrm{I}}\mathbf{2}\rangle$.  Multiplying and dividing by the propagator denominator, gives us
\begin{align}
    \mathcal{M}_t^{++} &=
    \frac{e^2 m_e x_{1,24}x_{13,2}\langle\mathbf{12}\rangle}{t-m_e^2} 
    \label{eq:M_t^++ initial x amp}
    \\
    \mathcal{M}_u^{++} &=
    -\frac{e^2 m_e x_{1,23}x_{14,2}\langle\mathbf{12}\rangle}{u-m_e^2}
    \ ,
    \label{eq:M_ueeA+A+ constructive initial}
\end{align}
where we have used the mass relation $\langle\mathbf{1P}_{\mathrm{I}}\rangle\langle\mathbf{P}^{\mathrm{I}}\mathbf{2}\rangle = -m_e\langle\mathbf{12}\rangle$.  We now need to replace the x factors using the techniques described in App.~\ref{sec:x = alternate}.  In this case, we have two x factors to replace but only one other diagram.  One of the x factors should include the propagator denominator from the other diagram, but we must determine what to do with the other.  As explained in App.~\ref{sec:x = alternate}, each x factor has two possible forms with the two choices of propagator denominator.  We find
\begin{align}
    x_{1,24} 
    &= \frac{\lbrack3\lvert p_1p_2\rvert3\rbrack}{m(u-m^2)}
    &=& \frac{\lbrack3\lvert p_1p_4\rvert3\rbrack}{m s}
    \label{eq:x_1,24}
    \\
    x_{13,2} 
    &= \frac{\lbrack4\lvert p_1p_2\rvert4\rbrack}{m(u-m^2)}
    &=& \frac{\lbrack4\lvert p_1p_3\rvert4\rbrack}{m s}
    \label{eq:x_13,2}
    \\
    x_{1,23} 
    &= \frac{\lbrack4\lvert p_1p_2\rvert4\rbrack}{m(t-m^2)}
    &=& \frac{\lbrack4\lvert p_1p_3\rvert4\rbrack}{m s}
    \label{eq:x_1,23}
    \\
    x_{14,2} 
    &= \frac{\lbrack3\lvert p_1p_2\rvert3\rbrack}{m(t-m^2)}
    &=& \frac{\lbrack3\lvert p_1p_4\rvert3\rbrack}{m s}
    \label{eq:x_14,2}
    \ .
\end{align}
However, the numerators are related by momentum conservation and the denominators are also if we take the propagator of the diagram to be on shell.  That is, the top two expressions are for the t-channel diagram where the on-shell condition is $t=m^2$, resulting in $s=-t-u+2m^2=-u+m^2$.  The last two are similar, since they are used in the u-channel diagram.  Therefore, we can use either the middle or the right expressions for $x$, since we can use the on-shell condition with momentum conservation during the simplification.  Furthermore, since there are two photons, we could also use forms of $x$ that use the other photon's helicity spinor as the reference spinor.  For example, 
\begin{align}
    x_{1,23} = -\frac{\langle3\lvert p_1\rvert4\rbrack}{m_e\langle34\rangle}
    \quad \mbox{and}\quad 
    x_{14,2} = \frac{\langle4\lvert p_2\rvert3\rbrack}{m_e\langle43\rangle}
    \ .
    \label{eq:x_1,23 and x_14,2 reference spinors}
\end{align}
However, since we prefer to get the angle brackets out of the denominator, we need to multiply the numerator and denominator by $\lbrack34\rbrack$.  Using the helicity-spinor relation $\lvert3\rbrack\langle3\rvert=p_3$, and similarly for $p_4$, we have
\begin{align}
    x_{1,23} = -\frac{\lbrack4\lvert p_3p_1\rvert4\rbrack}{m_e s}
    \quad \mbox{and}\quad 
    x_{14,2} = \frac{\lbrack3\lvert p_4p_2\rvert3\rbrack}{m_e s}
    \ ,
\end{align}
which, after a reversal of the spinor products and a use of momentum conservation are the same as the identies in Eqs.~(\ref{eq:x_1,23}) and (\ref{eq:x_14,2}).  Therefore, we can use any of these identities for the $x$ and they all work.  However, as in other amplitude calculations, some calculations are simpler one way than another.  

Since we will demonstrate the calculation with the t-channel diagram in App.~\ref{app:eeA+A+} using Eqs.~(\ref{eq:x_1,24}) and (\ref{eq:x_13,2}), we will show how the u-channel diagram works here and use the form of $x$ in Eq.~(\ref{eq:x_1,23 and x_14,2 reference spinors}).  Multiplying by $\lbrack34\rbrack$ in the numerator and denominator, but not simplifying the numerator yet, we have
\begin{equation}
\mathcal{M}_u^{++} =
    \frac{e^2 \langle4\lvert p_2\rvert3\rbrack \langle3\lvert p_1\rvert4\rbrack
    \lbrack34\rbrack^2\langle\mathbf{12}\rangle}{m_e s^2\left(u-m_e^2\right)}
    \ .
\end{equation}
Since the $\lbrack34\rbrack^2$ is already what we want in the final answer, we directly simplify $-\langle4\lvert p_2\rvert3\rbrack \langle3\lvert p_1\rvert4\rbrack = -\mbox{Tr}\left(p_4p_2p_3p_1\right) = -2p_1\cdot p_3p_2\cdot p_4 + 2p_1\cdot p_2p_3\cdot p_4 - 2p_1\cdot p_4p_2\cdot p_3 = -\frac{1}{2}\left(t-m_e^2\right)^2+\frac{1}{2}s\left(s-2m_e^2\right)-\frac{1}{2}\left(u-m_e^2\right)^2$.  Since the diagram is considered on shell, $u=m_e^2$ and $s=-(t-m_e^2)$, giving us $-\langle4\lvert p_2\rvert3\rbrack \langle3\lvert p_1\rvert4\rbrack = \frac{1}{2}s\left(s+t-3m_e^2\right)$.  Due to the on-shell condition, we have $s+t=m_e^2$, giving us finally, $-\langle4\lvert p_2\rvert3\rbrack \langle3\lvert p_1\rvert4\rbrack = -m_e^2 s$.  In the denominator, we cancel one $s$ and use $s=-(t-m_e^2)$ on the other.  We end with
\begin{equation}
   \mathcal{M}_u^{++} = 
    -\frac{e^2 m_e \langle\mathbf{12}\rangle \lbrack34\rbrack^2 }
    {\left(t-m_e^2\right)\left(u-m_e^2\right)} \ ,
\end{equation}
in agreement with Eq.~(\ref{eq:M_eeA+A+}).

\subsection{\label{app:eeA+A- no shift}$\mathbf{e,\bar{e},\gamma^{\pm},\gamma^{\mp}}$}
For our last process, we consider the photons having opposite helicity.  We will demonstrate the $+-$ case.  For the t-channel diagram, the vertices are $e x_{1,24}\langle\mathbf{1P}_{24\mathrm{I}}\rangle$ and $e \tilde{x}_{13,2}\lbrack\mathbf{P}_{13}^{\mathrm{I}}\mathbf{2}\rbrack$ while the vertices for the u-channel diagram are $e \tilde{x}_{1,23}\lbrack\mathbf{1P}_{23\mathrm{I}}\rbrack$ and $e x_{14,2}\langle\mathbf{P}_{14}^{\mathrm{I}}\mathbf{2}\rangle$, giving the preliminary amplitudes
\begin{align}
    \mathcal{M}_t^{+-} &=
    \frac{e^2 x_{1,24}\tilde{x}_{13,2}
    \left(m_e\lbrack\mathbf{12}\rbrack
    +\langle\mathbf{1}\lvert p_3\rvert\mathbf{2}\rbrack
    \right)}
    {t-m_e^2}
    \label{eq:M_eeA+A- initial constructive}
    \
    \\
    \mathcal{M}_u^{+-} &=
    \frac{e^2 \tilde{x}_{1,23}x_{14,2}
    \left(
    m_e\langle\mathbf{12}\rangle
    +\lbrack\mathbf{1}\lvert p_4\rvert \mathbf{2}\rangle
    \right)}
    {u-m_e^2}\ ,
\end{align}
where we have used the momentum relation $\lvert\mathbf{P}_{ij}^\mathrm{I}\rangle\lbrack\mathbf{P}_{ij\mathrm{I}}\rvert = p_i+p_j$ as well as the mass relations $\langle\mathbf{1}\rvert p_1 = m_e\lbrack\mathbf{1}\rvert$ and $\lbrack\mathbf{1}\rvert p_1 = m_e\langle\mathbf{1}\rvert$.  

As in the $++$-helicity case described in the previous subsection, when we replace $x$ and $\tilde{x}$, we have a choice of whether to replace both of them with the $u-m_e^2$ for the t-channel and $t-m_e^2$ for the u-channel diagram, or whether to replace one with this propagator denominator and the other with $s$.  As before, they are equivalent, on shell.  Moreover, as in the previous section, we can use a form for $x$ and $\tilde{x}$ that uses the other photon's helicity spinor as the reference spinor.  In this case, because we have one $x$ and one $\tilde{x}$, it turns out to be especially simple this way. We take $x_{1,24}=-\langle4\lvert p_1\rvert3\rbrack/m_e\langle43\rangle$ and $\tilde{x}_{13,2}=\lbrack3\lvert p_2\rvert4\rangle/m_e\lbrack34\rbrack$.  Their product is then
\begin{equation}
    x_{1,24}\tilde{x}_{13,2}
    =
    -\frac{\langle4\lvert p_1\rvert3\rbrack\lbrack3\lvert p_2\rvert4\rangle}{m_e^2\langle43\rangle\lbrack34\rbrack}
    \ ,
\end{equation}
or, using momentum conservation,
\begin{equation}
    x_{1,24}\tilde{x}_{13,2}
    =
    \frac{\lbrack3\lvert p_1\rvert4\rangle^2}{m_e^2 s}
    \ .
\end{equation}
Of course, on shell, we can replace $s=-(u-m_e^2)$ and find
\begin{equation}
    \mathcal{M}_t^{+-} =
    -\frac{e^2 \lbrack3\lvert p_1\rvert4\rangle^2
    \left(m_e\lbrack\mathbf{12}\rbrack
    +\langle\mathbf{1}\lvert p_3\rvert\mathbf{2}\rbrack
    \right)}
    {m_e^2\left(t-m_e^2\right)\left(u-m_e^2\right)}
    \ .
\end{equation}
We can see from Eq.~(\ref{eq:M_eeA+A-}) that we need one factor of $\lbrack3\lvert p_1\rvert4\rangle$, so we will just work on the rest of the numerator.  We perform a Schouten identity on the first term, $\lbrack3\lvert p_1\rvert4\rangle\lbrack\mathbf{12}\rbrack = \lbrack\mathbf{2}\lvert p_1\rvert4\rangle\lbrack\mathbf{1}3\rbrack - m_e\lbrack\mathbf{2}3\rbrack\langle\mathbf{1}4\rangle$, and on the second term, $\lbrack3\lvert p_1\rvert4\rangle\langle\mathbf{1}\lvert p_3\rvert\mathbf{2}\rbrack = - \lbrack\mathbf{2}3\rbrack\langle\mathbf{1}\lvert p_3p_1\rvert4\rangle$, where we used a mass identity in both.  We next interchange the order of the momenta in the second term to obtain $\lbrack3\lvert p_1\rvert4\rangle\langle\mathbf{1}\lvert p_3\rvert\mathbf{2}\rbrack = - 2p_1\cdot p_3\lbrack\mathbf{2}3\rbrack\langle\mathbf{1}4\rangle + m_e\lbrack\mathbf{2}3\rbrack\lbrack\mathbf{1}\lvert p_3\rvert4\rangle$, again using a mass identity.  The on-shell condition gives us $2p_1\cdot p_3 = (t-m_e^2) = 0$, so $\lbrack3\lvert p_1\rvert4\rangle\langle\mathbf{1}\lvert p_3\rvert\mathbf{2}\rbrack =  m_e\lbrack\mathbf{2}3\rbrack\lbrack\mathbf{1}\lvert p_3\rvert4\rangle$.  We perform another Schouten identity, $\lbrack3\lvert p_1\rvert4\rangle\langle\mathbf{1}\lvert p_3\rvert\mathbf{2}\rbrack =  m_e\langle4\lvert p_3\rvert\mathbf{2}\rbrack\lbrack\mathbf{1}3\rbrack$ and use momentum conservation, $\lbrack3\lvert p_1\rvert4\rangle\langle\mathbf{1}\lvert p_3\rvert\mathbf{2}\rbrack =  -m_e\langle4\lvert p_1\rvert\mathbf{2}\rbrack\lbrack\mathbf{1}3\rbrack+m_e^2\langle4\mathbf{2}\rangle\lbrack\mathbf{1}3\rbrack$, again while using mass identities.  Plugging these in, we have
\begin{equation}
    \mathcal{M}_t^{+-} =
    \frac{e^2 \lbrack3\lvert p_1\rvert4\rangle
    \left(
    \langle\mathbf{2}4\rangle\lbrack\mathbf{1}3\rbrack 
    + \langle\mathbf{1}4\rangle\lbrack\mathbf{2}3\rbrack
    \right)
    }
    {\left(t-m_e^2\right)\left(u-m_e^2\right)}
    \ ,
\end{equation}
agreeing with Eq.~(\ref{eq:M_eeA+A-}).  We have also done this calculation with the u-channel diagram and obtained the same result.  We have also done it with the larger expressions for $x$ and $\tilde{x}$ in Eqs.~(\ref{eq:x_1,24}) and (\ref{eq:x_13,2}) and obtained the same result.

\section{\label{app:analytic continuation}Analytic Continuation and the Spinor Shift}
Following the discovery of greatly simplified formulas for scattering amplitudes in gluodynamics \cite{Gastmans:1990xh,Dixon:1996wi,Parke:1986gb}, the
BCFW recursion relations were developed \cite{Britto:2005fq} allowing any tree-level helicity amplitude of gluons
to be calculated using a simple on-shell recursion relation based on the asymptotic behavior of
the complexified gluon momenta.
Very briefly, as it relates to the present work, the authors of \cite{Britto:2005fq} analytically continued two of the momenta $p_i$ and $p_j$ such that momentum conservation and the on-shell properties of the external particles were maintained.
\begin{eqnarray}
    \hat{p}_i &=& p_i + z q\\
    \hat{p}_j &=& p_j - z q\ ,
\end{eqnarray}
where $z$ is a complex number and $q$ must further satisfy momentum conservation ($\hat{p}_i+\hat{p}_j=p_i+p_j$) and the on-shell property for massless particles $i$ and $j$, namely
\begin{eqnarray}
    p_i\cdot q=p_j\cdot q = q^2 = 0\ .
\end{eqnarray}
This can be achieved, for example, in the center of momentum frame for particles 1 and 2, where
\begin{equation}
      p_1 = (E,0,0,p) \quad,\quad p_2 = (E,0,0,-p)
\end{equation}
and
\begin{equation}
      q = (0,1,i,0)\ .
\end{equation}

They further showed that, for gluon amplitudes, for some choice of $i$ and $j$, the amplitude vanished in the limit $z\to\infty$.  The reason this is important is that meromorphic functions that vanish at infinity are completely determined by
their poles. In particular, if we take a contour integral
of the amplitude divided by $(z^{\prime}-z)$ around the circle at complex infinity, we get zero. But this integral encloses
all its poles, therefore, it can be written as a sum of the residues divided by the poles.  The poles were understood to be Feynman propagator denominators and the residues, which occurred when the propagator denominator went on shell, split into a product of amplitudes on the two sides of the propagator.  This allowed the amplitude to be recursively split into products of smaller amplitudes until the 3-point amplitude was reached.  

In the massive case, there is no such proof and we do not know whether it is possible to build up all amplitudes recursively in the same way.  However, several examples have been done, and been verified with Feynman diagrams, in the all-massive case \cite{Christensen:2019mch} as well in App.~\ref{sec:Massive Photon} of this article.  It appears to us that if all the particles are massive, and no $x$ factors are involved, the amplitudes can be built up in the same way Feynman diagrams are built up, except with spinor vertices.  However, it does not appear to us, in this case, that any fewer diagrams are required compared to Feynman diagrams, even though the final expressions have been simpler than their Feynman counterparts.  On the other hand, in a partly massive theory, such as QED, we show in the present note that some amplitudes can be worked out with the current tools, including the amplitudes with external photons that we present here, and some that still cannot, including the amplitude with an internal photon that we also present here.  Further tools are still required to achieve success for the last.

In order to implement the BCFW recursion relations, in addition to analytically continuing the momenta $p_i$ and $p_j$, we also need to shift the helicity spinors associated with them in a consistent way.   We expect a similar requirement for massive theories.  Therefore, in App.~\ref{app:complexification 2}, we describe the associated shift of the spinors for all the relevant combinations of massless and massive particles, beginning with the all-massless case.  These shifts are then used in App.~\ref{app:Calc AHH Spinor Amps} where we calculate the spinor amplitudes obtained using the recursion-like rules for QED.  
In App.~\ref{app:large z behavior}, we analyze the asymptotic $z$ behavior of the correct amplitudes for each process and gain insight into when and why the present tools work for some amplitudes and not for others.

\subsection{\label{app:complexification 2}Momentum Complexification and the Spin-Spinor Shift}
In the all-massless case, the constructive technique relies on the internal lines being on shell \cite{Britto:2005fq}.  In this subsection, we consider the likely situation that the same is true in the partly-massless case with the hope of resolving the discrepancy between the x-factor amplitudes and those from Feynman diagrams.  
In order to put the internal line on shell while preserving momentum conservation and the on shell property for the external momenta, we must complexify the momenta of at least two external particles.  In this subsection, we attempt to generalize the momentum complexification of the all-massless case.  We note that the shift of the massive spinors has also been considered in \cite{Falkowski:2020aso,Herderschee:2019dmc,Aoude:2019tzn,Franken:2019wqr,Ballav:2020ese,Wu:2021nmq}.

\subsubsection{\label{app:[i,j> massless massless}$\lbrack i,j\rangle$ Shift with Two Massless Particles}
In the all-massless case, this is done through the shift $\lbrack i,j\rangle$ where
\begin{eqnarray}
    \lbrack\hat{i}\rvert &=& \lbrack i\rvert + z \lbrack j\rvert
    \label{eq:ihat massless}\\
    |\hat{j}\rangle &=& |j\rangle - z |i\rangle\ ,
    \label{eq:jhat massless}
\end{eqnarray}
with all other spinors unchanged.
It is easy to see that this preserves momentum conservation since
\begin{eqnarray}
    \hat{p}_i + \hat{p}_j &=& |i\rangle\lbrack\hat{i}| + |\hat{j}\rangle\lbrack j| \nonumber\\
    &=& |i\rangle\lbrack i| + z |i\rangle\lbrack j| + |j\rangle\lbrack j| - z |i\rangle\lbrack j| \nonumber\\
    &=& p_i + p_j\ .
\end{eqnarray}
We can also see that both particles remain on shell as in
\begin{eqnarray}
    2\hat{p}_i^2 &=& \langle ii\rangle\lbrack \hat{i}\hat{i}\rbrack = 0\\
    2\hat{p}_j^2 &=& \langle \hat{j}\hat{j}\rangle\lbrack jj\rbrack = 0\ ,
    \label{eq:app:[i,j> massless i on shell}
\end{eqnarray}
since $\langle ii\rangle=\lbrack jj\rbrack = 0$ for any massless particle.  These two properties are satisfied for any value of $z$, as can be seen in the previous discussion.  However, for a particular value of $z$, which we will call $z_p$, we can also set an internal line on shell.  For example, suppose there are two particles on the left side of the internal line, $i$ and $k$, and all the rest, including $j$, are on the other side.  Now, we consider the momentum squared of the internal line
\begin{eqnarray}
    (\hat{p}_i + p_k)^2 &=& 2\hat{p}_i\cdot p_k \nonumber\\
    &=& \langle i k\rangle\lbrack k\hat{i}\rbrack \nonumber\\
    &=& \langle ik\rangle\lbrack ki\rbrack + z_p\langle ik\rangle\lbrack kj\rbrack \nonumber\\
    &=& 0\ ,
    \label{eq:app:[i,j> massless pihat+pk on shell}
\end{eqnarray}
where the last line comes because we are demanding that the internal line is massless and on shell.  This is satisfied when
\begin{equation}
    z_p 
    = -\frac{\langle ik\rangle\lbrack ki\rbrack}{\langle ik\rangle\lbrack kj\rbrack}
    = -\frac{\lbrack ik\rbrack}{\lbrack jk\rbrack}\ ,
    \label{eq:app:zp=-[ik]/[jk]}
\end{equation}
or, similarly considering the momenta from the other side of the internal ine,
\begin{eqnarray}
    (\hat{p}_j + p_l)^2 &=& 2\hat{p}_j\cdot p_l \nonumber\\
    &=& \langle \hat{j} l\rangle\lbrack lj\rbrack \nonumber\\
    &=& \langle jl\rangle\lbrack lj\rbrack - z_p\langle il\rangle\lbrack lj\rbrack \nonumber\\
    &=& 0
\end{eqnarray}
is satisfied when
\begin{equation}
    z_p 
    = \frac{\langle jl\rangle\lbrack lj\rbrack}{\langle il\rangle\lbrack lj\rbrack}
    = \frac{\langle jl\rangle}{\langle il\rangle}\ .
    \label{eq:z=<jl>/<il> massless}
\end{equation}
We then use this $z_p$ in Eqs.~(\ref{eq:ihat massless}) and (\ref{eq:jhat massless}) in the amplitude to obtain the final form.

Before moving on to more general cases, it is convenient to rewrite $z$ in a notation that is closer to later expressions by combining $\lvert k\rangle\lbrack k\rvert = p_k$.  In that case, Eqs.~(\ref{eq:app:zp=-[ik]/[jk]}) and (\ref{eq:z=<jl>/<il> massless}) become
\begin{eqnarray}
    z_p 
    &= -\frac{\langle i\lvert p_k\rvert i\rbrack}{\langle i\lvert p_k\rvert j\rbrack}
    &= \frac{-s_{ik}}{\langle i\lvert p_k\rvert j\rbrack}
    \label{eq:app:zp=-<i|pk|i]/<i|pk|j]}
    \\
    z_p 
    &= \frac{\langle j\lvert p_l\rvert j\rbrack}{\langle i\lvert p_l\rvert j\rbrack}
    &= \frac{s_{jl}}{\langle i\lvert p_l\rvert j\rbrack}\ ,
    \label{eq:app:zp=<i|pl|i]/<i|pl|j]}
\end{eqnarray}
where we are introducing the shorthand $s_{ij}=(p_i+p_j)^2$ for compactness and clarity.
Moreover, we find convenient during our calculations to have formulas for the shifts that bypass the $z$ by plugging it directly into Eqs.~(\ref{eq:ihat massless}) and (\ref{eq:jhat massless}).  Doing this gives us,
\begin{eqnarray}
    \lbrack\hat{i}\rvert &=& \lbrack i\rvert 
    -\frac{s_{ik}}{\langle i\lvert p_k\rvert j\rbrack} \lbrack j\rvert
    \label{eq:ihat massless bypass z}\\
    \lvert\hat{j}\rangle &=& \lvert j\rangle 
    +\lvert i\rangle\frac{s_{ik}}{\langle i\lvert p_k\rvert j\rbrack} \ .
    \label{eq:jhat massless bypass z}
\end{eqnarray}

In addition to a purely massless theory, we might want to use this shift in a theory such as QED, where some particles are massless and others massive.  We will do several cases in the following subsections.  Before moving to those cases, we begin with the case where both external particles shifted are massless but the internal line is massive.  This is applicable to the process $e\bar{e}\gamma\gamma$ calculated in App.~\ref{app:eeA+A+} and \ref{app:eeA+A-}.  In this case, Eqs.~(\ref{eq:jhat massless}) through (\ref{eq:app:[i,j> massless i on shell}) are all unchanged.  However, Eqs.~(\ref{eq:app:[i,j> massless pihat+pk on shell}) through (\ref{eq:jhat massless bypass z}) become 
\begin{eqnarray}
    (\hat{p}_i + p_k)^2 =& 
    s_{ik} + z_p\langle i\lvert p_k\rvert j\rbrack
    &= M^2
    \nonumber\\
    (\hat{p}_j + p_l)^2 =& 
    s_{jl} - z_p\langle i\lvert p_l\rvert j\rbrack
    &= M^2\ ,
\end{eqnarray}
where $M$ is the mass of the particle on the internal line, giving us
\begin{eqnarray}
    z_p 
    &=& \frac{M^2-s_{ik}}{\langle i\lvert p_k\rvert j\rbrack}
    \label{eq:app:zp=(M^2-<i|pk|i])/<i|pk|j]}
    \\
    z_p 
    &=& \frac{s_{jl}-M^2}{\langle i\lvert p_l\rvert j\rbrack}\ ,
    \label{eq:app:zp=(<i|pl|i]-M^2)/<i|pl|j]}
\end{eqnarray}
and we also allow for the possibility that $m_k$ and $m_l$ are not zero.
Plugging these into Eqs.~(\ref{eq:ihat massless}) and (\ref{eq:jhat massless}), we have
\begin{eqnarray}
    \lbrack\hat{i}\rvert &=& \lbrack i\rvert 
    -\frac{\left(s_{ik}-M^2\right)}{\langle i\lvert p_k\rvert j\rbrack} \lbrack j\rvert
    \label{eq:ihat massless bypass z internal M}\\
    \lvert\hat{j}\rangle &=& \lvert j\rangle 
    +\lvert i\rangle\frac{\left(s_{ik}-M^2\right)}{\langle i\lvert p_k\rvert j\rbrack} \ .
    \label{eq:jhat massless bypass z internal M}
\end{eqnarray}
These are the expressions we will use in App.~\ref{app:eeA+A+} and \ref{app:eeA+A-}.

\subsubsection{\label{app:[i,j> massive massless}$\mathbf{\lbrack\mathbf{i},\mathrm{j}\rangle}$ Shift with One Massive and One Massless Particle}
We must now attempt to find a generalization of this prescription to the case where one of the external particles shifted is massive.  In this subsection, we will do the $\lbrack\mathbf{i},\mathrm{j}\rangle$ shift.  We take particle $i$ to be represented by a massive spin spinor and particle $j$ by a massless helicity spinor.  For an $\lbrack \mathbf{i},j\rangle$ shift, we will try
\begin{eqnarray}
    \lbrack\mathbf{\hat{i}}_\mathrm{I}| &=& \lbrack\mathbf{i}_\mathrm{I}| + z_\mathrm{I} \lbrack j| 
    \label{eq:app:|ihat] def 3}\\
    |\hat{j}\rangle &=& |j\rangle - z_\mathrm{I} |\mathbf{i}^\mathrm{I}\rangle\ ,
    \label{eq:app:|jhat> def 3}
\end{eqnarray}
where the $\mathrm{I}$ is a spin index and appears on the complex $z$ as well as on the spin spinor $\lbrack\mathbf{i}|$.  
We first see that this preserves momentum conservation.
\begin{eqnarray}
    \hat{p}_i+\hat{p}_j &=& |\mathbf{i}^\mathrm{I}\rangle\lbrack\hat{\mathbf{i}}_\mathrm{I}| + |\hat{j}\rangle\lbrack j|\nonumber\\
    &=& |\mathbf{i}^\mathrm{I}\rangle\lbrack\mathbf{i}_\mathrm{I}| + z_\mathrm{I}|\mathbf{i}^I\rangle\lbrack j|
    + |j\rangle\lbrack j| - z_\mathrm{I}|\mathbf{i}^\mathrm{I}\rangle\lbrack j|
    \nonumber\\
    &=& |\mathbf{i}^\mathrm{I}\rangle\lbrack\mathbf{i}_\mathrm{I}| 
    + |j\rangle\lbrack j|
    \nonumber\\
    &=& p_i + p_j\ .
\end{eqnarray}
We next need to satisfy the on-shell condition for particles $i$ and $j$.  We begin with particle $i$ giving us
\begin{eqnarray}
    2\hat{p}_i^2 &=& 
    \langle\mathbf{i}^\mathrm{I}\mathbf{i}^{\mathbf{J}}\rangle\lbrack\hat{\mathbf{i}}_{\mathrm{J}}\hat{\mathbf{i}}_{\mathrm{I}}\rbrack 
    \nonumber\\
    &=& \langle\mathbf{i}^\mathrm{I}\mathbf{i}^{\mathbf{J}}\rangle\lbrack\mathbf{i}_{\mathrm{J}}\mathbf{i}_{\mathrm{I}}\rbrack 
    + 2z_\mathrm{I}\langle\mathbf{i}^\mathrm{I}\mathbf{i}^{\mathbf{J}}\rangle\lbrack\mathbf{i}_{\mathrm{J}}j\rbrack 
    + z_{\mathrm{I}}z_{\mathrm{J}}\langle\mathbf{i}^\mathrm{I}\mathbf{i}^{\mathbf{J}}\rangle\lbrack jj\rbrack
    \nonumber\\
    &=& \langle\mathbf{i}^\mathrm{I}\mathbf{i}^{\mathbf{J}}\rangle\lbrack\mathbf{i}_{\mathrm{J}}\mathbf{i}_{\mathrm{I}}\rbrack = 2p_i^2=2m_i^2\ ,
\end{eqnarray}
where we have used $z_{\mathrm{J}}\langle\mathbf{i}^\mathrm{I}\mathbf{i}^{\mathbf{J}}\rangle\lbrack j\mathbf{i}_{\mathrm{I}}\rbrack = z_{\mathrm{I}}\langle\mathbf{i}^\mathrm{J}\mathbf{i}^{\mathbf{I}}\rangle\lbrack j\mathbf{i}_{\mathrm{J}}\rbrack = z_{\mathrm{I}}\langle\mathbf{i}^{\mathbf{I}}\mathbf{i}^\mathrm{J}\rangle\lbrack \mathbf{i}_{\mathrm{J}}j\rbrack$ in the second line and the third line is what we must obtain for the on-shell condition to be preserved.
We remember that $\lbrack jj\rbrack=0$ and we use the mass identies  $\langle\mathbf{i}^\mathrm{I}\mathbf{i}^{\mathbf{J}}\rangle=-m_i\epsilon^{\mathrm{IJ}}$ and $\lbrack\mathbf{i}_{\mathrm{L}}\mathbf{i}_{\mathrm{K}}\rbrack=-m_i\epsilon_{\mathrm{LK}}$ given in \cite{Christensen:2018zcq}.  Plugging these in, we must have
\begin{equation}
    z_\mathrm{I}\lbrack\mathbf{i}^{\mathrm{I}}j\rbrack  = 0\ .
    \label{2z+zzepseps 3}
\end{equation}
We must now perform the same calculation for particle $j$ 
\begin{eqnarray}
    2\hat{p}_j^2 &=& 
    \langle\hat{j}\hat{j}\rangle\lbrack jj\rbrack 
    = 0\ ,
\end{eqnarray}
which is trivially satisfied because particle $j$ is massless and $\lbrack jj\rbrack=0$.

We have one further on-shell constraint, that of the internal line.  For this, we assume, once again, that the internal line has momentum $(p_i+p_k)=-(p_j+p_l)$, where particle $i$ and $j$ are on opposite ends of the propagator.  $p_k$ and $p_l$ stand for the rest of the momenta on each side.  
\begin{eqnarray}
    (\hat{p}_i + p_k)^2 
    &=& m_i^2 + m_k^2 + 2\hat{p}_i\cdot p_k \nonumber\\
    &=& m_i^2 + m_k^2 + \langle \mathbf{i}^{\mathrm{I}} \lvert p_k\rvert\hat{\mathbf{i}}_{\mathrm{I}}\rbrack \nonumber\\
    &=& m_i^2 + m_k^2 + \langle \mathbf{i}^{\mathrm{I}} \lvert p_k\rvert\mathbf{i}_{\mathrm{I}}\rbrack + z_{\mathrm{I}}\langle \mathbf{i}^{\mathrm{I}} \lvert p_k\rvert j\rbrack \nonumber\\
    &=& M^2\ ,
\end{eqnarray}
where $M^2$ stands for the mass of the internal-line particle.  We must now solve for $z_{\mathrm{I}}$.  We begin by placing the $z$ on one side of the equation
\begin{equation}
    z_{\mathrm{I}}\langle \mathbf{i}^{\mathrm{I}} \lvert p_k\rvert j\rbrack =
    M^2 - s_{ik} \ .
    \label{eq:app:z<>[]=M2-mi2-... 3}
\end{equation}
We also consider the same line from the other direction.  We have
\begin{eqnarray}
    (\hat{p}_j + p_l)^2 
    &=& m_l^2 + 2\hat{p}_j\cdot p_l \nonumber\\
    &=& m_l^2 + \langle \hat{j} \lvert p_l\rvert j\rbrack \nonumber\\
    &=& m_l^2 + \langle j \lvert p_l\rvert j\rbrack - z_{\mathrm{I}}\langle \mathbf{i}^{\mathrm{I}} \lvert p_l\rvert j\rbrack \nonumber\\
    &=& M^2\ ,
\end{eqnarray}
giving us
\begin{equation}
    z_{\mathrm{I}}\langle \mathbf{i}^{\mathrm{I}} \lvert p_l\rvert j\rbrack =
    -M^2 + s_{jl} \ .
    \label{eq:app:z<>[]=-M2+mj2+... 3}
\end{equation}
We must now solve the system of two equations, Eqs.~(\ref{2z+zzepseps 3}) and either (\ref{eq:app:z<>[]=M2-mi2-... 3}) or (\ref{eq:app:z<>[]=-M2+mj2+... 3}).   Eqs. (\ref{eq:app:z<>[]=M2-mi2-... 3}) and (\ref{eq:app:z<>[]=-M2+mj2+... 3}) should be solvable by considering $A^{I}T_{I} = B$ and noting that the solution is $A^{I}=B C^{I}/(C^{K}T_{K})$ for some $C^{I}$.  Let's begin by solving Eq.~(\ref{eq:app:z<>[]=M2-mi2-... 3}).  It is not hard to see that
\begin{equation}
    z_I = \frac{\lbrack j\mathbf{i}_I\rbrack}{
    \lbrack j\lvert p_i  p_k\rvert j\rbrack}
    \left(M^2 - s_{ik}\right),
    \label{eq:app:z_I= 1}
\end{equation}
which is analogous to Eq.~(\ref{eq:app:zp=-[ik]/[jk]}).  We next check Eq.~(\ref{2z+zzepseps 3}), and find that it works since
\begin{eqnarray}
    \lbrack j\mathbf{i}_I\rbrack\lbrack\mathbf{i}^{\mathrm{I}}j\rbrack 
    = m_i\lbrack jj\rbrack 
    = 0\ .
\end{eqnarray}
We note that this has the right limit when the masses go to zero.  Setting all the masses to zero in Eq.~(\ref{eq:app:z_I= 1}) gives
\begin{equation}
    z_{\mathrm{I}} \to \frac{\lbrack ji\rbrack}{
    \lbrack ji\rbrack\langle i k\rangle\lbrack k j\rbrack}
    \left( - \langle i k\rangle\lbrack k i\rbrack\right)
    = -\frac{\lbrack ki\rbrack}{\lbrack kj\rbrack}\ ,
\end{equation}
which agrees with Eq.~(\ref{eq:app:zp=-[ik]/[jk]}).  For completeness, let's also solve Eq.~(\ref{eq:app:z<>[]=-M2+mj2+... 3}), obtaining
\begin{equation}
    z_\mathrm{I} = \frac{\lbrack j\mathbf{i}_\mathrm{I}\rbrack}{\lbrack j\lvert p_i p_l\rvert j\rbrack}
    \left(-M^2 + s_{jl}\right)\ .
    \label{eq:z_I = [ji_I]/[j|pipl|j](s-M^2)}
\end{equation}
This formulas is equivalent to Eq.~(\ref{eq:app:z_I= 1}) because $s_{jl}=s_{ik}$ and using momentum conservation $p_l=-p_i-p_j-p_k$ and the masslessness of particle $j$ turns one into the other.
We also note that Eq.~(\ref{2z+zzepseps 3}) is satisfied for the same reason as before ($\lbrack j\mathbf{i}_I\rbrack\lbrack\mathbf{i}^{\mathrm{I}}j\rbrack 
    = m_i\lbrack jj\rbrack 
    = 0$).  In the massless limit, this reduces to
\begin{equation}
    z_{\mathrm{I}}\to
    \frac{\lbrack ji\rbrack}{\lbrack ji\rbrack\langle i l\rangle\lbrack lj\rbrack}
    \left(\langle j l\rangle\lbrack lj\rbrack\right)
    = \frac{\langle jl\rangle}{\langle il\rangle}\ ,
\end{equation}
which exactly agrees with Eq.~(\ref{eq:z=<jl>/<il> massless}).

It will be convenient to plug our $z_{\mathrm{I}}$ into Eqs.~(\ref{eq:app:|ihat] def 3}) and (\ref{eq:app:|jhat> def 3}) to obtain implicit-spin-index shifts that bypass the intermediate $z_I$.  We obtain
\begin{eqnarray}
    \lbrack\mathbf{\hat{i}}| &=& \lbrack\mathbf{i}| + \frac{\lbrack \mathbf{i} j\rbrack\lbrack j\rvert}{
    \lbrack j\lvert p_i  p_k\rvert j\rbrack}
    \left(s_{ik}-M^2\right)  
    \label{eq:app:|ihat] def 3 index free}\\
    |\hat{j}\rangle &=& |j\rangle - \frac{p_i\lvert j\rbrack}{
    \lbrack j\lvert p_i  p_k\rvert j\rbrack}
    \left(s_{ik}-M^2\right) \ .
    \label{eq:app:|jhat> def 3 index free}
\end{eqnarray}

\subsubsection{\label{app:[i,j> massless massive}$\mathbf{\lbrack\mathrm{i},\mathbf{j}\rangle}$ Shift with One Massless and One Massive Particle}
In the last subsection, we shifted massive $\lbrack\mathbf{i}_{\mathrm{I}}\rvert$ and massless $\lvert j\rangle$.  In this subsection, we shift massless $\lbrack\mathrm{i}\rvert$ and massive $\lvert\mathbf{j}^{\mathrm{J}}\rangle$.  Since it is directly analogous to the last subsection, we will be much more terse in this subsection.  In the next subsection, we will consider the case where both particles shifted are massive.   For an $\lbrack \mathrm{i},\mathbf{j}\rangle$ shift, we will try
\begin{eqnarray}
    \lbrack\hat{i}| &=& \lbrack\mathrm{i}| + z^\mathrm{J} \lbrack \mathbf{j}_J| 
    \label{eq:[ihat|=[i|+z^J[j_J|}
    \\
    |\mathbf{\hat{j}}^{\mathrm{J}}\rangle &=& |\mathbf{j}^{\mathrm{J}}\rangle - z^\mathrm{J} \lvert i\rangle\ .
    \label{eq:|jhat^J>=|j^J>+z^J|i>}
\end{eqnarray}  
This preserves momentum conservation.
\begin{eqnarray}
    \hat{p}_i+\hat{p}_j &=& |i\rangle\lbrack\hat{i}| + |\hat{\mathbf{j}}^J\rangle\lbrack \mathbf{j}_J|\nonumber\\
    &=& |i\rangle\lbrack i| + z^\mathrm{J}|i\rangle\lbrack \mathbf{j}_J|
    + |\mathbf{j}^J\rangle\lbrack \mathbf{j}_J| - z^\mathrm{J}|i\rangle\lbrack \mathbf{j}_J|
    \nonumber\\
    &=& |i\rangle\lbrack i| +  |\mathbf{j}^J\rangle\lbrack \mathbf{j}_J| 
    \nonumber\\
    &=& p_i + p_j\ .
    \label{eq:[i,j> massive: pihat+pjhat=pi+pj}
\end{eqnarray}
The on-shell condition for particle $j$ is
\begin{eqnarray}
    2\hat{p}_j^2 &=& 
    \langle\hat{\mathbf{j}}^\mathrm{I}\hat{\mathbf{j}}^{\mathbf{J}}\rangle\lbrack\mathbf{j}_{\mathrm{J}}\mathbf{j}_{\mathrm{I}}\rbrack 
    \nonumber\\
    &=& \langle\mathbf{j}^\mathrm{I}\mathbf{j}^{\mathbf{J}}\rangle\lbrack\mathbf{j}_{\mathrm{J}}\mathbf{j}_{\mathrm{I}}\rbrack 
    - 2z^\mathrm{J}\langle\mathbf{j}^\mathrm{I}i\rangle\lbrack\mathbf{j}_{\mathrm{J}}\mathbf{j}_{\mathrm{I}}\rbrack 
    + z^{\mathrm{I}}z^{\mathrm{J}}\langle ii\rangle\lbrack \mathbf{j}_J\mathbf{j}_I\rbrack
    \nonumber\\
    &=& \langle\mathbf{j}^\mathrm{I}\mathbf{j}^{\mathbf{J}}\rangle\lbrack\mathbf{j}_{\mathrm{J}}\mathbf{j}_{\mathrm{I}}\rbrack = 2p_j^2=2m_j^2\ .
\end{eqnarray}
We use $\langle ii\rangle=0$ and the mass identies to get the condition
\begin{equation}
    z^\mathrm{J}\langle\mathbf{j}_{\mathrm{J}}i\rangle  = 0\ .
    \label{eq:z^J<j_Ji>=0}
\end{equation}
The on-shell condition for particle $i$, on the other hand, 
\begin{eqnarray}
    2\hat{p}_i^2 &=& 
    \langle ii\rangle\lbrack \hat{i}\hat{i}\rbrack 
    = 0\ ,
\end{eqnarray}
is trivially satisfied since particle $i$ is massless.

The on-shell condition for the internal line with momentum  $(p_i+p_k)=-(p_j+p_l)$, where particle $i$ and $j$ are on opposite ends of the propagator and $p_k$ and $p_l$ stand for the rest of the momenta on each side, is
\begin{eqnarray}
    (\hat{p}_i + p_k)^2 
    &=& m_i^2 + m_k^2 + 2\hat{p}_i\cdot p_k \nonumber\\
    &=& m_i^2 + m_k^2 + \langle i \lvert p_k\rvert\hat{i}\rbrack \nonumber\\
    &=& m_i^2 + m_k^2 + \langle i \lvert p_k\rvert i\rbrack + z^{\mathrm{J}}\langle i \lvert p_k\rvert \mathbf{j}_{\mathrm{J}}\rbrack \nonumber\\
    &=& M^2\ ,
\end{eqnarray}
where $M^2$ stands for the mass of the internal-line particle.  This gives us
\begin{equation}
    z^{\mathrm{J}}\langle i \lvert p_k\rvert \mathbf{j}_{\mathrm{J}}\rbrack =
    M^2 - s_{ik} \ .
    \label{eq:z^J<i|pk|j]=M2-sik}
\end{equation}
From the other direction,
\begin{eqnarray}
    (\hat{p}_j + p_l)^2 
    &=& m_l^2 + 2\hat{p}_j\cdot p_l \nonumber\\
    &=& m_l^2 + \langle \hat{\mathbf{j}}^{\mathrm{J}} \lvert p_l\rvert \mathbf{j}_{\mathrm{J}}\rbrack \nonumber\\
    &=& m_l^2 + \langle \mathbf{j}^{\mathrm{J}} \lvert p_l\rvert \mathbf{j}_{\mathrm{J}}\rbrack - z^{\mathrm{J}}\langle i \lvert p_l\rvert \mathbf{j}_{\mathrm{J}}\rbrack \nonumber\\
    &=& M^2\ ,
\end{eqnarray}
giving us
\begin{equation}
    z^{\mathrm{J}}\langle i \lvert p_l\rvert \mathbf{j}_{\mathrm{J}}\rbrack =
    -M^2 + s_{jl} \ .
    \label{eq:z^J<i|pl|j_J=-M2+sjl}
\end{equation}
Solving Eq.~(\ref{eq:z^J<i|pk|j]=M2-sik}) gives,
\begin{eqnarray}
    z^{\mathrm{J}} &=& \frac{\langle \mathbf{j}^{\mathrm{J}}i\rangle}{
    \langle i \lvert p_kp_j\rvert i\rangle}
    \left(M^2 - s_{ik}\right)\ .
    \nonumber\\
    \label{eq:z^J = <j^Ji>/...(M2-sik)}
\end{eqnarray}
Checking Eq.~(\ref{eq:z^J<j_Ji>=0}), 
\begin{eqnarray}
    \langle \mathbf{j}^{\mathrm{J}}i\rangle\langle\mathbf{j}_{\mathrm{J}}i\rangle
    = -m_j\langle ii\rangle
    = 0\ .
\end{eqnarray}
Eq.~(\ref{eq:z^J = <j^Ji>/...(M2-sik)}) has the right massless limit,
\begin{eqnarray}
    z^{\mathrm{J}} &\to& -\frac{\langle ji\rangle\langle ik\rangle\lbrack ki\rbrack}{
    \langle ik\rangle\lbrack k j\rbrack\langle ji\rangle}
     = -\frac{\lbrack ki\rbrack}{\lbrack kj\rbrack}\ ,
\end{eqnarray}
in agreement with Eq.~(\ref{eq:app:zp=-[ik]/[jk]}).  Solving  Eq.~(\ref{eq:z^J<i|pl|j_J=-M2+sjl}), gives
\begin{equation}
    z^\mathrm{J} = \frac{\langle \mathbf{j}^{\mathrm{J}}i\rangle}{
    \langle i \lvert p_lp_j\rvert i\rangle}
    \left(-M^2 + s_{jl}\right)\ .
\end{equation}
Once again, this is equivalent to Eq.~(\ref{eq:z^J = <j^Ji>/...(M2-sik)}) using momentum conservation by replacing $p_l=-p_i-p_j-p_k$ due to the masslessness of particle $i$.
This satisfies Eq.~(\ref{eq:z^J<j_Ji>=0}) in the same way as before ($\langle \mathbf{j}^{\mathrm{J}}i\rangle\langle\mathbf{j}_{\mathrm{J}}i\rangle
    = -m_j\langle ii\rangle
    = 0$).  In the massless limit, 
\begin{equation}
    z^\mathrm{J} \to \frac{\langle ji\rangle\langle jl\rangle\lbrack lj\rbrack}{
    \langle i l\rangle\lbrack l j\rbrack\langle ji\rangle}
    = \frac{\langle jl\rangle}{\langle il\rangle}\ ,
\end{equation}
agreeing with Eq.~(\ref{eq:z=<jl>/<il> massless}).

We also create formulas that bypass $z^{\mathrm{J}}$ and have implicit spin indices, as we did in the previous subsection, by plugging our $z^{\mathrm{J}}$ into Eqs.~(\ref{eq:[ihat|=[i|+z^J[j_J|}) and (\ref{eq:|jhat^J>=|j^J>+z^J|i>}), giving us,
\begin{eqnarray}
    \lbrack\hat{i}| &=& \lbrack\mathrm{i}| - \frac{\langle i \rvert p_j}{
    \langle i \lvert p_jp_k\rvert i\rangle}
    \left(s_{ik}-M^2\right) 
    \label{eq:[ihat|=[i|+z^J[j_J| spin-index free}
    \\
    |\mathbf{\hat{j}}\rangle &=& |\mathbf{j}\rangle + \frac{\lvert i\rangle\langle i \mathbf{j}\rangle}{
    \langle i \lvert p_jp_k\rvert i\rangle}
    \left(s_{ik}-M^2\right) \ .
    \label{eq:|jhat^J>=|j^J>+z^J|i> spin-index free}
\end{eqnarray}

\subsubsection{\label{app:[i,j> both massive}$\mathbf{\lbrack\mathbf{i},\mathbf{j}\rangle}$ Shift with Two Massive Spinors and a Massless Internal Line}
In this case, we take both particle $i$ and $j$ to be massive and of different masses (although they can be set equal at the end if appropriate) and with a massless internal line.  We have not seen any evidence that we need a momentum shift in the all-massive case (where all external and all internal lines are massive).  On the other hand, we do have diagrams where both shifted particles are massive but either the internal line is massless or one of the other external particles (that is unshifted) are massless.  We will begin with the case where the internal line is massless, which is potentially appropriate for an internal photon (or other massless particle) that connects fermions on its ends.  We will consider the case where the internal line is massive and one of the external particles is massless in the next subsection.  For an $\lbrack \mathbf{i},\mathbf{j}\rangle$ shift, we will try
\begin{eqnarray}
    \lbrack\mathbf{\hat{i}}_\mathrm{I}| &=& \lbrack\mathbf{i}_\mathrm{I}| + z^{\ \mathrm{J}}_\mathrm{I} \lbrack\mathbf{j}_\mathrm{J}| 
    \label{eq:app:|ihat] def 2}\\
    |\mathbf{\hat{j}}^\mathrm{J}\rangle &=& |\mathbf{j}^\mathrm{J}\rangle - z^{\ \mathrm{J}}_\mathrm{I} |\mathbf{i}^\mathrm{I}\rangle\ .
    \label{eq:app:|jhat> def 2}
\end{eqnarray}
We first see that this preserves momentum conservation.
\begin{eqnarray}
    \hat{p}_i+\hat{p}_j &=& |\mathbf{i}^\mathrm{I}\rangle\lbrack\hat{\mathbf{i}}_\mathrm{I}| + |\hat{\mathbf{j}}^\mathrm{I}\rangle\lbrack\mathbf{j}_\mathrm{I}|\nonumber\\
    &=& |\mathbf{i}^\mathrm{I}\rangle\lbrack\mathbf{i}_\mathrm{I}| + z_\mathrm{I}^{\ \mathrm{J}}|\mathbf{i}^I\rangle\lbrack\mathbf{j}_\mathrm{J}|
    + |\mathbf{j}^\mathrm{I}\rangle\lbrack\mathbf{j}_\mathrm{I}| - z^{\ \mathrm{I}}_\mathrm{J}|\mathbf{i}^\mathrm{J}\rangle\lbrack\mathbf{j}_\mathrm{I}|
    \nonumber\\
    &=& |\mathbf{i}^\mathrm{I}\rangle\lbrack\mathbf{i}_\mathrm{I}| 
    + |\mathbf{j}^\mathrm{I}\rangle\lbrack\mathbf{j}_\mathrm{I}|
    \nonumber\\
    &=& p_i + p_j\ .
\end{eqnarray}
We next need to satisfy the on-shell condition for particles $i$ and $j$.  We begin with particle $i$ giving us
\begin{eqnarray}
    2\hat{p}_i^2 &=& 
    \langle\mathbf{i}^\mathrm{I}\mathbf{i}^{\mathbf{J}}\rangle\lbrack\hat{\mathbf{i}}_{\mathrm{J}}\hat{\mathbf{i}}_{\mathrm{I}}\rbrack 
    \nonumber\\
    &=& \langle\mathbf{i}^\mathrm{I}\mathbf{i}^{\mathbf{J}}\rangle\lbrack\mathbf{i}_{\mathrm{J}}\mathbf{i}_{\mathrm{I}}\rbrack 
    + 2z_\mathrm{I}^{\ \mathrm{K}}\langle\mathbf{i}^\mathrm{I}\mathbf{i}^{\mathbf{J}}\rangle\lbrack\mathbf{i}_{\mathrm{J}}\mathbf{j}_{\mathrm{K}}\rbrack 
    + z_{\mathrm{I}}^{\ \mathrm{K}}z_{\mathrm{J}}^{\ \mathrm{L}}\langle\mathbf{i}^\mathrm{I}\mathbf{i}^{\mathbf{J}}\rangle\lbrack\mathbf{j}_{\mathrm{L}}\mathbf{j}_{\mathrm{K}}\rbrack
    \nonumber\\
    &=& \langle\mathbf{i}^\mathrm{I}\mathbf{i}^{\mathbf{J}}\rangle\lbrack\mathbf{i}_{\mathrm{J}}\mathbf{i}_{\mathrm{I}}\rbrack = 2p_i^2=2m_i^2\ .
\end{eqnarray}
However, we can now use the mass identities  $\langle\mathbf{i}^\mathrm{I}\mathbf{i}^{\mathrm{J}}\rangle=-m_i\epsilon^{\mathrm{IJ}}$ and $\lbrack\mathbf{i}_{\mathrm{L}}\mathbf{i}_{\mathrm{K}}\rbrack=-m_i\epsilon_{\mathrm{LK}}$.  Plugging these in, we must have
\begin{equation}
    2z_\mathrm{I}^{\ \mathrm{K}}m_i\lbrack\mathbf{i}^{\mathrm{I}}\mathbf{j}_{\mathrm{K}}\rbrack 
    + z_{\mathrm{I}}^{\ \mathrm{K}}z_{\mathrm{J}}^{\ \mathrm{L}}m_im_j\epsilon^{IJ}\epsilon_{LK} = 0\ ,
    \label{2z+zzepseps}
\end{equation}
or
\begin{equation}
    z_\mathrm{I}^{\ \mathrm{J}}\left(
    2\lbrack\mathbf{i}^{\mathrm{I}}\mathbf{j}_{\mathrm{J}}\rbrack 
    - z_{\ \mathrm{J}}^{\mathrm{I}}m_j
    \right)= 0\ .
\end{equation}
We must now perform the same calculation for particle $j$ which gives
\begin{eqnarray}
    2\hat{p}_j^2 &=& 
    \langle\hat{\mathbf{j}}^\mathrm{I}\hat{\mathbf{j}}^{\mathbf{J}}\rangle\lbrack\mathbf{j}_{\mathrm{J}}\mathbf{j}_{\mathrm{I}}\rbrack 
    \nonumber\\
    &=& \langle\mathbf{j}^\mathrm{I}\mathbf{j}^{\mathbf{J}}\rangle\lbrack\mathbf{j}_{\mathrm{J}}\mathbf{j}_{\mathrm{I}}\rbrack 
    - 2z^{\ \mathrm{J}}_{\mathrm{K}}\langle\mathbf{j}^\mathrm{I}\mathbf{i}^{\mathbf{K}}\rangle\lbrack\mathbf{j}_{\mathrm{J}}\mathbf{j}_{\mathrm{I}}\rbrack 
    + z^{\ \mathrm{I}}_{\mathrm{K}}z^{\ \mathrm{J}}_{\mathrm{L}}\langle\mathbf{i}^\mathrm{K}\mathbf{i}^{\mathbf{L}}\rangle\lbrack\mathbf{j}_{\mathrm{J}}\mathbf{j}_{\mathrm{I}}\rbrack
    \nonumber\\
    &=& \langle\mathbf{j}^\mathrm{I}\mathbf{j}^{\mathbf{J}}\rangle\lbrack\mathbf{j}_{\mathrm{J}}\mathbf{j}_{\mathrm{I}}\rbrack = 2p_j^2 = 2m_j^2\ ,
\end{eqnarray}
giving us the relation
\begin{equation}
    2z^{\ \mathrm{K}}_{\mathrm{I}}m_j\langle\mathbf{i}^{\mathbf{I}}\mathbf{j}_\mathrm{K}\rangle
    + z^{\ \mathrm{K}}_{\mathrm{I}}z^{\ \mathrm{L}}_{\mathrm{J}}m_im_j\epsilon^{IJ}\epsilon_{LK}
    = 0\ ,
    \label{2z+zzepseps 2}
\end{equation}
or
\begin{equation}
    z^{\ \mathrm{J}}_{\mathrm{I}}\left(
    2\langle\mathbf{i}^{\mathbf{I}}\mathbf{j}_\mathrm{J}\rangle
    - z^{\mathrm{I}}_{\ \mathrm{J}}m_i
    \right) = 0\ .
\end{equation}
Subtracting them, we find
\begin{equation}
    z_\mathrm{I}^{\ \mathrm{J}}
    \left(m_i\lbrack\mathbf{i}^{\mathrm{I}}\mathbf{j}_{\mathrm{J}}\rbrack 
    -
    m_j\langle\mathbf{i}^{\mathbf{I}}\mathbf{j}_\mathrm{J}\rangle\right)
    = 0\ ,
    \label{z(m[]-m<>)=0:eqD67}
\end{equation}
which is linear in $z$.  We must satisfy this as well as either Eq.~(\ref{2z+zzepseps}) or (\ref{2z+zzepseps 2}).

We have one further on-shell constraint, that of the internal line.  Although we have considered the case where the internal line is massive, we will only be required to complexify two massive particle momenta when we have a massless internal line.  Moreover, the solution with a massless internal line is significantly simpler.  Therefore, we assume a massless internal line in this case.  We assume, once again, that the internal line has momentum $(p_i+p_k)=-(p_j+p_l)$, where particle $i$ and $j$ are on opposite ends of the propagator.  $p_k$ and $p_l$ stand for the rest of the momenta on each side.  
\begin{eqnarray}
    (\hat{p}_i + p_k)^2 
    &=& m_i^2 + m_k^2 + 2\hat{p}_i\cdot p_k \nonumber\\
    &=& m_i^2 + m_k^2 + \langle \mathbf{i}^{\mathrm{I}} \lvert p_k\rvert\hat{\mathbf{i}}_{\mathrm{I}}\rbrack \nonumber\\
    &=& m_i^2 + m_k^2 + \langle \mathbf{i}^{\mathrm{I}} \lvert p_k\rvert\mathbf{i}_{\mathrm{I}}\rbrack + z_{\mathrm{I}}^{\ \mathrm{J}}\langle \mathbf{i}^{\mathrm{I}} \lvert p_k\rvert\mathbf{j}_{\mathrm{J}}\rbrack \nonumber\\
    &=& 0\ ,
\end{eqnarray}
where $m_k$ is the invariant mass for $p_k$, if it represents more than one particle.  We must now solve for $z_{\mathrm{I}}^{\ \mathrm{J}}$.  We begin by placing the $z$ on one side of the equation
\begin{equation}
    z_{\mathrm{I}}^{\ \mathrm{J}}\langle \mathbf{i}^{\mathrm{I}} \lvert p_k\rvert\mathbf{j}_{\mathrm{J}}\rbrack =
     -s_{ik}\ ,
    \label{eq:app:z<>[]=M2-mi2-...}
\end{equation}
We also consider the same line from the other direction.  We have
\begin{eqnarray}
    (\hat{p}_j + p_l)^2 
    &=& m_j^2 + m_l^2 + 2\hat{p}_j\cdot p_l \nonumber\\
    &=& m_j^2 + m_l^2 + \langle \hat{\mathbf{j}}^{\mathrm{J}} \lvert p_l\rvert\mathbf{j}_{\mathrm{J}}\rbrack \nonumber\\
    &=& m_j^2 + m_l^2 + \langle \mathbf{j}^{\mathrm{J}} \lvert p_l\rvert\mathbf{j}_{\mathrm{J}}\rbrack - z_{\mathrm{I}}^{\ \mathrm{J}}\langle \mathbf{i}^{\mathrm{I}} \lvert p_l\rvert\mathbf{j}_{\mathrm{J}}\rbrack \nonumber\\
    &=& 0\ ,
\end{eqnarray}
giving us
\begin{equation}
    z_{\mathrm{I}}^{\ \mathrm{J}}\langle \mathbf{i}^{\mathrm{I}} \lvert p_l\rvert\mathbf{j}_{\mathrm{J}}\rbrack =
    s_{jl} \ .
    \label{eq:app:z<>[]=-M2+mj2+...}
\end{equation}
We must now solve the system of three equations, Eqs.~(\ref{2z+zzepseps 2}), (\ref{z(m[]-m<>)=0:eqD67}) and (\ref{eq:app:z<>[]=M2-mi2-...}) or (\ref{eq:app:z<>[]=-M2+mj2+...}).   Eqs. (\ref{eq:app:z<>[]=M2-mi2-...}) and (\ref{eq:app:z<>[]=-M2+mj2+...}) should be solvable by considering $A^{IJ}T_{IJ} = B$ and noting that the solution is $A^{IJ}=B C^{IJ}/(C^{KL}T_{KL})$ for some $C^{IJ}$.  Then, we are just left with Eqs.~(\ref{2z+zzepseps 2}) and (\ref{z(m[]-m<>)=0:eqD67}).  Let's begin by solving Eq.~(\ref{eq:app:z<>[]=M2-mi2-...}), obtaining
\begin{eqnarray}
    z_I^{\ J} &=& -\frac{C_I^{\ J}s_{ik}}{C_K^{\ L}\langle \mathbf{i}^{\mathrm{K}} \lvert p_k\rvert\mathbf{j}_{\mathrm{L}}\rbrack}\ ,
    \label{eq:z_I^J Intermediate}
\end{eqnarray}
for any $C_I^{\ J}$.  We next apply Eqs.~(\ref{z(m[]-m<>)=0:eqD67}) and (\ref{2z+zzepseps 2}) giving us
\begin{eqnarray}
    m_iC_I^{\ J}
    \lbrack\mathbf{i}^{\mathrm{I}}\mathbf{j}_{\mathrm{J}}\rbrack 
    -
    m_jC_I^{\ J}\langle\mathbf{i}^{\mathbf{I}}\mathbf{j}_\mathrm{J}\rangle 
    = 0
    \label{eq:CIJ([ij]-<ij>)=0}
\end{eqnarray}
and
\begin{equation}
    2C^{\ \mathrm{J}}_{\mathrm{I}}\langle\mathbf{i}^{\mathbf{I}}\mathbf{j}_\mathrm{J}\rangle C_K^{\ L}\langle \mathbf{i}^{\mathrm{K}} \lvert p_k\rvert\mathbf{j}_{\mathrm{L}}\rbrack
    + C^{\ \mathrm{K}}_{\mathrm{I}}C^{\mathrm{I}}_{\ \mathrm{K}}m_is_{ik}
    = 0
    \label{eq:CIJ<ij>CKL<ipkj]+CIKCIK=0}
    \ .
\end{equation}
The indices on $C_I^{\ J}$ must come from the spinors for particle $i$ and $j$, therefore it must be a linear combination of $\langle \mathbf{i}_{\mathrm{I}}\mathbf{j}^{\mathrm{J}}\rangle, \lbrack \mathbf{i}_{\mathrm{I}}\mathbf{j}^{\mathrm{J}} \rbrack, \langle \mathbf{i}_{\mathrm{I}}\lvert p_k\rvert \mathbf{j}^{\mathrm{J}}\rbrack$ and $\lbrack \mathbf{i}_{\mathrm{I}}\lvert p_k\rvert \mathbf{j}^{\mathrm{J}}\rangle$.  It turns out that the only solution that requires only two of these is
\begin{equation}
    C_I^{\ J} = \lbrack \mathbf{i}_{\mathrm{I}}\mathbf{j}^{\mathrm{J}} \rbrack + \frac{1}{m_i}\langle \mathbf{i}_{\mathrm{I}}\lvert p_k\rvert \mathbf{j}^{\mathrm{J}}\rbrack\ .
\end{equation}
If we keep all four possible terms, the solution is quite complicated.
We begin by showing Eq.~(\ref{eq:CIJ([ij]-<ij>)=0}) is satisfied.  The left-hand side is
\begin{eqnarray}
    -2m_j\left(m_i^2+p_i\cdot p_j\right)
    -2m_j\left(p_i\cdot p_k+p_j\cdot p_k\right)\ ,
    \nonumber
\end{eqnarray}
but, $p_i\cdot p_j=-p_i^2-p_i\cdot p_k-p_i\cdot p_l$ and $p_j\cdot p_k=p_i\cdot p_l$, by momentum conservation, where we are also assuming $m_i=m_k$ and $m_j=m_l$.  This is true for the four-point amplitudes we are considering here, but will need to be revisited for more complex higher-point amplitudes.  The left-hand side now equals
\begin{eqnarray}
    2m_j\left(p_i\cdot p_k+p_i\cdot p_j\right)
    -2m_j\left(p_i\cdot p_k+p_i\cdot p_l\right) = 0\ .
    \nonumber
\end{eqnarray}
For Eq.~(\ref{eq:CIJ<ij>CKL<ipkj]+CIKCIK=0}), let's look at each piece.  
\begin{eqnarray}
    C^{\ \mathrm{J}}_{\mathrm{I}}\langle\mathbf{i}^{\mathbf{I}}\mathbf{j}_{\mathbf{J}}\rangle &=& 
    2p_i\cdot p_j+2p_j\cdot p_k 
    \nonumber\\
    2C_K^{\ L}\langle \mathbf{i}^{\mathrm{K}} \lvert p_k\rvert\mathbf{j}_{\mathrm{L}}\rbrack &=&
    -4m_i^2m_j-4m_jp_i\cdot p_k
    \nonumber\\
    C^{\ \mathrm{K}}_{\mathrm{I}}C^{\mathrm{I}}_{\ \mathrm{K}}m_i &=&
    -4m_i^2m_j-4m_jp_i\cdot p_k
    \nonumber\\
    s_{ik} &=& 2m_i^2+2p_i\cdot p_k\ .
    \nonumber
\end{eqnarray}
However, focusing on the top line, once again we use $p_i\cdot p_j=-p_i^2-p_i\cdot p_k-p_i\cdot p_l$ and $p_j\cdot p_k=p_i\cdot p_l$ to obtain
\begin{eqnarray}
    C^{\ \mathrm{J}}_{\mathrm{I}}\langle\mathbf{i}^{\mathbf{I}}\mathbf{j}_{\mathbf{J}}\rangle &=& -2m_i^2-2p_i\cdot p_k\ ,
\end{eqnarray}
and we see that all the terms in Eq.~(\ref{eq:CIJ<ij>CKL<ipkj]+CIKCIK=0}) cancel and the constraint is satisfied.  Therefore, 
\begin{equation}
    z_I^{\ J} = -\frac{\left(m_i\lbrack \mathbf{i}_{\mathrm{I}}\mathbf{j}^{\mathrm{J}} \rbrack + \langle \mathbf{i}_{\mathrm{I}}\lvert p_k\rvert \mathbf{j}^{\mathrm{J}}\rbrack\right)s_{ik}}{\left(m_i\lbrack \mathbf{i}_{\mathrm{K}}\mathbf{j}^{\mathrm{L}} \rbrack + \langle \mathbf{i}_{\mathrm{K}}\lvert p_k\rvert \mathbf{j}^{\mathrm{L}}\rbrack\right)\langle \mathbf{i}^{\mathrm{K}} \lvert p_k\rvert\mathbf{j}_{\mathrm{L}}\rbrack}\ ,
    \label{eq:Z_I^J Intermediate 2}
\end{equation}
which can be simplified to
\begin{equation}
    z_I^{\ J} = \frac{m_i\lbrack \mathbf{i}_{\mathrm{I}}\mathbf{j}^{\mathrm{J}} \rbrack + \langle \mathbf{i}_{\mathrm{I}}\lvert p_k\rvert \mathbf{j}^{\mathrm{J}}\rbrack}{m_im_j}\ .
    \label{eq:Z_I^J}
\end{equation}
In the massless limit, we begin with Eq.~(\ref{eq:Z_I^J Intermediate 2}) to obtain
\begin{eqnarray}
    z_I^{\ J} &\to&
    -\frac{\langle i\lvert p_k\rvert j\rbrack \langle i\lvert p_k\rvert i\rbrack}
    {\langle i\lvert p_k\rvert j\rbrack\langle i \lvert p_k\rvert j\rbrack}
    \nonumber\\
    &&= -\frac{ \langle ik\rangle\lbrack ki\rbrack}
    {\langle ik\rangle\lbrack kj\rbrack}
    \nonumber\\
    &&= -\frac{ \lbrack ki\rbrack}
    {\lbrack kj\rbrack}\ ,
\end{eqnarray}
in agreement with Eq.~(\ref{eq:app:zp=-[ik]/[jk]}).

We can follow a similar line of logic using Eq.~(\ref{eq:app:z<>[]=-M2+mj2+...}) to obtain
\begin{eqnarray}
    z_I^{\ J} &=& 
    \frac{\left(m_j\langle \mathbf{i}_{\mathrm{I}}\mathbf{j}^{\mathrm{J}} \rangle - \langle \mathbf{i}_{\mathrm{I}}\lvert p_l\rvert \mathbf{j}^{\mathrm{J}}\rbrack\right)s_{jl}}
    {\left(m_j\langle \mathbf{i}_{\mathrm{K}}\mathbf{j}^{\mathrm{L}} \rangle - \langle \mathbf{i}_{\mathrm{K}}\lvert p_l\rvert \mathbf{j}^{\mathrm{L}}\rbrack\right)\langle \mathbf{i}^{\mathrm{K}} \lvert p_l\rvert\mathbf{j}_{\mathrm{L}}\rbrack}\\
    &=& \frac{m_j\langle \mathbf{i}_{\mathrm{I}}\mathbf{j}^{\mathrm{J}} \rangle - \langle \mathbf{i}_{\mathrm{I}}\lvert p_l\rvert \mathbf{j}^{\mathrm{J}}\rbrack}
    {m_im_j}\ .
    \label{eq:Z_I^J 2}
\end{eqnarray}
Taking the massless limit gives
\begin{eqnarray}
    z_I^{\ J} &\to&
    \frac{\langle i\lvert p_l\rvert j\rbrack \langle j\lvert p_l\rvert j\rbrack}
    {\langle i\lvert p_l\rvert j\rbrack\langle i \lvert p_l\rvert j\rbrack}
    \nonumber\\
    &&= \frac{ \langle jl\rangle}
    {\langle il\rangle}\ ,
\end{eqnarray}
in agreement with Eq.~(\ref{eq:z=<jl>/<il> massless}).

Before leaving, let us note that the spinor shift can be simplified in the massive case using the mass and momentum identities.  If we plug our solution for $z_{\mathrm{I}}^{\mathrm{J}}$ in Eq.~(\ref{eq:Z_I^J}) back into Eqs.~(\ref{eq:app:|ihat] def 2}) and (\ref{eq:app:|jhat> def 2}), we obtain
\begin{eqnarray}
    \lbrack\mathbf{\hat{i}}_\mathrm{I}| &=& \lbrack\mathbf{i}_\mathrm{I}| + \frac{m_i\lbrack \mathbf{i}_{\mathrm{I}}\mathbf{j}^{\mathrm{J}} \rbrack\lbrack\mathbf{j}_\mathrm{J}| + \langle \mathbf{i}_{\mathrm{I}}\lvert p_k\rvert \mathbf{j}^{\mathrm{J}}\rbrack\lbrack\mathbf{j}_\mathrm{J}|}{m_im_j} 
    \nonumber\\
    &=& \lbrack\mathbf{i}_\mathrm{I}| + \frac{-m_im_j\lbrack \mathbf{i}_{\mathrm{I}}\rvert -m_j \langle \mathbf{i}_{\mathrm{I}}\lvert p_k}{m_im_j}
    \nonumber\\
    &=& - \frac{1}{m_i}\langle \mathbf{i}_{\mathrm{I}}\lvert p_k\ ,
\end{eqnarray}
or, with implicit spin indices, we have
\begin{equation}
    \lbrack\mathbf{\hat{i}}| =
    - \frac{1}{m_i}\langle \mathbf{i}\lvert p_k\ .
    \label{eq:[ihat|=-1/mi<i|pk}
\end{equation}
Following the analogous steps for $\lvert\hat{\mathbf{j}}^{\mathrm{J}}\rangle$, we have
\begin{eqnarray}
    |\mathbf{\hat{j}}^\mathrm{J}\rangle &=& |\mathbf{j}^\mathrm{J}\rangle - \frac{m_i|\mathbf{i}^\mathrm{I}\rangle\lbrack \mathbf{i}_{\mathrm{I}}\mathbf{j}^{\mathrm{J}} \rbrack + |\mathbf{i}^\mathrm{I}\rangle\langle \mathbf{i}_{\mathrm{I}}\lvert p_k\rvert \mathbf{j}^{\mathrm{J}}\rbrack}{m_im_j} 
    \nonumber\\
    &=& 
    |\mathbf{j}^\mathrm{J}\rangle - \frac{1}{m_j}\left(p_i+p_k\right)\lvert\mathbf{j}^{\mathrm{J}} \rbrack
    \nonumber\\
    &=& 
    |\mathbf{j}^\mathrm{J}\rangle + \frac{1}{m_j}\left(p_j+p_l\right)\lvert\mathbf{j}^{\mathrm{J}} \rbrack
    \nonumber\\
    &=& 
    \frac{1}{m_j}p_l\lvert\mathbf{j}^{\mathrm{J}} \rbrack\ ,
\end{eqnarray}
and, with implicit spin indices,
\begin{equation}
    |\mathbf{\hat{j}}\rangle
    =
    \frac{1}{m_j}p_l\lvert\mathbf{j} \rbrack\ .
    \label{eq:|jhat>=1/mjpl|j>}
\end{equation}
Both of these results are exactly the same if we instead use $z_{\mathrm{I}}^{\ \mathrm{J}}$ from Eq.~(\ref{eq:Z_I^J 2}).  This allows us to bypass the $z_{\mathrm{I}}^{\ \mathrm{J}}$ altogether in the massive-massive case.  In fact, with this choice for the pole $z_{\mathrm{I}}^{\mathrm{J}}$, the modification of the momentum is not complex.
\begin{eqnarray}
    \hat{p}_i &=&
    \lvert\mathbf{i}^{\mathrm{I}}\rangle\lbrack\hat{\mathbf{i}}_{\mathrm{I}}\rvert
    \nonumber\\
    &=& -\frac{1}{m_i}\lvert\mathbf{i}^{\mathrm{I}}\rangle\langle\mathbf{i}_{\mathrm{I}}\rvert p_k
    \nonumber\\
    &=& -p_k\ ,
    \label{eq:pi=-pk complexification}
\end{eqnarray}
and similarly,
\begin{eqnarray}
    \hat{p}_j &=& 
    \lvert\hat{\mathbf{j}}^{\mathrm{J}}\rangle\lbrack\mathbf{j}_{\mathrm{J}}\rvert
    \nonumber\\
    &=& \frac{1}{m_j}p_l\lvert\mathbf{j}^{\mathrm{J}}\rbrack\lbrack\mathbf{j}_{\mathrm{J}}\rvert
    \nonumber\\
    &=& -p_l\ .
    \label{eq:pj=-pl complexification}
\end{eqnarray}
All the on-shell conditions are still met, as they must be. For particle $i$, we have
\begin{eqnarray}
    \hat{p}_i^2 &=& 
    p_j^2 = m_i^2
    \\
    \hat{p}_k^2 &=&
    p_l^2 = m_k^2\ ,
\end{eqnarray}
since $m_j=m_i$ and $m_l=m_k$.
We can also check the internal lines,
\begin{eqnarray}
(\hat{p}_i + p_k)^2 
    =& (-p_k+p_k)^2
    &= 0
    \\
(\hat{p}_j + p_l)^2 
    =& (-p_l+p_l)^2
    &= 0\ .
\end{eqnarray}
This does not, necessarily, mean that the complexification step is unnecessary.  It might be that it is important in the proof of this method in analogy with the proof for the massless case \cite{Britto:2005fq}.  Since, as we show in other sections of this paper, this is still not sufficient to get agreement with Feynman diagrams, it might be that there are other ingredients necessary that will clarify this point.  For now, we simply note this property and see where it takes us.

\subsubsection{\label{app:[i,j> both massive massive internal line}$\mathbf{\lbrack\mathbf{i},\mathbf{j}\rangle}$ Shift with Two Massive Spinors and a Massive Internal Line}
This case applies to an internal fermion line with a massless particle on an external line, such as a photon and a Higgs connected to an electron line or similar.
In this case, we take both particle $i$ and $j$ to be massive and of the same mass and the same mass as the internal line.  (If we try to do this case in complete generality, the solution becomes very complex and unilluminating.) 
As in the previous subsection, for an $\lbrack \mathbf{i},\mathbf{j}\rangle$ shift, we will try
\begin{eqnarray}
    \lbrack\mathbf{\hat{i}}_\mathrm{I}| &=& \lbrack\mathbf{i}_\mathrm{I}| + z^{\ \mathrm{J}}_\mathrm{I} \lbrack\mathbf{j}_\mathrm{J}| 
    \label{eq:app:|ihat] def 2}\\
    |\mathbf{\hat{j}}^\mathrm{J}\rangle &=& |\mathbf{j}^\mathrm{J}\rangle - z^{\ \mathrm{J}}_\mathrm{I} |\mathbf{i}^\mathrm{I}\rangle\ .
    \label{eq:app:|jhat> def 2}
\end{eqnarray}
We saw in Eq.~(\ref{eq:[i,j> massive: pihat+pjhat=pi+pj}) that this preserves momentum conservation.
We next need to satisfy the on-shell condition for particles $i$ and $j$ which gives us
\begin{eqnarray}
    z_\mathrm{I}^{\ \mathrm{J}}\left(
    2\lbrack\mathbf{i}^{\mathrm{I}}\mathbf{j}_{\mathrm{J}}\rbrack 
    - z_{\ \mathrm{J}}^{\mathrm{I}}m
    \right) &=& 0
    \label{eq:[i,j> massive: on-shell 1}
    \\
    z^{\ \mathrm{J}}_{\mathrm{I}}\left(
    2\langle\mathbf{i}^{\mathbf{I}}\mathbf{j}_\mathrm{J}\rangle
    - z^{\mathrm{I}}_{\ \mathrm{J}}m
    \right) &=& 0\ .
    \label{eq:[i,j> massive: on-shell 2}
\end{eqnarray}
Subtracting them, we find
\begin{equation}
    z_\mathrm{I}^{\ \mathrm{J}}
    \left(\lbrack\mathbf{i}^{\mathrm{I}}\mathbf{j}_{\mathrm{J}}\rbrack 
    -
    \langle\mathbf{i}^{\mathbf{I}}\mathbf{j}_\mathrm{J}\rangle\right)
    = 0\ ,
    \label{z(m[]-m<>)=0}
\end{equation}
which is linear in $z$.  We must satisfy this as well as either Eq.~(\ref{eq:[i,j> massive: on-shell 1}) or (\ref{eq:[i,j> massive: on-shell 2}).

We must also satisfy the on-shell condition for the internal line, as before, 
\begin{eqnarray}
    z_{\mathrm{I}}^{\ \mathrm{J}}\langle \mathbf{i}^{\mathrm{I}} \lvert p_k\rvert\mathbf{j}_{\mathrm{J}}\rbrack &=&
     m^2-s_{ik}
    \label{eq:app:all three massive:z<>[]=M2-mi2-...}
    \\
    z_{\mathrm{I}}^{\ \mathrm{J}}\langle \mathbf{i}^{\mathrm{I}} \lvert p_l\rvert\mathbf{j}_{\mathrm{J}}\rbrack &=&
    s_{jl}-m^2 \ .
    \label{eq:app:all 3 massive:z<>[]=-M2+mj2+...}
\end{eqnarray}
Solving the first of these gives
\begin{eqnarray}
    z_I^{\ J} &=& \frac{C_I^{\ J}\left(m^2-s_{ik}\right)}{C_K^{\ L}\langle \mathbf{i}^{\mathrm{K}} \lvert p_k\rvert\mathbf{j}_{\mathrm{L}}\rbrack}\ ,
    \label{eq:all 3 massive:z_I^J Intermediate}
\end{eqnarray}
for any $C_I^{\ J}$.  We next apply Eqs.~(\ref{z(m[]-m<>)=0}) and (\ref{eq:[i,j> massive: on-shell 2}) giving us
\begin{eqnarray}
    C_I^{\ J}
    \lbrack\mathbf{i}^{\mathrm{I}}\mathbf{j}_{\mathrm{J}}\rbrack 
    -
    C_I^{\ J}\langle\mathbf{i}^{\mathbf{I}}\mathbf{j}_\mathrm{J}\rangle 
    &=& 0
    \nonumber\\
    \label{eq:all 3 massive:CIJ([ij]-<ij>)=0}
    \\
    2C^{\ \mathrm{J}}_{\mathrm{I}}\langle\mathbf{i}^{\mathbf{I}}\mathbf{j}_\mathrm{J}\rangle C_K^{\ L}\langle \mathbf{i}^{\mathrm{K}} \lvert p_k\rvert\mathbf{j}_{\mathrm{L}}\rbrack
    + C^{\ \mathrm{K}}_{\mathrm{I}}C^{\mathrm{I}}_{\ \mathrm{K}}m \left(s_{ik}-m^2\right)
    &=& 0\ .
    \nonumber\\
    \label{eq:all 3 massive:CIJ<ij>CKL<ipkj]+CIKCIK=0}
\end{eqnarray}
As before, the $C_I^{\ J}$ must be a linear combination of $\langle \mathbf{i}_{\mathrm{I}}\mathbf{j}^{\mathrm{J}}\rangle, \lbrack \mathbf{i}_{\mathrm{I}}\mathbf{j}^{\mathrm{J}} \rbrack, \langle \mathbf{i}_{\mathrm{I}}\lvert p_k\rvert \mathbf{j}^{\mathrm{J}}\rbrack$ and $\lbrack \mathbf{i}_{\mathrm{I}}\lvert p_k\rvert \mathbf{j}^{\mathrm{J}}\rangle$.  In this case, there are no solutions with only two of these spinor products.  There are solutions with only three of them.  However, they are very complicated unless we set either $m_k=0$ or $m_l=0$.  In our case, we have a massless photon as an external state and, in fact, the only time we see that this shift would be needed is when there is a massless external particle.  So, we will only show the solution when $m_k=0$.  It turns out that there are two equally simple solutions.  One contains $\langle \mathbf{i}_{\mathrm{I}}\mathbf{j}^{\mathrm{J}}\rangle, \langle \mathbf{i}_{\mathrm{I}}\lvert p_k\rvert \mathbf{j}^{\mathrm{J}}\rbrack$ and $\lbrack \mathbf{i}_{\mathrm{I}}\lvert p_k\rvert \mathbf{j}^{\mathrm{J}}\rangle$ and the other contains $\lbrack \mathbf{i}_{\mathrm{I}}\mathbf{j}^{\mathrm{J}} \rbrack, \langle \mathbf{i}_{\mathrm{I}}\lvert p_k\rvert \mathbf{j}^{\mathrm{J}}\rbrack$ and $\lbrack \mathbf{i}_{\mathrm{I}}\lvert p_k\rvert \mathbf{j}^{\mathrm{J}}\rangle$.  We will do the latter. 
\begin{eqnarray}
    C_I^{\ J} &=& m\left(s_{ij}-m_l^2\right)\left(s_{il}-m^2\right)\lbrack \mathbf{i}_{\mathrm{I}}\mathbf{j}^{\mathrm{J}} \rbrack 
    \nonumber\\
    &&+ \left(s_{ij}s_{il}-m^2m_l^2\right)\langle \mathbf{i}_{\mathrm{I}}\lvert p_k\rvert \mathbf{j}^{\mathrm{J}}\rbrack
    \nonumber\\
    &&+m^2\left(s_{ij}-m_l^2\right)\lbrack \mathbf{i}_{\mathrm{I}}\lvert p_k\rvert \mathbf{j}^{\mathrm{J}}\rangle
    \ .
\end{eqnarray}
However, when we plug this into the denominator of $z$ from Eq.~(\ref{eq:all 3 massive:z_I^J Intermediate}), it vanishes, so this is not a satisfactory solution.  The other solution with $\langle \mathbf{i}_{\mathrm{I}}\mathbf{j}^{\mathrm{J}}\rangle$ also vanishes.  In fact, this is not difficult to see if we consider a general form for $C_{\mathrm{I}}^{\ \mathrm{J}}$ and multiply by $\langle \mathbf{i}^{\mathrm{I}} \lvert p_k\rvert\mathbf{j}_{\mathrm{J}}\rbrack$, we get
\begin{eqnarray}
    \left(
    A\langle \mathbf{i}_{\mathrm{I}}\mathbf{j}^{\mathrm{J}}\rangle
    +B\lbrack \mathbf{i}_{\mathrm{I}}\mathbf{j}^{\mathrm{J}} \rbrack
    +C\langle \mathbf{i}_{\mathrm{I}}\lvert p_k\rvert \mathbf{j}^{\mathrm{J}}\rbrack
    +D\lbrack \mathbf{i}_{\mathrm{I}}\lvert p_k\rvert \mathbf{j}^{\mathrm{J}}\rangle
    \right)
    \langle \mathbf{i}^{\mathrm{I}} \lvert p_k\rvert\mathbf{j}_{\mathrm{J}}\rbrack
    \nonumber\\
    =
    A m\left(m^2-s_{il}\right)
    +B m\left(m^2-s_{ik}\right)
    \nonumber\\
    +D\left(m^2-s_{ik}\right)\left(m^2-s_{il}\right)\ .
    \nonumber\\
\end{eqnarray}
As we can see, $\langle \mathbf{i}_{\mathrm{I}}\lvert$ does not contribute to the denominator of $z$ because that term is proportional to $p_k^2=0$.  Furthermore, if we look for a solution with $A=0$, we get a solution of the form $B=D(s_{il}-m^2)/m$, which causes the denominator to vanish.  A similar thing happens if we set $B=0$ instead.  The denominator of $z$ vanishes because the solution is of the form $A=D(s_{ik}-m^2)/m$.  It is possible to find a non-trivial solution if we keep all of $A, B, C$ and $D$ non-zero, however, the solution is quite complicated, not illuminating and not necessary.  We already have success obtaining the amplitude for $e\gamma\bar{e}h$ using other shifts where one external particle is massive and the other is massless.  So, we will not pursue this case further.  We might also consider this shift for the process $e\bar{e}\gamma\gamma$, where we set both $m_k=m_l=0$.  However, we run into exactly the same problem.  All simple solutions make the denominator of $z$ vanish.  All valid solutions are too complicated to be valuable here, especially when we have shifts with at least one shifted particle massless.

\subsection{\label{app:large z behavior}Large $\mathbf{z}$ Behavior}

In this subsection, for each correct amplitude, we will consider every possible momentum shift and analyze the large $z$ behavior to determine whether it vanishes in the limit $z\to\infty$.  We have not achieved success for any shifts when the amplitude does not vanish in this limit and the photon is massless from the beginning.  On the other hand, in every case where the amplitude does vanish in the large $z$ limit and where we have a simple momentum shift to apply, we have had successfully obtained agreement with Feynman diagrams.  One possible exception is the $\lbrack2,\mathbf{4}\rangle$ shift for $e\gamma\bar{e}h$ process (see App.~\ref{sec:Constructive Massless heAe}), where we were unable to simplify to the correct form.  However, we were successful for this process with other shifts.

\subsubsection{\label{app:eemm large z}$e\bar{e}\mu\bar{\mu}$ and $e\bar{e}e\bar{e}$}
The correct amplitude for this process was given in Eq.~(\ref{eq:M_eemm}), and is
\begin{equation}
    \mathcal{M} = \frac{e^2
    \left(\langle\mathbf{13}\rangle\lbrack\mathbf{24}\rbrack
    +\langle\mathbf{14}\rangle\lbrack\mathbf{23}\rbrack
    +\lbrack\mathbf{13}\rbrack\langle\mathbf{24}\rangle
    +\lbrack\mathbf{14}\rbrack\langle\mathbf{23}\rangle
    \right)}{(p_1+p_2)^2}\ .
\end{equation}
We can already see what will happen here by inspection.  The choice of the two momenta must always be split across the propagator, therefore we can analytically continue $p_1$ and $p_3$, $p_1$ and $p_4$, $p_2$ and $p_3$ or $p_2$ and $p_4$.  Every one of these choices will either extend $p_1$ or $p_2$ but not both.  Therefore, since the denominator is squared, the denominator will go as $z^2$ in the large $z$ limit for every choice.  For the numerator, the only possible shifts (associated with the momentum shifts) are $\lbrack\mathbf{1,3}\rangle, \lbrack\mathbf{3,1}\rangle, \lbrack\mathbf{1,4}\rangle, \lbrack\mathbf{4,1}\rangle, \lbrack\mathbf{2,3}\rangle, \lbrack\mathbf{3,2}\rangle, \lbrack\mathbf{2,4}\rangle$ or $ \lbrack\mathbf{4,2}\rangle$.  In every case, the square index for one particle, $\lvert\mathbf{i}\rbrack$, is shifted and the angle index for one other particle $\lvert\mathbf{j}\rangle$, is shifted.  But, no matter what our choice of $i$ and $j$, there is always a term with one of each.  Therefore, the numerator also grows as $z^2$ in the large $z$ limit.  Together with the denominator, then, the amplitude approaches a constant in the large $z$ limit and does not vanish.  

Let's do an example to see how this works.  Consider the shift $\lbrack\mathbf{1,3}\rangle$.  In this case, we see from Eqs.~(\ref{eq:app:|ihat] def 2}) and (\ref{eq:app:|jhat> def 2}) that
\begin{eqnarray}
    \lbrack\mathbf{\hat{1}}_\mathrm{I}| &=& \lbrack\mathbf{1}_\mathrm{I}| + z^{\ \mathrm{J}}_\mathrm{I} \lbrack\mathbf{3}_\mathrm{J}| 
    \label{eq:large z:[1hat|}\\
    |\mathbf{\hat{3}}^\mathrm{J}\rangle &=& |\mathbf{3}^\mathrm{J}\rangle - z^{\ \mathrm{J}}_\mathrm{I} |\mathbf{1}^\mathrm{I}\rangle\ .
\end{eqnarray}
As a result, 
\begin{eqnarray}
\hat{p}_1 &=&
    \lvert\mathbf{1}^{\mathrm{I}}\rangle\lbrack\hat{\mathbf{1}}_{\mathrm{I}}\rvert
    \nonumber\\
    &=& p_1 + z_{\mathrm{I}}^{\ \mathrm{J}}\lvert\mathbf{1}^{\mathrm{I}}\rangle\lbrack\mathbf{3}_{\mathrm{J}}\rvert
    \nonumber\\
    &=& p_1 + z_{\mathrm{I}}^{\ \mathrm{J}}q_{\ \mathrm{J}}^{\mathrm{I}}
    \\
    \hat{p}_3 &=& p_3 - z_{\mathrm{I}}^{\ \mathrm{J}}q_{\ \mathrm{J}}^{\mathrm{I}}
    \ .
    \label{eq:large z:p3hat}
\end{eqnarray}
The shifted amplitude is
\begin{eqnarray}
    \hat{\mathcal{M}}
    &=& \frac{e^2
    \left(\langle\mathbf{1}\hat{\mathbf{3}}\rangle\lbrack\mathbf{24}\rbrack
    +\langle\mathbf{14}\rangle\lbrack\mathbf{23}\rbrack
    +\lbrack\hat{\mathbf{1}}\mathbf{3}\rbrack\langle\mathbf{24}\rangle
    +\lbrack\hat{\mathbf{1}}\mathbf{4}\rbrack\langle\mathbf{2}\hat{\mathbf{3}}\rangle
    \right)}{(\hat{p}_1+p_2)^2}
    \nonumber\\
\end{eqnarray}
In the numerator, we have terms with no shifted spinors and terms with one shifted spinor and one term with two shifted spinors.  We will always find this pattern for this amplitude.  For compactness, in the following, we will only expose the spin indices on particles $1$ and $3$ and we will lower the spin index for particle $1$.  Plugging the shifted spinors in, we have
\begin{eqnarray}
    \hat{\mathcal{M}}_{\mathrm{I}}^{\mathrm{J}}
    &=& \frac{e^2}{(p_1+z_{\mathrm{M}}^{\ \mathrm{N}}q_{\ \mathrm{N}}^{\mathrm{M}}+p_2)^2}
    \Big(\langle\mathbf{1}_{\mathrm{I}}\mathbf{3}^{\mathrm{J}}\rangle\lbrack\mathbf{24}\rbrack
    - z_{\mathrm{K}}^{\ \mathrm{J}}\langle\mathbf{1}_{\mathrm{I}}\mathbf{1}^{\mathrm{K}}\rangle\lbrack\mathbf{24}\rbrack
    \nonumber\\
    &&\hspace{-0.1in}+\langle\mathbf{14}\rangle\lbrack\mathbf{23}\rbrack
    +\lbrack\mathbf{1}\mathbf{3}\rbrack\langle\mathbf{24}\rangle
    +z_{\mathrm{I}}^{\ \mathrm{L}}\lbrack\mathbf{3}_{\mathrm{L}}\mathbf{3}^{\mathrm{J}}\rbrack\langle\mathbf{24}\rangle
    +\lbrack\mathbf{1}_{\mathrm{I}}\mathbf{4}\rbrack\langle\mathbf{2}\mathbf{3}^{\mathrm{J}}\rangle
    \nonumber\\
    &&\hspace{-0.25in}-z_{\mathrm{K}}^{\ \mathrm{J}}\lbrack\mathbf{1}_{\mathrm{I}}\mathbf{4}\rbrack\langle\mathbf{2}\mathbf{1}^{\mathrm{K}}\rangle
    +z_{\mathrm{I}}^{\ \mathrm{L}}\lbrack\mathbf{3}_{\mathrm{L}}\mathbf{4}\rbrack\langle\mathbf{2}\mathbf{3}^{\mathrm{J}}\rangle
    -z_{\mathrm{K}}^{\ \mathrm{J}}z_{\mathrm{I}}^{\ \mathrm{L}}\lbrack\mathbf{3}_{\mathrm{L}}\mathbf{4}\rbrack\langle\mathbf{2}\mathbf{1}^{\mathrm{K}}\rangle
    \Big).
    \nonumber\\
\end{eqnarray}
Now, when we take the large $z$ limit, we find
\begin{equation}
    \hat{\mathcal{M}}_{\lbrack\mathbf{1,3}\rangle}^{ee\mu\mu}\xrightarrow[z \to \infty]{} \sim \frac{e^2\lbrack\mathbf{3}\mathbf{4}\rbrack\langle\mathbf{2}\mathbf{1}\rangle}{q^2}\ ,
\end{equation}
where the $\sim$ is to represent that we have dropped the indices to give the generic structure in the limit.  The details will depend on how we take $z$ to $\infty$, but the final result is that it does not generically vanish in this limit.  We can see that every choice for a shift has the same property.
\begin{equation}
    \hat{\mathcal{M}}_{\lbrack\mathbf{i,j}\rangle}^{ee\mu\mu}\xrightarrow[z \to \infty]{} \mathcal{O}\left(z^0\right)\ ,
\end{equation}
where by $\mathcal{O}\left(z^0\right)$, we mean a non-zero constant term.

We might wonder what happens for the process $e\bar{e}e\bar{e}$.  Perhaps, due to a cancellation between the two diagrams, it vanishes for some choice of analytic continuation.  The amplitude is given in Eq.~(\ref{eq:M_eeee}) and is
\begin{align}
    \mathcal{M} &= \frac{e^2
    \left(\langle\mathbf{13}\rangle\lbrack\mathbf{24}\rbrack
    +\langle\mathbf{14}\rangle\lbrack\mathbf{23}\rbrack
    +\lbrack\mathbf{13}\rbrack\langle\mathbf{24}\rangle
    +\lbrack\mathbf{14}\rbrack\langle\mathbf{23}\rangle
    \right)}{(p_1+p_2)^2}
    \nonumber\\
    &- \frac{e^2
    \left(\langle\mathbf{12}\rangle\lbrack\mathbf{34}\rbrack
    -\langle\mathbf{14}\rangle\lbrack\mathbf{23}\rbrack
    +\lbrack\mathbf{12}\rbrack\langle\mathbf{34}\rangle
    -\lbrack\mathbf{14}\rbrack\langle\mathbf{23}\rangle
    \right)}{(p_1+p_3)^2}\ .
    \label{eq:large z:M_eeee}
\end{align}
If we choose any shift that involves momenta on opposite sides of one diagram but on the same side of the other diagram, then the amplitude will grow as $z^2$ since the denominator of the other diagram will not contain $z$ at all.  For example, consider the shift $\lbrack\mathbf{1,3}\rangle$.  The shifts of the spinors and momenta are given in Eqs.~(\ref{eq:large z:[1hat|}) through (\ref{eq:large z:p3hat}).  We see that $\hat{p}_1+\hat{p}_3=p_1+p_3$ so that the denominator of the second term in Eq.~(\ref{eq:large z:M_eeee}) will not depend on $z$.  On the other hand, the last term of the numerator contains $\lbrack\mathbf{14}\rbrack\langle\mathbf{23}\rangle$, which will grow as $z^2$.  Therefore, in this case,
\begin{eqnarray}
    \hat{\mathcal{M}}_{\lbrack\mathbf{1,3}\rangle}^{eeee}&\xrightarrow[z \to \infty]{}& \mathcal{O}\left(z^2\right)
    \ .
\end{eqnarray}
The same will be true for the shifts $\lbrack\mathbf{3,1}\rangle, \lbrack\mathbf{1,2}\rangle$ and $\lbrack\mathbf{2,1}\rangle$.  On the other hand, a shift that has momenta on opposite sides of both diagrams, such as $\lbrack\mathbf{1,4}\rangle$ will have the property that both diagrams approach constant values, so we should look at the details.  For example, 
\begin{eqnarray}
\hat{\mathcal{M}}_{\lbrack\mathbf{1,4}\rangle}^{eeee} &\xrightarrow[z \to \infty]{} \frac{e^2
    \lbrack\mathbf{13}\rbrack\langle\mathbf{24}\rangle
    }{q^2}
    - \frac{e^2
    \lbrack\mathbf{12}\rbrack\langle\mathbf{34}\rangle
    }{q^2}\ .
\end{eqnarray}
However, the terms are different and, once again, do not cancel.  So, we have
\begin{eqnarray}
    \hat{\mathcal{M}}_{\lbrack\mathbf{1,4}\rangle}^{eeee}&\xrightarrow[z \to \infty]{}& \mathcal{O}\left(z^0\right)
    \ ,
\end{eqnarray}
and the same is true for the shifts $\lbrack\mathbf{4,1}\rangle, \lbrack\mathbf{2,3}\rangle$ and $\lbrack\mathbf{3,2}\rangle$.

\subsubsection{\label{app:large z:eAeh}$e\gamma\bar{e} h$}
The correct amplitude for the process $e\gamma^+\bar{e} h$ is given in Eq.~(\ref{eq:M_eA+eh}) and is
\begin{eqnarray}
    \mathcal{M}^{e\gamma^+eh} &=& 
    -\frac{e m_e}{v\left(s-m_e^2\right)\left(u-m_e^2\right)}
      \Big( m_h ^{2} \lbrack\mathbf{1}2\rbrack\lbrack2\mathbf{3}\rbrack 
      \nonumber\\
    &&- m_e \lbrack\mathbf{1}2\rbrack\lbrack2\lvert p_{4} \rvert \mathbf{3}\rangle 
    + m_e \lbrack2\mathbf{3}\rbrack \langle\mathbf{1}\lvert p_{4} \rvert 2\rbrack
    \nonumber\\
    &&- \langle\mathbf{1}\mathbf{3}\rangle \lbrack2\lvert p_{3} p_{4} \rvert 2\rbrack
    \Big)\ .
\end{eqnarray}
We can see multiple differences in this process that will support a vanishing amplitude in the $z\to\infty$ limit.  The first is that this amplitude contains the product of two Feynman propagator denominators.  Therefore, for some choice of spinor shifts, the denominator will grow as $z^4$.  The second is that, since the Higgs boson is spinless, there are no spinors for particle $4$.  Therefore, any shift that contains $\lvert\mathbf{4}\rbrack$ or $\lvert\mathbf{4}\rangle$ will not have contributions directly from their shifts.  However, on the other hand, the momentum $p_4$ does appear in the numerator and it will grow as $z$.  We could use momentum conservation to change it to the other momenta, but we would still get the same large $z$ behavior from the other momentum in the shift after use of momentum conservation.  So, that won't help us.  Finally, we have that the external photon is massless and only comes in one helicity or the other.  This means that the amplitude has either angle helicity spinors $\lvert2\rangle$ if the photon has negative helicity or square helicity spinors $\lvert2\rbrack$ if the photon has positive helicity.  In this case, our amplitude has only $\lvert2\rbrack$ and no $\lvert2\rangle$.  Therefore, choosing a shift of the form $\lbrack\mathbf{i},2\rangle$ will maximize the fall off at large $z$.
As we can see, we have several features that lead to a potential shift choice that leads to a vanishing amplitude in the limit.  In order to obtain the maximum growth potential of the denominator with a shift in $\lvert2\rangle$, we should use the shift $\lbrack\mathbf{4},2\rangle$.  If we do this, the denominator will grow as $z^4$ and the numerator will grow as $z$ for a total amplitude falloff as $1/z^3$, more than sufficient to vanish in the large $z$ limit.  Let's do this case in a little detail.  From Eqs.~(\ref{eq:app:|ihat] def 3}) and (\ref{eq:app:|jhat> def 3}), we have
\begin{eqnarray}
    \lbrack\mathbf{\hat{4}}_\mathrm{I}| &=& \lbrack\mathbf{4}_\mathrm{I}| + z_\mathrm{I} \lbrack 2| 
    \label{eq:app:large z:|4hat]}\\
    |\hat{2}\rangle &=& |2\rangle - z_\mathrm{I} |\mathbf{4}^\mathrm{I}\rangle\ .
    \label{eq:app:large z:|2hat>}
\end{eqnarray}
As we can see, neither of these spinors appear explicitly in the amplitude, therefore, the effect comes solely through the momenta,
\begin{eqnarray}
    \hat{p}_2 &=& p_2 - z^Jq_J\\
    \hat{p}_4 &=& p_4 + z^Jq_J
    \ .
\end{eqnarray}
Plugging this in, we find
\begin{eqnarray}
    \hat{\mathcal{M}}^{e\gamma^+eh}_{\lbrack\mathbf{4},2\rangle}
    &\xrightarrow[z \to \infty]{}
    \sim&
    \frac{em_e}{vz^3q^4}
    \Big(m_e \lbrack\mathbf{1}2\rbrack\lbrack2\lvert q \rvert \mathbf{3}\rangle
    - m_e \lbrack2\mathbf{3}\rbrack \langle\mathbf{1}\lvert q \rvert 2\rbrack 
    \nonumber\\
    &&+ \langle\mathbf{1}\mathbf{3}\rangle \lbrack2\lvert p_{3} q \rvert 2\rbrack
       \Big)\ ,
    \nonumber\\
\end{eqnarray}
or, in other words,
\begin{eqnarray}
    \hat{\mathcal{M}}^{e\gamma^+eh}_{\lbrack\mathbf{4},2\rangle}
    &\xrightarrow[z \to \infty]{}
    \mathcal{O}\left(\frac{1}{z^3}\right)\ .
    \label{eq:large z:M_eA+eh [4,2>}
\end{eqnarray}
This case was engineered to have the steepest falloff at large z and falls off faster than necessary, so we might expect that there are other shifts that could work and, of course, some that don't.  Here are a few cases,
\begin{eqnarray}
    \hat{\mathcal{M}}^{e\gamma^+eh}_{\lbrack\mathbf{1},\mathbf{3}\rangle,\lbrack\mathbf{3},\mathbf{1}\rangle}
    &\xrightarrow[z \to \infty]{}
    \mathcal{O}\left(\frac{1}{z^2}\right)
    \\
    \hat{\mathcal{M}}^{e\gamma^+eh}_{\lbrack2,\mathbf{4}\rangle,\lbrack\mathbf{3},2\rangle}
    &\xrightarrow[z \to \infty]{}
    \mathcal{O}\left(\frac{1}{z}\right)
    \\
    \hat{\mathcal{M}}^{e\gamma^+eh}_{\lbrack\mathbf{1},2\rangle}
    &\xrightarrow[z \to \infty]{}
    \mathcal{O}\left(\frac{1}{z}\right)
    \\
    \hat{\mathcal{M}}^{e\gamma^+eh}_{\lbrack\mathbf{3},\mathbf{4}\rangle,\lbrack\mathbf{4},\mathbf{3}\rangle}
    &\xrightarrow[z \to \infty]{}
    \mathcal{O}\left(z^0\right)
    \\
    \hat{\mathcal{M}}^{e\gamma^+eh}_{\lbrack\mathbf{1},\mathbf{4}\rangle,\lbrack\mathbf{4},\mathbf{1}\rangle}
    &\xrightarrow[z \to \infty]{}
    \mathcal{O}\left(z^0\right)
    \\
    \hat{\mathcal{M}}^{e\gamma^+eh}_{\lbrack2,\mathbf{1}\rangle}
    &\xrightarrow[z \to \infty]{}
    \mathcal{O}\left(z\right)
    \\
    \hat{\mathcal{M}}^{e\gamma^+eh}_{\lbrack2,\mathbf{3}\rangle}
    &\xrightarrow[z \to \infty]{}
    \mathcal{O}\left(z^2\right)
    \label{eq:large z:M_eA+eh [2,3>}
    \ .
\end{eqnarray}
In App.~\ref{sec:Constructive Massless heAe}, we use these shifts to find the correct amplitude.  Interestingly, as we show and discuss there, we only need one of the diagrams for each shift to obtain the full result.

\subsubsection{\label{app:large z:eeAA}$e\bar{e}\gamma\gamma$}
The correct amplitude for the process $e\bar{e}\gamma^+\gamma^+$ is given by Eq.~(\ref{eq:M_eeA+A+}) and is
\begin{equation}
\mathcal{M}^{ee\gamma^+\gamma^+} = \frac{e^2M_e\lbrack34\rbrack^2\langle\mathbf{12}\rangle}{(t-M_e^2)(u-M_e^2)} \ .
 \label{eq:large z:M_eeA+A+}
\end{equation}
Similarly to the last subsection, we can see there are are many choices for momenta shift that will create a asymptotically vanishing amplitude.  This amplitude is simple enough that they can easily be read off.  Among the choices are 
\begin{eqnarray}
    \hat{\mathcal{M}}^{ee\gamma^+\gamma^+}_{\lbrack\mathbf{1,2}\rangle,\lbrack\mathbf{2,1}\rangle}
    &\xrightarrow[z \to \infty]{}
    \mathcal{O}\left(\frac{1}{z^3}\right)
    \\
    \hat{\mathcal{M}}^{ee\gamma^+\gamma^+}_{\lbrack3,4\rangle,\lbrack4,3\rangle}
    &\xrightarrow[z \to \infty]{}
    \mathcal{O}\left(\frac{1}{z^2}\right)
    \\
    \hat{\mathcal{M}}^{ee\gamma^+\gamma^+}_{\lbrack\mathbf{1},3\rangle,\lbrack\mathbf{1},4\rangle}
    &\xrightarrow[z \to \infty]{}
    \mathcal{O}\left(\frac{1}{z^2}\right)
    \\
    \hat{\mathcal{M}}^{ee\gamma^+\gamma^+}_{\lbrack\mathbf{2},3\rangle,\lbrack\mathbf{2},4\rangle}
    &\xrightarrow[z \to \infty]{}
    \mathcal{O}\left(\frac{1}{z^2}\right)
    \\
    \hat{\mathcal{M}}^{ee\gamma^+\gamma^+}_{\lbrack3,\mathbf{1}\rangle,\lbrack3,\mathbf{2}\rangle}
    &\xrightarrow[z \to \infty]{}
    \mathcal{O}\left(z\right)
    \\
    \hat{\mathcal{M}}^{ee\gamma^+\gamma^+}_{\lbrack4,\mathbf{1}\rangle,\lbrack4,\mathbf{2}\rangle}
    &\xrightarrow[z \to \infty]{}
    \mathcal{O}\left(z\right)
    \ ,
\end{eqnarray}
giving us potentially eight shifts that should enable us to obtain the correct amplitudes (see App.~\ref{app:eeA+A+}.)

For the process $e\bar{e}\gamma^+\gamma^-$, the correct amplitude is given by Eq.~(\ref{eq:M_eeA+A-}), and is
\begin{equation}
\mathcal{M}^{+-} = \frac{e^2\left(\lbrack\mathbf{1}3\rbrack\langle\mathbf{2}4\rangle+\langle\mathbf{1}4\rangle\lbrack\mathbf{2}3\rbrack\right)\lbrack3\lvert p_1\rvert4\rangle}{(t-M_e^2)(u-M_e^2)} \ .
\label{eq:large z:M_eeA+A-}
\end{equation}
Asymptotically vanishing shifts can be found, but are slightly different combinations this time, namely
\begin{eqnarray}
    \hat{\mathcal{M}}^{ee\gamma^+\gamma^-}_{\lbrack4,3\rangle}
    &\xrightarrow[z \to \infty]{}
    \mathcal{O}\left(\frac{1}{z^4}\right)
    \\
    \hat{\mathcal{M}}^{ee\gamma^+\gamma^-}_{\lbrack\mathbf{2},3\rangle,\lbrack4,\mathbf{2}\rangle}
    &\xrightarrow[z \to \infty]{}
    \mathcal{O}\left(\frac{1}{z}\right)
    \\
    \hat{\mathcal{M}}^{ee\gamma^+\gamma^-}_{\lbrack\mathbf{1,2}\rangle,\lbrack\mathbf{2,1}\rangle}
    &\xrightarrow[z \to \infty]{}
    \mathcal{O}\left(\frac{1}{z}\right)
    \\
    \hat{\mathcal{M}}^{ee\gamma^+\gamma^-}_{\lbrack3,4\rangle}
    &\xrightarrow[z \to \infty]{}
    \mathcal{O}\left(z^0\right)
    \\
    \hat{\mathcal{M}}^{ee\gamma^+\gamma^-}_{\lbrack\mathbf{1},3\rangle,\lbrack4,\mathbf{1}\rangle}
    &\xrightarrow[z \to \infty]{}
    \mathcal{O}\left(z^0\right)
    \\
    \hat{\mathcal{M}}^{ee\gamma^+\gamma^-}_{\lbrack\mathbf{2},4\rangle,\lbrack3,\mathbf{2}\rangle}
    &\xrightarrow[z \to \infty]{}
    \mathcal{O}\left(z\right)
    \\
    \hat{\mathcal{M}}^{ee\gamma^+\gamma^-}_{\lbrack\mathbf{1},4\rangle,\lbrack3,\mathbf{1}\rangle}
    &\xrightarrow[z \to \infty]{}
    \mathcal{O}\left(z^2\right)
    \ ,
\end{eqnarray}
giving us five choices (see App.~\ref{app:eeA+A-}.)

\section{\label{app:Calc AHH Spinor Amps}Constructive Diagram Calculations With a Momentum and Spinor Shift}
In App.~\ref{app:Calc AHH Spinor Amps No Shifts}, we calculated the amplitudes constructively, but did not directly use a momentum or spinor shift.  Without the complexification of the momenta, we were unable to obtain the correct result for the process $e\bar{e}\mu\bar{\mu}$.  However, we did find the correct amplitude for the processes with an external photon, namely $e\gamma\bar{e}h$ and $e\bar{e}\gamma\gamma$.  In App.~\ref{app:analytic continuation}, we developed the analytic continuation of the momenta and the resulting shifts of the spinors.  In this appendix, we would like to reanalyze the constructive amplitudes and see what effect the shifts have on these amplitudes.  Unfortunately, we will still not find agreement with Feynman diagrams for $e\bar{e}\mu\bar{\mu}$, with the internal photon, but we will still find agreement when the photon is external in $e\gamma\bar{e}h$ and $e\bar{e}\gamma\gamma$.  Although we already have the correct amplitude from massless-photon constructive calculations for the last two, we perform the shift anyway to gain confidence in the shifts and to gain a greater appreciation of how they work.

\subsection{\label{sec:Constructive Massless eemm}$\mathbf{e,\bar{e},\mu,\bar{\mu}}$}

Since we already began this calculation in App.~\ref{sec:Constructive Massless eemm no Spinor Shift}, we begin with Eqs.~(\ref{eq:M_eemm-constructive [][]}) and (\ref{eq:M_eemm-constructive <><>}).  As we might expect, it does not matter which of these forms we use when we apply the momentum shift.  That is, given a momentum shift, we get the same final amplitude for either of the final forms of App.~\ref{sec:Constructive Massless eemm no Spinor Shift}.  It does matter, on the other hand, which momentum shift we use.  
However, the final amplitudes fall into two classes, which are conjugates of each other, so that the squares are the same for all the shifts.  As we pointed out in App.~\ref{app:eemm large z}, none of the momentum shifts lead to a vanishing amplitude in the large $z$ limit, therefore, we should not expect that any of these shifts will lead to the correct form of the amplitude.  Indeed, we have tried all of them on all of the forms from App.~\ref{sec:Constructive Massless eemm no Spinor Shift} and none of them lead to agreement with Feynman diagram and they all grow quadratically at high energy.  Nevertheless, we will give the details of two examples that lead to the two classes.

For the momentum complexification step, we must choose two momenta to complexify, say $p_i$ and $p_j$, and then we must choose whether to do the shift $\lbrack\mathbf{i},\mathbf{j}\rangle$ or $\lbrack\mathbf{j},\mathbf{i}\rangle$.  In the all-massless case, there are rules that determine which complexifications to use, beyond the fact that they have to be separated by the propagator.  However, in our massive case, we do not have clear guidance, therefore,  we will consider all possible momentum complexifications and analyze the results.  Our only restriction is that the momenta must be separated by the propagator.  Let us illustrate with a $\lbrack\mathbf{1,3}\rangle$ shift and perform it on Eq.~(\ref{eq:M_eemm-constructive [][]}), which, in shifted form, is
\begin{align}
    \mathcal{M} =
    &\ \frac{e^2}{2m_em_\mu s}  \Big[
    (\hat{u}-t+2m_e^2+2m_\mu^2)
    \lbrack\hat{\mathbf{1}}\mathbf{2}\rbrack 
    \lbrack\mathbf{3}\mathbf{4}\rbrack
    \nonumber\\
    &+2\left(
    \lbrack\hat{\mathbf{1}}\mathbf{2}\rbrack \lbrack\mathbf{3}\lvert p_2\hat{p}_1\rvert\mathbf{4}\rbrack+
    \lbrack\hat{\mathbf{1}}\lvert p_4\hat{p}_3\rvert\mathbf{2}\rbrack \lbrack\mathbf{34}\rbrack 
    \right)
    \Big]\ .
\end{align}
According to Eqs.~(\ref{eq:[ihat|=-1/mi<i|pk}) and (\ref{eq:|jhat>=1/mjpl|j>}),
\begin{eqnarray}
    \lbrack\hat{\mathbf{1}}\rvert &=&
    -\frac{1}{m_e}\langle\mathbf{1}\rvert p_2
    \\
    \lvert\hat{\mathbf{3}}\rangle &=&
    \frac{1}{m_\mu}p_4\lvert\mathbf{3}\rbrack\ .
\end{eqnarray}
All other spinors are unchanged.  Moreover, as we find in Eqs.~(\ref{eq:pi=-pk complexification}) and (\ref{eq:pj=-pl complexification}), 
\begin{eqnarray}
    \hat{p}_1 &=& - p_2
    \\
    \hat{p}_3 &=& -p_4
    \ .
\end{eqnarray}
This is the end result of following the complexification procedure and plugging in the final value for $z$.

We now need to apply this complexification to both the momenta in $u$ and in the spinor products.  Beginning with $u$,
\begin{eqnarray}
    \hat{u} &=&
    m_e^2+m_\mu^2+2\hat{p}_1\cdot p_4
    \nonumber\\
    &=& m_e^2+m_\mu^2 - 2p_2\cdot p_4
    \nonumber\\
    &=& 2m_e^2+2m_\mu^2 - t
    \nonumber\\
    &=& s+u
    \ ,
\end{eqnarray}
where we used $s+t+u=2m_e^2+2m_\mu^2$ in the last line.  If we further use the on-shell condition, $s=0$, we have $\hat{u}=u$.  We next work on $\lbrack\hat{\mathbf{1}}\mathbf{2}\rbrack$, 
\begin{align}
    \lbrack\hat{\mathbf{1}}\mathbf{2}\rbrack &= 
    -\frac{1}{m_e}\langle\mathbf{1}\lvert p_2\rvert\mathbf{2}\rbrack
    \nonumber\\
    &=
    \langle\mathbf{12}\rangle\ .
\end{align}
Following this, we note
\begin{align}
    \lbrack\hat{\mathbf{1}}\lvert p_4\hat{p}_3\rvert\mathbf{2}\rbrack 
    &= \frac{1}{m_e}\langle\mathbf{1}\lvert p_2 p_4^2\rvert\mathbf{2}\rbrack
    \nonumber\\
    &= -m_\mu^2\langle\mathbf{12}\rangle\ .
\end{align}
Finally,
\begin{align}
    \lbrack\mathbf{3}\lvert p_2\hat{p}_1\rvert\mathbf{4}\rbrack
    &= - \lbrack\mathbf{3}\lvert p_2^2\rvert\mathbf{4}\rbrack
    \nonumber\\
    &= -m_e^2\lbrack\mathbf{34}\rbrack\ .
\end{align}
Plugging these all in, we have
\begin{align}
    \mathcal{M}_{\lbrack\mathbf{1,3}\rangle} =
    &\frac{e^2}{2m_em_\mu s}  \Big[
    \left(u-t+2m_e^2+2m_\mu^2\right)
    \langle\mathbf{1}\mathbf{2}\rangle
    \lbrack\mathbf{3}\mathbf{4}\rbrack
    \nonumber\\
    &-2\left(m_\mu^2\langle\mathbf{12}\rangle\lbrack\mathbf{34}\rbrack
    +m_e^2\langle\mathbf{12}\rangle\lbrack\mathbf{34}\rbrack\right)
    \Big]\ .
\end{align}
However, we can now see that all the spinor products are the same and we are left with
\begin{align}
    \mathcal{M}_{\lbrack\mathbf{1,3}\rangle,\lbrack\mathbf{2,4}\rangle,\lbrack\mathbf{1,4}\rangle,\lbrack\mathbf{2,3}\rangle} =
    &\frac{e^2(u-t)}{2m_em_\mu s}  
    \langle\mathbf{1}\mathbf{2}\rangle
    \lbrack\mathbf{3}\mathbf{4}\rbrack
    \ .
    \label{eq:M_eemm [1,3> Shift}
\end{align}
We get the same final form for the shifts $\lbrack\mathbf{2,4}\rangle,\lbrack\mathbf{1,4}\rangle$ and $\lbrack\mathbf{2,3}\rangle$.

On the other hand, if we do a $\lbrack\mathbf{3,1}\rangle$ shift, we begin with
\begin{align}
    \mathcal{M} =
    &\ \frac{e^2}{2m_em_\mu s}  \Big[
    (\hat{u}-t+2m_e^2+2m_\mu^2)
    \lbrack\mathbf{1}\mathbf{2}\rbrack 
    \lbrack\hat{\mathbf{3}}\mathbf{4}\rbrack
    \nonumber\\
    &+2\left(
    \lbrack\mathbf{12}\rbrack \lbrack\hat{\mathbf{3}}\lvert p_2\hat{p}_1\rvert\mathbf{4}\rbrack+
    \lbrack\mathbf{1}\lvert p_4\hat{p}_3\rvert\mathbf{2}\rbrack \lbrack\hat{\mathbf{3}}\mathbf{4}\rbrack 
    \right)
    \Big]\ .
\end{align}
For this shift, we have
\begin{align}
    \lbrack\hat{\mathbf{3}}\rvert &= -\frac{1}{m_\mu}\langle\mathbf{3}\rvert p_4
    \\
    \lvert\hat{\mathbf{1}}\rangle &= \frac{1}{m_e}p_2\lvert\mathbf{1}\rbrack
    \\
    \hat{p}_3 &= -p_4
    \\
    \hat{p}_1 &= -p_2\ .
\end{align}
As we can see this is very similar to the $\lbrack\mathbf{1,3}\rangle$ shift, but switching square brackets and angle brackets.  
Plugging these in, we have $\hat{u}=u$ (assuming $s=0$ again).  We also have
\begin{align}
    \lbrack\hat{\mathbf{3}}\mathbf{4}\rbrack
    &= -\frac{1}{m_\mu}\langle\mathbf{3}\lvert p_4\rvert\mathbf{4}\rbrack
    \nonumber\\
    &= \langle\mathbf{34}\rangle
\end{align}
and
\begin{align}
    \lbrack\mathbf{1}\lvert p_4\hat{p}_3\rvert\mathbf{2}\rbrack
    &= -\lbrack\mathbf{1}\lvert p_4^2\rvert\mathbf{2}\rbrack
    \nonumber\\
    &= -m_\mu^2\lbrack\mathbf{12}\rbrack
    \\
    \lbrack\hat{\mathbf{3}}\lvert p_2\hat{p}_1\rvert\mathbf{4}\rbrack
    &= \frac{1}{m_\mu}\langle\mathbf{3}\lvert p_4p_2^2\rvert\mathbf{4}\rbrack
    \nonumber\\
    &= -m_e^2\langle\mathbf{34}\rangle
    \ .
\end{align}
Plugging this all in, we see we once again only have one form of the spinor products, leaving us with our final result
\begin{align}
    \mathcal{M}_{\lbrack\mathbf{3,1}\rangle,\lbrack\mathbf{4,2}\rangle,\lbrack\mathbf{4,1}\rangle,\lbrack\mathbf{3,2}\rangle} =
    &\frac{e^2(u-t)}{2m_em_\mu s}  
    \lbrack\mathbf{12}\rbrack
    \langle\mathbf{34}\rangle
    \ .
    \label{eq:M_eemm [3,1> Shift}
\end{align}
This is the conjugate of Eq.~(\ref{eq:M_eemm [1,3> Shift}).  Both amplitudes clearly grow quadratically with energy and violate perturbative unitarity and do not agree with Feynman diagrams.  As we describe in App.~\ref{app:eemm large z}, the large-$z$ limit of this diagram does not vanish for any of the shifts.  Therefore, in some sense, it is not surprising that this procedure is unsuccessful.  New ingredients are apparently necessary to achieve this amplitude with a massless photon.

It is interesting to note that, although the result given for the amplitude in Eq.~(5.44) of \cite{Arkani-Hamed:2017jhn} is different than our result before shifting the momenta, it turns out that they are the same after shifting.  To see this, we recall that their result is $\mathcal{M} = e^2/s*(u-t)/(2m_em_\mu)    *\langle\mathbf{12}\rangle\langle\mathbf{34}\rangle$ and consider the shift $\lbrack\mathbf{1,3}\rangle$.  Their amplitude becomes $\hat{\mathcal{M}} = e^2/s*(\hat{u}-t)/(2m_em_\mu) *\langle\mathbf{12}\rangle\langle\hat{\mathbf{3}}\mathbf{4}\rangle$.  However, we have already shown that $\hat{u}=u$ (on shell) and $\langle\hat{\mathbf{3}}\mathbf{4}\rangle = \lbrack\mathbf{34}\rbrack$, leaving us with Eq.~(\ref{eq:M_eemm [1,3> Shift}).  In fact, whichever shift we perform on their result, we end with Eqs.~(\ref{eq:M_eemm [1,3> Shift}) and (\ref{eq:M_eemm [3,1> Shift}), the same as if we begin with our amplitudes.  Moreover, the same would be true whether they ended with angle brackets $\langle\mathbf{12}\rangle\langle\mathbf{34}\rangle$, square brackets $\lbrack\mathbf{12}\rbrack\lbrack\mathbf{34}\rbrack$, or mixed brackets $\langle\mathbf{12}\rangle\lbrack\mathbf{34}\rbrack$ or $\lbrack\mathbf{12}\rbrack\langle\mathbf{34}\rangle$.  No matter which of these forms we begin with in their amplitude, we would end with Eqs.~(\ref{eq:M_eemm [1,3> Shift}) and (\ref{eq:M_eemm [3,1> Shift}) after a momentum shift.

\subsection{\label{sec:Constructive Massless heAe}$\mathbf{e,\gamma^+,\bar{e},h}$}
We began this calculation in App.~\ref{sec:Constructive Massless heAe no Shift}.  We found the initial amplitude was given by
\begin{align}
    \mathcal{M}_s^{+} &=
    \frac{em_e}{v}\frac{x_{1,34}
    (2m_e\langle\mathbf{13}\rangle
    -\langle\mathbf{1}\lvert p_4\rvert\mathbf{3}\rbrack)
    }
    {s-m_e^2}
    \\
    \mathcal{M}^+_u 
    &= \frac{em_e}{v}\frac{x_{14,3}
    (2m_e\langle\mathbf{13}\rangle+\lbrack\mathbf{1}\lvert p_4\rvert\mathbf{3}\rangle)
    }
    {u-m_e^2}
    \ .
\end{align}
This time we will replace $x_{1,34}$ and $x_{14,3}$ with Eqs.~(\ref{eq:x_1,34=}) and (\ref{eq:x_14,3=}) described in App.~\ref{sec:x = alternate}.  We include them here for convenience.
\begin{eqnarray}
    x_{1,34} &=&         \frac{\lbrack2|p_1p_3|2\rbrack}{m_e\left(u-m_e^2\right)}
    \\
    x_{14,3} &=& \frac{\lbrack2\lvert p_1p_3\rvert2\rbrack}{m_e(s-m_e^2)}\ .
\end{eqnarray}
As we can see, these forms of the $x$ factor immediately expose the positive helicity spinors of the photon.  Plugging these in, we obtain,
\begin{align}
    \mathcal{M}_s^{+} &=
    \frac{e}{v}\frac{\lbrack2\lvert p_1p_3\rvert2\rbrack
    (2m_e\langle\mathbf{13}\rangle
    -\langle\mathbf{1}\lvert p_4\rvert\mathbf{3}\rbrack)
    }
    {\left(s-m_e^2\right)\left(u-m_e^2\right)}
    \\
    \mathcal{M}^+_u 
    &= \frac{e}{v}\frac{\lbrack2\lvert p_1p_3\rvert2\rbrack
    (2m_e\langle\mathbf{13}\rangle+\lbrack\mathbf{1}\lvert p_4\rvert\mathbf{3}\rangle)
    }
    {\left(s-m_e^2\right)\left(u-m_e^2\right)}
    \ .
\end{align}

At this point, we discuss complexification of the momenta and the associated spinor shifts.  If we analytically continue two of the momenta as described in App.~\ref{app:complexification 2}, then asymptotic behavior of this amplitude is given in App.~\ref{app:large z:eAeh} by Eqs.~(\ref{eq:large z:M_eA+eh [4,2>}) through (\ref{eq:large z:M_eA+eh [2,3>}).  As described at the beginning of App.~\ref{app:analytic continuation}, if the amplitude vanishes in the large $z$ limit, then we should get the correct amplitude after performing the momentum and spinor shifts.  We can see that the amplitude vanishes if we choose any of the shifts given by $\lbrack\mathbf{4},2\rangle, \lbrack\mathbf{1},\mathbf{3}\rangle, \lbrack\mathbf{3},\mathbf{1}\rangle, \lbrack2,\mathbf{4}\rangle, \lbrack\mathbf{3},2\rangle$ or $ \lbrack\mathbf{1},2\rangle$, written in order of how quickly they fall off with increasing $z$.

 In these calculations, we run into some apriori ambiguity and unknowns in the way we apply momentum complexification.  First of all, when we complexify momenta, the only momenta that are usually not complexified are the momenta in the original propagator denominator.  So, in the first diagram, we would not complexify $(s-m_e^2)$ in the denominator and in the second diagram, we would not complexify $(u-m_e^2)$.  However, normally, this does not apply to any other part of the diagram, so in the present context, we might wonder whether we should complexify the momenta in the propagator denomaintor coming from the other diagram, $(u-m_e^2)$ in the first diagram and $(s-m_e^2)$ in the second diagram.  Moreover, we know that $(s-m_e^2)(u-m_e^2)$ in the final denominator is correct and therefore, we cannot ruin either propagator denominator.  We might imagine that we need to choose different complexifications for each diagram that don't ruin the other propagator denominator.  But, it turns out to be simpler than this.  We find that we obtain the correct amplitude if we simply hold \textit{both} propagator denominators unchanged by the complexification and we only shift the numerator, including the part of the numerator that came with the other propagator denominator.  We further find that we only need one of the diagrams and not the other to obtain the correct amplitude.  We can not prove this is a general feature.  However, we do expect it to generalize beyond these amplitudes.
 
Of the spinor shifts that lead to a vanishing amplitude, we only have simple formulas for the shifts in the case of $\lbrack\mathbf{4},2\rangle, \lbrack2,\mathbf{4}\rangle, \lbrack\mathbf{3},2\rangle$ and $ \lbrack\mathbf{1},2\rangle$ (see App.~\ref{app:[i,j> both massive massive internal line} for a discussion of $\lbrack\mathbf{1},\mathbf{3}\rangle$ and $\lbrack\mathbf{3},\mathbf{1}\rangle$.)  Of these, we have had success with all but $\lbrack2,\mathbf{4}\rangle$ (we were not able to simplify the expression to the correct form.)  That leaves us with $\lbrack\mathbf{4},2\rangle, \lbrack\mathbf{3},2\rangle$ and $ \lbrack\mathbf{1},2\rangle$.  We will describe the calculation with $\lbrack\mathbf{4},2\rangle$ and do the s-channel diagram in detail.  The steps for the u-channel are nearly identical and the other shifts are similar and end with the same final result.
\begin{equation}
    \hat{\mathcal{M}}_s^{+} =
    \frac{e}{v}\frac{\lbrack2\lvert p_1p_3\rvert2\rbrack
    (2m_e\langle\mathbf{13}\rangle
    -\langle\mathbf{1}\lvert \hat{p}_4\rvert\mathbf{3}\rbrack)
    }
    {\left(s-m_e^2\right)\left(u-m_e^2\right)}
    \ .
\end{equation}
As we can see, only $\hat{p}_4$ is shifted, and we have left the entire denominator unchanged, making this a relatively simple case to describe.  Focusing for the moment on the term with $\hat{p}_4$, using Eq.~(\ref{eq:app:|ihat] def 3 index free}), we find
\begin{align}
    \langle\mathbf{1}\lvert \hat{p}_4\rvert\mathbf{3}\rbrack &=
    \langle\mathbf{14}^{\mathrm{I}}\rangle\lbrack\hat{\mathbf{4}}_{\mathrm{I}}\mathbf{3}\rbrack
    \nonumber\\
    &= \langle\mathbf{1}\lvert p_4\rvert\mathbf{3}\rbrack
    + \langle\mathbf{1}\lvert p_4\rvert 2\rbrack\lbrack2\mathbf{3}\rbrack
    \frac{(s-m_e^2)}{\lbrack 2\lvert p_4  p_3\rvert 2\rbrack}\ .
\end{align}
If we plug this in to the numerator, and use momentum conservation to give $\lbrack2\lvert p_1p_3\rvert2\rbrack=-\lbrack2\lvert p_4p_3\rvert2\rbrack=\lbrack2\lvert p_3p_4\rvert2\rbrack$, we get
\begin{align}
    \frac{\textrm{num}}{e} &=
    2m_e\langle\mathbf{13}\rangle\lbrack2\lvert p_3p_4\rvert2\rbrack
    - \langle\mathbf{1}\lvert p_4\rvert\mathbf{3}\rbrack\lbrack2\lvert p_3p_4\rvert2\rbrack
    \nonumber\\
    &+ 2p_3\cdot p_4 \langle\mathbf{1}\lvert p_4\rvert 2\rbrack\lbrack2\mathbf{3}\rbrack 
    + m_h^2 \langle\mathbf{1}\lvert p_4\rvert 2\rbrack\lbrack2\mathbf{3}\rbrack\ .
%    \nonumber\\
\end{align}
We next perform a Schouten identity on the second term, 
\begin{align}
    \langle\mathbf{1}\lvert p_4\rvert\mathbf{3}\rbrack\lbrack2\lvert p_3p_4\rvert2\rbrack &= -\lbrack2\mathbf{3}\rbrack\lbrack2\lvert p_3p_4p_4\rvert\mathbf{1}\rangle + \lbrack2\lvert p_4\rvert\mathbf{1}\rangle\lbrack2\lvert p_3p_4\rvert\mathbf{3}\rbrack 
    \nonumber\\
    &= -m_h^2\lbrack2\mathbf{3}\rbrack\lbrack2\lvert p_3\rvert\mathbf{1}\rangle 
    + 2p_3\cdot p_4\lbrack2\lvert p_4\rvert\mathbf{1}\rangle\lbrack2\mathbf{3}\rbrack 
    \nonumber\\
    &+ m_e \lbrack2\lvert p_4\rvert\mathbf{1}\rangle\lbrack2\lvert p_4\rvert\mathbf{3}\rangle\ .
\end{align}
Plugging this into the numerator, we have
\begin{align}
    \frac{\textrm{num}}{e} &=
    2m_e\langle\mathbf{13}\rangle\lbrack2\lvert p_3p_4\rvert2\rbrack
    - m_e \lbrack2\lvert p_4\rvert\mathbf{1}\rangle\lbrack2\lvert p_4\rvert\mathbf{3}\rangle
    \nonumber\\
    &+m_h^2\lbrack2\mathbf{3}\rbrack\lbrack2\lvert p_3\rvert\mathbf{1}\rangle  
    + m_h^2 \lbrack2\mathbf{3}\rbrack\lbrack2\lvert p_4\rvert\mathbf{1}\rangle  \ .
\end{align}
The last two terms can be combined and $p_3+p_4=-p_1-p_2$ can be used along with the mass identities to bring it to
\begin{align}
    \frac{\textrm{num}}{e} &=
    2m_e\langle\mathbf{13}\rangle\lbrack2\lvert p_3p_4\rvert2\rbrack
    - m_e \lbrack2\lvert p_4\rvert\mathbf{1}\rangle\lbrack2\lvert p_4\rvert\mathbf{3}\rangle
    \nonumber\\
    &-m_em_h^2\lbrack\mathbf{1}2\rbrack\lbrack2\mathbf{3}\rbrack   \ .
\end{align}
Finally, we Schouten transform half of the first term using $-m_e\langle\mathbf{13}\rangle\lbrack2\lvert p_3p_4\rvert2\rbrack = m_e\lbrack2\lvert p_4\rvert\mathbf{3}\rangle\lbrack2\lvert p_3\rvert\mathbf{1}\rangle -m_e \lbrack2\lvert p_4\rvert\mathbf{1}\rangle\lbrack2\lvert p_3\rvert\mathbf{3}\rangle$.  We next use momentum conservation on the first of these terms $p_3=-p_1-p_2-p_4$ followed by the mass identities to obtain $-m_e\langle\mathbf{13}\rangle\lbrack2\lvert p_3p_4\rvert2\rbrack = m_e^2\lbrack2\lvert p_4\rvert\mathbf{3}\rangle\lbrack2\mathbf{1}\rbrack -m_e\lbrack2\lvert p_4\rvert\mathbf{3}\rangle\lbrack2\lvert p_4\rvert\mathbf{1}\rangle +m_e^2 \lbrack2\lvert p_4\rvert\mathbf{1}\rangle\lbrack2\mathbf{3}\rbrack$.  Plugging this in, we have
\begin{align}
    \frac{\textrm{num}}{e} &=
    m_e\langle\mathbf{13}\rangle\lbrack2\lvert p_3p_4\rvert2\rbrack
    + m_e^2\lbrack\mathbf{1}2\rbrack \lbrack2\lvert p_4\rvert\mathbf{3}\rangle
    \nonumber\\
    &-m_e^2 \lbrack2\mathbf{3}\rbrack\lbrack2\lvert p_4\rvert\mathbf{1}\rangle
    -m_em_h^2\lbrack\mathbf{1}2\rbrack\lbrack2\mathbf{3}\rbrack   \ .
\end{align}
This agrees exactly with Eq.~(\ref{eq:M_eA+eh}).

\subsection{\label{app:eeA+A+}$\mathbf{e,\bar{e},\gamma^{\pm},\gamma^{\pm}}$}
We began this calculation in App.~\ref{app:eeA+A+ no shift} and found the initial amplitude to be given by
\begin{align}
    \mathcal{M}_t^{++} &=
    \frac{e^2 m_e x_{1,24}x_{13,2}\langle\mathbf{12}\rangle}{t-m_e^2} 
    \label{eq:M_t^++ initial x amp}
    \\
    \mathcal{M}_u^{++} &=
    -\frac{e^2 m_e x_{1,23}x_{14,2}\langle\mathbf{12}\rangle}{u-m_e^2}
    \ .
\end{align}
This time, for demonstration purposes, we choose to replace $x$ from among
\begin{align}
    x_{1,24} 
    &= \frac{\lbrack3\lvert p_1p_2\rvert3\rbrack}{m(u-m^2)}
    &=& \frac{\lbrack3\lvert p_1p_4\rvert3\rbrack}{m s}
    \label{eq:x_1,24}
    \\
    x_{13,2} 
    &= \frac{\lbrack4\lvert p_1p_2\rvert4\rbrack}{m(u-m^2)}
    &=& \frac{\lbrack4\lvert p_1p_3\rvert4\rbrack}{m s}
    \label{eq:x_13,2}
    \\
    x_{1,23} 
    &= \frac{\lbrack4\lvert p_1p_2\rvert4\rbrack}{m(t-m^2)}
    &=& \frac{\lbrack4\lvert p_1p_3\rvert4\rbrack}{m s}
    \label{eq:x_1,23}
    \\
    x_{14,2} 
    &= \frac{\lbrack3\lvert p_1p_2\rvert3\rbrack}{m(t-m^2)}
    &=& \frac{\lbrack3\lvert p_1p_4\rvert3\rbrack}{m s}
    \label{eq:x_14,2}
    \ .
\end{align}
We can see that the numerators are related by momentum conservation and the denominators are also if we take the propagator of the diagram to be on shell.  That is, the top two expressions are for the t-channel diagram where the on-shell condition is $t=m^2$, resulting in $s=-t-u+2m^2=-u+m^2$.  The last two are similar, since they are used in the u-channel diagram.  Therefore, we can use either the middle or the right expressions for $x$, since we can use the on-shell condition with momentum conservation during the simplification.

As described in App.~\ref{app:large z:eeAA}, the shifts that result in vanishing large $z$ behavior are $\lbrack\mathbf{1,2}\rangle, \lbrack\mathbf{2,1}\rangle, \lbrack\mathbf{1},3\rangle, \lbrack\mathbf{2},3\rangle, \lbrack\mathbf{1},4\rangle, \lbrack\mathbf{2},4\rangle, \lbrack3,4\rangle$ and $\lbrack4,3\rangle$.  Of these, we only have a simple momentum shift for the last six.  We have succeeded in obtaining agreement with Feynman diagrams with all six of these momentum shifts, however, each is simplest with its own replacement for $x$.  The shifts $\lbrack\mathbf{1},3\rangle$ and $\lbrack\mathbf{2},4\rangle$ only work for the u-channel diagram while the shifts $\lbrack\mathbf{1},4\rangle$ and $\lbrack\mathbf{2},3\rangle$ only work for the t-channel diagram, because the two momenta shifted must be on opposite sides of the diagrams propagator.  The shifts $\lbrack3,4\rangle$ and $\lbrack4,3\rangle$, on the other hand, work for either diagram and we have gotten the same agreeing result using either.

Although we could use any of these shifts, some are certainly simpler to carry out than others.  Namely, the fewer things that are shifted, the less algebraic simplification required to obtain the final result.  We will only demonstrate one of these, namely the $\lbrack\mathbf{1},4\rangle$ shift on the t-channel diagram.  After replacing $x$ and shifting, we have
\begin{equation}
    \hat{\mathcal{M}}_t^{++} = 
    \frac{e^2 \langle\mathbf{12}\rangle \lbrack3\lvert \hat{p}_1p_2\rvert3\rbrack\lbrack4\lvert \hat{p}_1p_2\rvert4\rbrack }{m_e\left(t-m_e^2\right)\left(u-m_e^2\right)^2} \ ,
\end{equation}
where, once again, we only shift the numerator after replacing $x$ (whether we do this shift or another.)   Next, we use the momentum shift for particle $1$ from Eq.~(\ref{eq:app:|ihat] def 3 index free}),
\begin{equation}
    \hat{p}_1 =
    \lvert\hat{\mathbf{1}}_\mathrm{I}\rbrack\langle\mathbf{1}^{\mathrm{I}}\rvert 
    = 
    p_1 - \frac{ \left(t-m_e^2\right) }{
    \lbrack 4\lvert p_1  p_3\rvert 4\rbrack}
    \lvert 4\rbrack\lbrack 4\lvert p_1\ .
\end{equation}
When we plug this into $\lbrack4\lvert \hat{p}_1p_2\rvert4\rbrack$, the second term vanishes due to $\lbrack44\rbrack=0$, so we are only left with the change to $\lbrack3\lvert \hat{p}_1p_2\rvert3\rbrack$.  When we plug this in, we will also use the identity $\left(t-m_e^2\right) = 2p_1\cdot p_3$, giving
\begin{align}
    \lbrack3\lvert \hat{p}_1p_2\rvert3\rbrack\lbrack4\lvert \hat{p}_1p_2\rvert4\rbrack
    =& 
    \lbrack3\lvert p_1p_2\rvert3\rbrack\lbrack4\lvert p_1p_2\rvert4\rbrack
    \nonumber\\
    &- \frac{ 2p_1\cdot p_3 }{
    \lbrack 4\lvert p_1  p_3\rvert 4\rbrack}
    \lbrack34\rbrack\lbrack 4\lvert p_1p_2\rvert3\rbrack\lbrack 4\lvert p_1p_2\rvert4\rbrack .
\end{align}
Using momentum conservation and the masslessness of the photon, $\lbrack 4\lvert p_1p_2\rvert4\rbrack = -\lbrack 4\lvert p_1p_3\rvert4\rbrack$, allowing the cancellation of the denominator in the right term.  We can also use momentum conservation  to obtain $\lbrack 4\lvert p_1p_2\rvert3\rbrack = -m_e^2\lbrack43\rbrack-2p_1\cdot p_4 \lbrack 43\rbrack = u\lbrack34\rbrack$, leaving us with
\begin{align}
    \lbrack3\lvert \hat{p}_1p_2\rvert3\rbrack\lbrack4\lvert \hat{p}_1p_2\rvert4\rbrack
    =& 
    \lbrack3\lvert p_1p_2\rvert3\rbrack\lbrack4\lvert p_1p_2\rvert4\rbrack
    \nonumber\\
    &+  2p_1\cdot p_3
    u \lbrack34\rbrack^2.
\end{align}
We can use a Schouten identities on the first term, $\lbrack3\lvert p_1p_2\rvert3\rbrack\lbrack4\lvert p_1p_2\rvert4\rbrack = \lbrack43\rbrack\lbrack4\lvert p_1p_2p_2p_1\rvert3\rbrack - \lbrack4\lvert p_2p_1\rvert3\rbrack\lbrack4\lvert p_1p_2\rvert3\rbrack = m_e^4\lbrack34\rbrack^2 + 2p_1\cdot p_2 u \lbrack34\rbrack^2 + u^2\lbrack34\rbrack^2.$  Combining, and using $2p_1\cdot p_2+2p_1\cdot p_3 = -2m_e^2-2p_1\cdot p_4 = -u-m_e^2$, we have
\begin{align}
    \lbrack3\lvert \hat{p}_1p_2\rvert3\rbrack\lbrack4\lvert \hat{p}_1p_2\rvert4\rbrack
    &= \lbrack34\rbrack^2
    \left[m_e^4-\left(u+m_e^2\right)u + u^2\right]
    \nonumber\\
    &= -m_e^2\left(u-m_e^2\right)\lbrack34\rbrack^2.
\end{align}
Plugging this in, the extra $(u-m_e^2)$ cancels and we end with
\begin{equation}
   \mathcal{M}_t^{++} = 
    -\frac{e^2 m_e \langle\mathbf{12}\rangle \lbrack34\rbrack^2 }
    {\left(t-m_e^2\right)\left(u-m_e^2\right)} \ ,
\end{equation}
in agreement with Eq.~(\ref{eq:M_eeA+A+}).

\subsection{\label{app:eeA+A-}$\mathbf{e,\bar{e},\gamma^{\pm},\gamma^{\mp}}$}
We began this amplitude in App.~\ref{app:eeA+A- no shift}, where we found
\begin{align}
    \mathcal{M}_t^{+-} &=
    \frac{e^2 x_{1,24}\tilde{x}_{13,2}
    \left(m_e\lbrack\mathbf{12}\rbrack
    +\langle\mathbf{1}\lvert p_3\rvert\mathbf{2}\rbrack
    \right)}
    {t-m_e^2}
    \label{eq:M_eeA+A- initial constructive}
    \
    \\
    \mathcal{M}_u^{+-} &=
    \frac{e^2 \tilde{x}_{1,23}x_{14,2}
    \left(
    m_e\langle\mathbf{12}\rangle
    +\lbrack\mathbf{1}\lvert p_4\rvert \mathbf{2}\rangle
    \right)}
    {u-m_e^2}\ .
\end{align}
As in the $++$-helicity case described in the previous subsection, when we replace $x$ and $\tilde{x}$, we have a choice of whether to replace both of them with the $u-m_e^2$ for the t-channel and $t-m_e^2$ for the u-channel diagram, or whether to replace one with this propagator denominator and the other with $s$.  If we look at App.~\ref{app:large z:eeAA}, we see that this amplitude vanishes for large $z$ if we do any of the shifts $\lbrack4,3\rangle, \lbrack\mathbf{2},3\rangle, \lbrack4,\mathbf{2}\rangle, \lbrack\mathbf{1,2}\rangle$ and $\lbrack\mathbf{2,1}\rangle$.  Of these, we have simple shift formulas for the first three.  
We have succeeded in obtaining agreement with Feynman diagrams (and with each other) using any of these shifts.  For our demonstration, we will consider the $\lbrack4,3\rangle$ shift of the u-channel diagram, therefore, we will replace $x$ and $\tilde{x}$ to obtain
\begin{equation}
    \hat{\mathcal{M}}_u^{+-} =
    \frac{e^2 
    \lbrack3\lvert p_1p_2\rvert3\rbrack
    \langle4\lvert p_1\hat{p}_3\rvert4\rangle
    \left(
    m_e\langle\mathbf{12}\rangle
    +\lbrack\mathbf{1}\lvert \hat{p}_4\rvert \mathbf{2}\rangle
    \right)}
    {m_e^2\ s \left(t-m_e^2\right)\left(u-m_e^2\right)}
    \ ,
\end{equation}
where the replacements of $\tilde{x}$ are exactly the same as for $x$ except that square brackets are replaced with angle brackets (the derivations follow exactly the same steps) and, therefore, Eqs.~(\ref{eq:x_1,23}) and (\ref{eq:x_14,2}) can be used.  

Since we are doing a shift with both external particles massless but a massive internal line, we use Eqs.~(\ref{eq:ihat massless bypass z internal M}) and (\ref{eq:jhat massless bypass z internal M}) for $\hat{p}$.  The shift of $\hat{p}_3$ in this amplitude is particularly simple since $\langle4\lvert p_1\hat{p}_3\rvert4\rangle = \langle4\lvert p_1p_3\rvert4\rangle+\langle4\lvert p_1\rvert3\rbrack\langle44\rangle\left(u-m_e^2\right)/\langle4\lvert p_1\rvert3\rbrack = \langle4\lvert p_1p_3\rvert4\rangle$ since $\langle44\rangle=0$.  On the other hand, $\lbrack\mathbf{1}\lvert \hat{p}_4\rvert \mathbf{2}\rangle = \lbrack\mathbf{1}\lvert p_4\rvert \mathbf{2}\rangle - \lbrack\mathbf{1}3\rbrack\langle4\mathbf{2}\rangle 2p_1\cdot p_4/\langle4\lvert p_1\rvert3\rbrack$.  In order to cancel the $\langle4\lvert p_1\rvert3\rbrack$ in the denominator, we first use momentum conservation, $\lbrack3\lvert p_1p_2\rvert3\rbrack \langle4\lvert p_1p_3\rvert4\rangle = -\lbrack3\lvert p_1p_2\rvert3\rbrack \langle4\lvert p_1p_2\rvert4\rangle$, followed by Schouten transformation, $\lbrack3\lvert p_1p_2\rvert3\rbrack \langle4\lvert p_1p_2\rvert4\rangle = \langle4\lvert p_1\rvert3\rbrack\langle4\lvert p_1p_2p_1\rvert3\rbrack + \langle4\lvert p_1\rvert3\rbrack \langle4\lvert p_1p_2p_2\rvert3\rbrack$, where we replaced $p_2=-p_1-p_3-p_4$ and used masslessness in the first term.  We further use $p_1p_2=2p_1\cdot p_2-p_2p_1$ in the first term to obtain $\lbrack3\lvert p_1p_2\rvert3\rbrack \langle4\lvert p_1p_3\rvert4\rangle = -\langle4\lvert p_1\rvert3\rbrack\left(2p_1\cdot p_2 \langle4\lvert p_1\rvert3\rbrack-m_e^2\langle4\lvert p_2\rvert3\rbrack+m_e^2\langle4\lvert p_1\rvert3\rbrack\right)$.  Using momentum conservation in the middle term gives finally, $\lbrack3\lvert p_1p_2\rvert3\rbrack \langle4\lvert p_1p_3\rvert4\rangle = -\langle4\lvert p_1\rvert3\rbrack^2\left(2p_1\cdot p_2 +2m_e^2\right) = -\langle4\lvert p_1\rvert3\rbrack^2 s$, which also gives us the $s$ to cancel this factor in the denominator.  Plugging this in, we have
\begin{align}
    \frac{\mbox{num}\left(\mathcal{M}_u^{+-}\right)}{e^2\lbrack3\lvert p_1\rvert4\rangle} &=
    -m_e\langle\mathbf{12}\rangle\lbrack3\lvert p_1\rvert4\rangle
    -\lbrack\mathbf{1}\lvert p_4\rvert \mathbf{2}\rangle\lbrack3\lvert p_1\rvert4\rangle 
    \nonumber\\
    &+ 2p_1\cdot p_4\lbrack\mathbf{1}3\rbrack\langle4\mathbf{2}\rangle 
    \ .
\end{align}
We Schouten transform the second term, $\lbrack\mathbf{1}\lvert p_4\rvert \mathbf{2}\rangle\lbrack3\lvert p_1\rvert4\rangle = -\langle4\mathbf{2}\rangle\lbrack3\lvert p_1p_4\rvert\mathbf{1}\rbrack$, followed by reodering the momenta giving $\lbrack\mathbf{1}\lvert p_4\rvert \mathbf{2}\rangle\lbrack3\lvert p_1\rvert4\rangle = -2p_1\cdot p_4 \langle4\mathbf{2}\rangle\lbrack3\mathbf{1}\rbrack -m_e\langle4\mathbf{2}\rangle\lbrack3\lvert p_4\rvert\mathbf{1}\rangle$.  Using momentum conservation on this second term, $-m_e\langle4\mathbf{2}\rangle\lbrack3\lvert p_4\rvert\mathbf{1}\rangle = -m_e^2\langle4\mathbf{2}\rangle\lbrack3\mathbf{1}\rbrack + m_e\langle4\mathbf{2}\rangle\lbrack3\lvert p_2\rvert\mathbf{1}\rangle$.  Applying a Schouten identity to this last term gives $ m_e\langle4\mathbf{2}\rangle\lbrack3\lvert p_2\rvert\mathbf{1}\rangle = m_e\langle\mathbf{12}\rangle\lbrack3\lvert p_2\rvert4\rangle + m_e^2\langle\mathbf{1}4\rangle\lbrack3\mathbf{2}\rbrack$.   Finally, using momentum conservation on the first of these and plugging everything in, we get
\begin{align}
    \frac{\mbox{num}\left(\mathcal{M}_u^{+-}\right)}{e^2\lbrack3\lvert p_1\rvert4\rangle} &=
    m_e^2\langle4\mathbf{2}\rangle\lbrack3\mathbf{1}\rbrack 
    - m_e^2\langle\mathbf{1}4\rangle\lbrack3\mathbf{2}\rbrack
    \ .
\end{align}
The full amplitude is now given by
\begin{equation}
    \mathcal{M}_u^{+-} =
    \frac{e^2 \lbrack3\lvert p_1\rvert4\rangle
    \left(
    \langle\mathbf{2}4\rangle\lbrack\mathbf{1}3\rbrack 
    + \langle\mathbf{1}4\rangle\lbrack\mathbf{2}3\rbrack
    \right)
    }
    {\left(t-m_e^2\right)\left(u-m_e^2\right)}
    \ ,
\end{equation}
agreeing with Eq.~(\ref{eq:M_eeA+A-}).

\end{document}